\documentclass[%
preprint,
aps,
pra,
amsmath,
amssymb,
floatfix,
]{revtex4-2}
\usepackage{afterpage}
\usepackage{graphicx}
\usepackage{dcolumn}
\usepackage{bm}
\usepackage{hyperref}
\usepackage{braket}
\usepackage{siunitx}
\usepackage[normalem]{ulem}
\usepackage{hyperref}

\hypersetup{
    colorlinks=true,
    linkcolor=blue,
    filecolor=magenta,      
    urlcolor=cyan,
    pdftitle={Overleaf Example},
    }

\newcommand{\RN}[1]{%
  \textup{\uppercase\expandafter{\romannumeral#1}}%
  }
\usepackage{tikz}

\newcommand{\dbarw}[1]{d\hspace*{-0.08em}\bar{}\hspace*{0.1em}\omega_{#1}}
\DeclareMathOperator{\sinc}{sinc}
\newcommand{\sbar}[1]{\bar{\sigma}_{#1}(t)}
\newcommand{\fbar}[2]{\bar{F}^{#1}_{#2}(t)}
\newcommand{\fhat}[2]{\hat{F}^{#1}_{#2}(t)}
\newcommand{\sbari}[1]{\bar{\sigma}_{#1}(t_1)}
\newcommand{\fhati}[2]{\hat{F}^{#1}_{#2}(t_1)}
\newcommand{\1}{\omega_\mathrm{I}}
\newcommand{\2}{\omega_\mathrm{II}}
\newcommand{\one}{_\mathrm{I}}
\newcommand{\two}{_\mathrm{II}}

\usepackage{comment}

\numberwithin{equation}{section}
\begin{document}

\title{Fluorescence driven by nonclassical light}
\author{C. Drago}
\email{christian.drago@mail.utoronto.ca}
\author{J. E. Sipe}
\email{sipe@physics.utoronto.ca}

\affiliation{Department of Physics, University of Toronto, 60 St. George Street, Toronto, Ontario, Canada, M5S 1A7}
\date{\today}

\begin{abstract}
We investigate whether or not irradiation by squeezed light can provide an enhancement of the two-photon excitation of a system over irradiation by classical light. Our emphasis 
is not only on whether or not there is such an enhancement, but also on whether or not any enhancement  
can be reasonably detected in an experiment. We begin by developing a model that includes radiative and nonradiative broadening to calculate the total scattered and absorbed energy. As an example calculation, we consider cesium atoms in a magneto-optical trap, and evaluate the fluorescence emission when driven by non-degenerate classical and squeezed light, in both the continuous-wave and pulsed regimes. We find that squeezed light can provide an enhancement in both regimes under ideal circumstances. 
These enhancements are in principle detectable. However, we stress that they are moderate at best compared to recently reported values for molecular systems. 
\end{abstract}

\maketitle
\section{Introduction \label{sec:Introduction}}
Entangled photon pairs have proved to be useful for various sensing applications \cite{pirandola2018advances, kolobov2007quantum} as well as a resource for quantum communication \cite{kaiser2016fully}, quantum cryptography \cite{grosshans2002continuous}, and quantum computation \cite{bourassa2021blueprint}. With advances in the development of nonlinear materials in both bulk and integrated devices \cite{arnbak2019compact,quesada2022beyond, dutt2024nonlinear}, squeezed states -- which include many photon pairs -- are now experimentally accessible, and it is natural to ask what advantage they might bring when applied to other applications. 

One recent topic of interest is the use of squeezed light to drive nonlinear optical processes. In the 1980s, it was first suggested that squeezed light with a large spectral-temporal correlation would lead to a linear scaling of a two-photon transition with the incident number of photons, and would provide an enhancement over the use of classical light where the scaling is quadratic \cite{klyshko1982transverse, gea1989two, javanainen1990linear}; shortly after this was presumed confirmed by experiments with Cesium in a magneto-optical-trap \cite{georgiades1995nonclassical}. Later, other applications were considered, including using squeezed light to coherently control a two-photon transition as well as to drive sum frequency generation; both were experimentally successful \cite{dayan2004two, dayan2005nonlinear}.  Other applications involve the generation of high harmonics \cite{gorlach2023high}.

Recently, attention has shifted back to driving two-photon transitions with squeezed light, but for the excitation of fluorescent molecules in solution \cite{schlawin2018entangled, schlawin2017entangled, dorfman2016nonlinear, dayan2007theory, fei1997entanglement, schlawin2013photon, oka2018two, lee2006entangled}. The renewed interest has been motivated by the possibility that the enhancement 
squeezed light might provide could be useful in various bio-sensing applications, such as two-photon fluorescent microscopy \cite{ashworth2007two} or photo-dynamic cancer therapies \cite{bhawalkar1997two}, where high intensity light could destroy or injure the sample \cite{podgorski2016brain}. Indeed, different groups predict a large enhancement and purport to have verified it experimentally \cite{tabakaev2021energy, tabakaev2022spatial, villabona2018two, li2020squeezed, villabona2017entangled, varnavski2017entangled}; in two publications they have gone as far as proposing to apply this phenomenon to fluorescence microscopy \cite{varnavski2020two, varnavski2022quantum}.

Despite the experimental claims for enhanced fluorescence from dyes in the low photon number regime \cite{schlawin2018entangled, schlawin2017entangled, dorfman2016nonlinear, dayan2007theory, fei1997entanglement, schlawin2013photon, oka2018two, lee2006entangled, tabakaev2021energy, tabakaev2022spatial, villabona2018two, li2020squeezed, villabona2017entangled, varnavski2017entangled}, recently a growing number of experimental groups either have not measured a signal, or can provide alternative explanations to describe the signal they have measured \cite{PhysRevApplied.15.044012, mikhaylov2022hot, hickam2022single, landes2024limitations, Raymer:21}; this is in agreement with recent theory calculations in the appropriate regime \cite{drago2023two, raymer2022theory, landes2021quantifying}. The conflicting results of different experimental groups is due to many different factors. For example, initial theoretical work \cite{schlawin2018entangled, schlawin2017entangled, dorfman2016nonlinear, dayan2007theory, fei1997entanglement, schlawin2013photon} focused on calculating the ratio between excitation probabilities due to the excitation by 
squeezed and classical light in the low photon number limit, but did not calculate the resulting fluorescence emission rates for a specific sample; these rates decrease as the incident photon number is reduced, thereby making the enhancement difficult to detect. Further, typical fluorescent dyes are large complex molecules with many degrees of freedom, which makes it challenging to properly model and experimentally analyze their response to optical excitation even in the linear regime. Interactions 
between molecules in solution may cause aggregation, which modifies cross-sections and could be leading to observed enhancements \cite{rocha2024reviewing}. An added experimental complication is that the usual transmission (or attenuation) experiments can be inappropriately used to infer a linear scaling of two-photon absorption, because other relevant processes also scale linearly. For example, one-photon transitions that can scatter light also scale linearly, and are often the dominant contribution \cite{hickam2022single, PhysRevA.106.023115}. To really distinguish the linear scaling of two-photon absorption from linear one-photon loss mechanisms, a careful measurement of the output photon statistics must be carried out \cite{Mazurek:19,martinez2023witnessing}. 

Given the controversy in this field, we feel it is useful to ``go back to basics,'' and in Section \ref{sec:model Hamiltonian} we begin with a model Hamiltonian of an arbitrary system interacting with the quantized electromagnetic field. When light interacts with the system -- for example, an atom, molecule, or quantum dot -- excited states of the system will be involved, and the excitation will generally depend on the incident field's energy, center frequency, and bandwidth, as well as the system's linewidth and other parameters. Ultimately the system can relax to its ground state or to any other lower energy state. For a single system isolated from its environment, the linewidth is solely due to the radiative contribution -- often referred to as the ``natural linewidth’’ that leads to the ``Einstein A coefficient'' if we consider single photon excitation -- where energy is emitted back into the electromagnetic field. This is referred to as \textit{scattered light}, and if 
one is interested in detecting some of it, typically over a particular frequency range, that light of interest is often referred to as \textit{fluorescence}.  

In reality we deal with a large number of systems -- such as a collection of atoms in a gas -- and other degrees of freedom of the system besides that associated with the optical excitation are relevant. The resulting linewidth consists of two types of contributions: homogeneous and inhomogeneous broadening. Excitation decay is responsible for homogeneous broadening, and it includes both the above-mentioned radiative decay and decay processes that are non-radiative in nature. The latter convert energy to degrees of freedom not involving the electromagnetic field; in its strictest use of the word, this is referred to as \textit{absorption}, and we adopt this usage here. Such broadening mechanisms are present even when the system is isolated from its environment due to the conversion of energy to vibrational and/or rotational degrees of freedom.

On the other hand, inhomogeneous broadening is due to some statistical fluctuations of the system. This can be Doppler broadening, transit time broadening, or broadening due to elastic collisions or dipole-dipole interactions \cite{milonni2010laser}. Such broadening mechanisms are present even when the system of interest does not lose energy. Whatever the root cause, broadening mechanisms can significantly affect the results of experiments, and often cannot be removed \cite{drago2023two}. 

Our goal here is to calculate results that could be observed in the laboratory, and the focus of experiments involving excitation by squeezed light is often the resulting fluorescence. Thus our aim is to determine not only if squeezed light provides an enhancement, but crucially if that enhancement can be reasonably expected to be experimentally detectable. Given the present difficulties in measuring two-photon excitation driven by squeezed light, we consider one example of a system where inhomogeneous broadening mechanisms can be largely neglected: atoms in a Magneto-Optical Trap (MOT) \cite{georgiades1994two}.

To accomplish our goal, we include in our Hamiltonian terms to model homogeneous broadening effects -- but neglect inhomogeneous contributions -- and in Section \ref{sec:absorption(s)} we discuss the inclusion of radiation reaction to consistently model the total scattered and absorbed energy. Then in sections \ref{sec:Input-output theory} - \ref{sec:perturbative solution} we introduce the interaction picture in which we work, focus on a four level system (referred to as the ``atom’’), and explicitly calculate the expected fluorescence.

When considering classical and squeezed light two-photon transitions, the center frequencies of the incident light can be in various regimes: the optical excitation can be either off resonance or on resonance, and it can be either degenerate or non-degenerate. When comparing excitation by classical and squeezed light, the consequences of being in one or another of these regimes can be very different. For example, CW squeezed light may be very broadband (on the order of THz) and so even if the center frequencies are very far detuned and off resonance, resonant contributions are still possible. As well, while resonant contributions may increase the rate of two-photon transitions for squeezed light  \cite{PhysRevA.106.023115}, they can also increase the usual single photon scattering events that are often much larger and also scale linearly with photon number \cite{hickam2022single, PhysRevA.106.023115}. Further, the photon statistics of incident squeezed light depend on whether the excitation is degenerate or non-degenerate, and this can impact the results of experiments. Finally, there are other factors that modify the excitation rate, including the spatial properties of the incident light \cite{schlawin2024two} and the dispersion it suffers \cite{raymer2021entangled}.

In our analysis we explore a regime not often considered, which is excitation in the resonant non-degenerate limit. In Section \ref{sec:perturbative solution} we apply perturbation theory to calculate the fluorescence count for a distribution of ``fixed atoms'' excited by an incident Gaussian spatial mode of either classical or squeezed light. In past theoretical treatments the focus has been primarily on CW squeezed light in the low photon rate limit (i.e., pairs of photons), but interesting physics exists beyond this regime, and it may restrict the fluorescence enhancement that could be observed. It is well known that in the high photon rate limit (i.e., squeezed states) there exists two contribution that drive the excitation -- they are named the ``coherent'' and ``incoherent'' contributions -- and they both scale quadratically with the incident photon rate; this was originally showed by Dayan, and more recently considered by others \cite{dayan2007theory,raymer2021entangled, PhysRevA.106.023115}.  The ``coherent'' contribution is the generalization of the low photon rate result and involves correlated photons, while the ``incoherent'' contribution involves photons that are not correlated, and it resembles chaotic (or thermal) excitation of the system. It is often assumed that because the incoherent contribution involves uncorrelated photons it is negligible, but recent work on molecular systems has shown the contrary \cite{PhysRevA.106.023115}; indeed, in this work we show that the magnitude of the incoherent contribution is very system dependent. For experiments this poses a problem in the high photon number limit, because one cannot easily separate the two contributions. The pulsed regime, in which both contributions are also present, is also interesting because the sought-after correlation between photons is actually \emph{reduced} in the high photon number limit, thereby increasing the “incoherent” contribution. 

To explore all these regimes, in Sections \ref{sec:CW Coherent state} and \ref{sec:Entangled States of Light} we derive equations for the fluorescence count when the system is excited with classical and squeezed light for both CW and pulsed sources; our results are valid in both the low and high photon number regimes. To determine the viability of measuring the squeezed light excitation in an experiment, in Section \ref{sec:example system} we focus on atomic Cesium in a MOT as an example system, and in Section \ref{sec:Classical and squeezed light fluorescence} we calculate the fluorescence emission count and provide an analysis of the results. We end with concluding remarks in Section \ref{sec:conculsion}; some technical details are presented in appendices. 

\section{model Hamiltonian \label{sec:model Hamiltonian}}
We begin by modeling an arbitrary system, with a Hamiltonian denoted by $H_\text{S}$, interacting with an electromagnetic field and a quantum reservoir. The free electromagnetic field Hamiltonian, $H_\text{EM}$, is given by
\begin{equation}
    \label{eq:HEM}
    H_\text{EM} =  \sum_{s}\int d\boldsymbol{k} \hbar\omega_{k} a_s^\dagger(\boldsymbol{k})a_s(\boldsymbol{k}),
\end{equation}
with $k = |\boldsymbol{k}|$, where 
\begin{equation}
    [a_s(\boldsymbol{k}),a_{s^\prime}^\dagger(\boldsymbol{k}^\prime)] = \delta(\boldsymbol{k} - \boldsymbol{k}^\prime)\delta_{ss^\prime};
\end{equation}
here the subscript `$s$' labels the polarization. We treat the coupling between the system and the field within the electric dipole approximation, neglecting certain terms that lead to renormalization effects \cite{craig1998molecular}, and take the interaction Hamiltonian  to be
\begin{equation}
    H_\text{S-EM} = -\boldsymbol{\mu} \cdot \boldsymbol{E}(\boldsymbol{r}_0),
    \label{eq:H_M-EM v1}
\end{equation}
where $\boldsymbol{\mu}$ is the dipole moment operator and $\boldsymbol{E}(\boldsymbol{r}_0)$ is the electric field operator at the position $\boldsymbol{r}_0$ of the system. These two operators commute, so we are free to order them in any way. In what follows we will often drop the position dependence of the electric field operator for convenience; but in Section \ref{sec:solving the equations of motion}, when the system response is explicitly calculated, we will return to the spatial dependence of the electric field operator. 

We expand each operator in terms of its positive and negative frequency components $\boldsymbol{E} = \boldsymbol{E_+} + \boldsymbol{E_-}$ and $\boldsymbol{\mu} = \boldsymbol{\mu_+} + \boldsymbol{\mu_-}$, with $(\boldsymbol{E_+})^\dagger = \boldsymbol{E_-}$, $(\boldsymbol{\mu_+})^\dagger = \boldsymbol{\mu_-}$ and assume the relevant incident field frequency components are close enough to resonances that the rotating-wave approximation (RWA) is valid; within this approximation the interaction is given by
\begin{equation}
    H_{\text{S-EM}} = -\boldsymbol{\mu_-}\cdot\boldsymbol{E_+} - \boldsymbol{E_-}\cdot \boldsymbol{\mu_+}.
    \label{eq:H_M-EM v3}
\end{equation}
In expanding $H_{\text{S-EM}}$ we have chosen the normally ordered form, which is the simplest to work with; however, the results we derive are independent of the order chosen \cite{PhysRevA.106.023115}.

In most experiments, the system is not isolated from the environment, and interactions lead to subsequent energy loss in the form of non-radiative decay of optically excited electronic states, resulting in a loss of energy from the electromagnetic field, as well as in a loss of coherence. To model the system interacting with a reservoir that leads to a loss of energy and coherence, we include the Hamiltonian $H_\text{R}$ for the reservoir and $H_\text{S-R}$ for the interaction between the system and reservoir. Both of these are taken to be general at this point; a concrete model is provided in later sections. Although here we neglect the inclusion of inhomogeneous broadening, its effects have been treated within the formalism we present here \cite{PhysRevA.106.023115}, and more extensively in the broader literature \cite{raymer2021entangled, mukamel1983nonimpact}.

Then the full Hamiltonian for the system-field-reservoir is 
\begin{equation}
    \label{eq:Hfull}
    H = H_0 + V,
\end{equation}
where
\begin{equation}
    \label{eq:H_0}
    H_0 = H_\text{S}  + H_\text{EM} + H_\text{R},
\end{equation}
is the free Hamiltonian and 
\begin{equation}
    \label{eq:V(t)}
    V = H_\text{S-EM} + H_\text{S-R},
\end{equation}
includes the interaction terms. 

\section{absorption(s)}
\label{sec:absorption(s)}
The term \emph{absorption} is used in a number of different ways in optics. In a simple perturbative calculation of the interaction of a pulse of light with a system initially in its ground state, the probability that the system would be found in an excited state after the pulse passes is sometimes referred to as the \emph{probability of absorption}. But the system will eventually return to its ground state, and that might involve the emission of light. In another usage, \emph{absorption} is taken to refer \emph{only} to the removal of energy from the electromagnetic field; any redistribution of that light with respect to direction or frequency is then referred to as \emph{scattering}. We adopt this stricter usage here, where we address absorption and the associated processes of scattering and extinction.

To set the stage, consider first a classical problem where a charge $e$ is harmonically bound to an origin, subject to an incident electric field $\boldsymbol{E}_0(t)$, and as well to a ``damping'' force $-\Gamma^\prime \boldsymbol{v}$, where $\boldsymbol{v}$ is the velocity of the charge \cite{jackson2021classical}. In seeking the dynamics of the dipole moment, $\boldsymbol{\mu}(t) = e\boldsymbol{r}(t)$, the charge must be allowed to respond to the total electric field, $\boldsymbol{E}(t)$, which includes the ``radiation reaction'' field due to the dipole itself,
\begin{equation}
    \label{eq: full E classical}
    \boldsymbol{E}(t) = \boldsymbol{E}_0(t) + \frac{1}{6\pi\epsilon_0c^3}\frac{d^3\boldsymbol{\mu}(t)}{dt^3},
\end{equation}
where the second term is the electric field of the dipole at the position of the dipole, neglecting divergent contributions that in a more sophisticated calculation would be associated with renormalization. The use of Eq. \eqref{eq: full E classical} in formulating the dynamics of $\boldsymbol{\mu}(t)$ leads to the so-called
Abraham-Lorentz equation \cite{jackson2021classical}. When the response
of $\boldsymbol{\mu}(t)$ to an incident field $\boldsymbol{E}_0(t)$ is then calculated for incident fields at different frequencies, the linewidth of the dipole response acquires a contribution both from $\Gamma^\prime$ and from the radiation reaction term, the latter providing the radiative contribution to the linewidth.

Since the rate at which a force $\boldsymbol{F}$ does work on a particle
with velocity $\boldsymbol{v}$ is $\boldsymbol{F}\cdot \boldsymbol{v}$, in our classical sketch the rate at which the electromagnetic field does work on the dipole is given by $\boldsymbol{E}\cdot e\boldsymbol{v} = \boldsymbol{E}\cdot d\boldsymbol{\mu}/dt$, and thus the work
done on the dipole by the electric field \eqref{eq: full E classical} from time
$t_I$ to time $t_F$ is given by
\begin{equation}
\label{eq: classical absorption}
    \mathcal{A} = \int\limits_{t_I}^{t_F}\boldsymbol{E}(t)\cdot \frac{d\boldsymbol{\mu}(t)}{dt}dt.
\end{equation}
We have labelled this by $\mathcal{A}$ because, since $\boldsymbol{E}(t)$ is the \emph{full} electric field, this should be the total energy removed
from the electromagnetic field, i.e., the \emph{absorption}. And indeed, if a calculation of Eq. \eqref{eq: classical absorption} is made with $\Gamma^\prime = 0$,
we find $\mathcal{A} = 0$.

Now taking the full expression for the electric field in Eq. \eqref{eq: full E classical} we expand the right-hand side of the absorption in Eq. \eqref{eq: classical absorption} and find the absorption is given by $\mathcal{A} = \mathcal{E} - \mathcal{S}$ where 
\begin{equation}
    \mathcal{E} = \int\limits_{t_I}^{t_F} \boldsymbol{E}_0(t)\cdot \frac{d\boldsymbol{\mu}(t)}{dt}dt,
\end{equation}
and
\begin{equation}
    \mathcal{S} = \frac{1}{6\pi\epsilon_0 c^3}\int\limits_{t_I}^{t_F}\frac{d^2\boldsymbol{\mu}(t)}{dt^2}\cdot \frac{d^2\boldsymbol{\mu}(t)}{dt^2}dt.
\end{equation}
The first term $\mathcal{E}$ is the work that the \emph{incident} field does on the dipole. If the incident field is idealized as a plane
wave in the region of the dipole it is easy to confirm \cite{hulst1981light} that $\mathcal{E}$ is the energy removed from the incident field, and thus characterizes its \emph{extinction}. Absorption can contribute to this, of course, because it removes energy from the incident electromagnetic field. But energy that is still in the electromagnetic field but scattered in different directions can contribute as well, and indeed $\mathcal{S}$ is the total energy that is radiated in all directions by a dipole between $t_I$ and $t_F$. Thus, rearranging our expression to $\mathcal{E} = \mathcal{A} + \mathcal{S}$, we confirm that the energy removed from the incident field is equal to the energy absorbed and the energy scattered, as expected from energy conservation.

This sketch has dealt with a linear system and proceeded classically, but if we consider a quantum treatment with our Hamiltonian $H = H_0 + V$ we find that the energy absorbed is given by an expression similar to the classical Eq. \eqref{eq: classical absorption}. Working in the Heisenberg picture within the RWA, the energy the system absorbs is given by
\begin{equation}
    \label{eq: quantum absorption}
    \mathcal{A} = \int\limits_{t_I}^{t_F}\bra{\Psi(t_0)}\boldsymbol{E}_-^H(t)\cdot\frac{d\boldsymbol{\mu}_+^H(t)}{dt}\ket{\Psi(t_0)}dt + \text{c.c.},
\end{equation}
where $\ket{\Psi(t_0)}$ is the initial ket ($t_0 < t_I$) of the system-field-reservoir, and the superscript $H$ indicates that the operators evolve according to the full Hamiltonian. We note that because we chose the normal ordering of the interaction term in Eq. \eqref{eq:H_M-EM v1} the end result involves the expectation value of normally ordered operators; however, the results for the operators symmetrically ordered, normally ordered, and anti-normally ordered are all identical \cite{PhysRevA.106.023115}.

To solve for $\boldsymbol{E}_\pm^H(t)$ there are two strategies. The first is to work with an expansion of the transverse electromagnetic field operators in modes of the system, solve the Heisenberg picture equations for the mode operators, and assemble the results to find an expression for $\boldsymbol{E}_\pm^H(t)$. This was done by Dalibard et al. \cite{dalibard1982vacuum}, and in their treatment they had to employ a renormalization scheme to avoid divergences. When this was done a ``quantum Abraham-Lorentz equation" was derived.

A second strategy is to note that the quantized Maxwell equations take the same form as their classical versions, and the radiation reaction term in Eq. \eqref{eq: full E classical} arises from the transverse part of the electromagnetic field at the site of the system that is out-of-phase with the dipole moment. So if the other, divergent terms are ignored \emph{ab initio} we have
\begin{equation}
    \label{eq:E^H(t)}
    \boldsymbol{E}_\pm^H(t) = \boldsymbol{E}_\pm^{H_0}(t) + \frac{1}{6\pi\epsilon_0c^3}\frac{d^3\boldsymbol{\mu}_\pm^H(t)}{dt^3},
\end{equation}
where the superscript $H_0$ and $H$ denote the time evolution in the respective Heisenberg pictures. This agrees with the results of Dalibard et al. \cite{dalibard1982vacuum} when the renormalization terms they encounter are ignored.

Inputting the solution of the field operator (Eq. \eqref{eq:E^H(t)}) into Eq. \eqref{eq: quantum absorption}  for the absorption, and following what was done in the classical sketch, we can write $\mathcal{A} = \mathcal{E} - \mathcal{S}$ where
\begin{equation}
    \label{eq: quantum extinction}
    \mathcal{E} = \int\limits_{t_I}^{t_F}\bra{\Psi(t_0)}\boldsymbol{E}_-^{H_0}(t)\cdot\frac{d\boldsymbol{\mu}_+^H(t)}{dt}\ket{\Psi(t_0)}dt + \text{c.c.},
\end{equation}
and
\begin{equation}
\label{eq: quantum scattering}
    \mathcal{S} = \frac{1}{6\pi\epsilon_0 c^3}\int\limits_{t_I}^{t_F}\bra{\Psi(t_0)}\frac{d^2\boldsymbol{\mu}_-^H(t)}{dt^2}\cdot\frac{d^2\boldsymbol{\mu}_+^H(t)}{dt^2}\ket{\Psi(t_0)}dt + \text{c.c.}.
\end{equation}
Again we can immediately identify $\mathcal{E}$ and $\mathcal{S}$ as the extinction and scattering respectively, and by rearranging we have $\mathcal{E} = \mathcal{A} + \mathcal{S}$ as expected.

We point out that each equation for $\mathcal{E}, \mathcal{A}$ and $\mathcal{S}$ is exact within the RWA and the neglect of divergent terms that can be removed by renormalization, although we have yet to solve for $\boldsymbol{\mu}_\pm^H(t)$ and $\boldsymbol{E}_\pm^H(t)$. To solve for them we will resort to perturbation theory, but first we present the form of the interaction picture we will use.

\section{Interaction Picture \label{sec:Input-output theory}}
In our full Hamiltonian \eqref{eq:Hfull}, $V$ will always be written as a function of a set of Schr\"{o}dinger operators $\left\{ O_{\alpha}\right\}$, $V\left(\left\{ O_{\alpha}\right\}\right)$; these operators may correspond to either the system, field, or reservoir. The time evolution of a ket from $t_{a}$ to $t_{b}$ in the absence of $V\left(\left\{ O_{\alpha}\right\}\right)$ is 
described by the action of the unitary evolution operator $\mathcal{U}_{0}(t_{b},t_{a})$ that satisfies the Schr\"{o}dinger equation
\begin{equation}
  i\hbar\frac{d}{dt_{b}}\mathcal{U}_{0}(t_{b},t_{a})=H_{0}\mathcal{U}_{0}(t_{b},t_{a}),
\end{equation}
and has a solution 
\begin{equation}
  \mathcal{U}_{0}(t_{b},t_{a})=e^{-iH_{0}(t_{b}-t_{a})/\hbar}\,
\end{equation}
that satisfies the usual initial conditions of interest, $\mathcal{U}_{0}(t_{a},t_{a})=\hat{1}$ for all $t_{a}$. Including $V\left(\left\{ O_{\alpha}\right\}\right)$, the full
evolution operator $\mathcal{U}(t_{b},t_{a})$ satisfies the equation
\begin{equation}
  i\hbar\frac{d}{dt_{b}}\mathcal{U}(t_{b},t_{a})=H\mathcal{U}(t_{b},t_{a}),
\end{equation}
with the initial conditions $\mathcal{U}(t_{a},t_{a})=\hat{1}$ for
all $t_{a}$.

Now consider times $t_\text{min}$ and $t_\text{max}$ such that for $t \leq t_\text{min}$ or $t \geq t_\text{max}$ the interaction between the electromagnetic field and the system vanishes. In general, of course, such times do not exist; even when there is no pulse of light, if initially the system were in its ground state -- or any other eigenstate of $H_\text{S}$ -- it would interact with the quantized electromagnetic field. However, for the RWA form of the Hamiltonian we adopt, such ``vacuum state'' interactions vanish \cite{PhysRevA.106.023115}. So for times $t_{a},t_{b}\leq t_\text{min}$ or $t_{a},t_{b}\geq t_\text{max}$ we have
\begin{equation}
  \mathcal{U}(t_{a},t_{b})=\mathcal{U}_{0}(t_{a},t_{b}).\label{eq:UUnought}
\end{equation}

It is then useful to introduce the operator 
\begin{equation}
  U(t_{b},t_{a})=\mathcal{U}_{0}(0,t_{b})\mathcal{U}(t_{b},t_{a})\mathcal{U}_{0}(t_{a},0),\label{eq:Udef}
\end{equation}
which satisfies all the properties of a unitary time evolution operator, including the condition that $U(t_{a},t_{a})=\hat{1}$ for all $t_{a}$. We can then write 
\begin{equation}
  \left|\Psi(t)\right\rangle =\mathcal{U}_{0}(t,0)U(t,t_0)\left|\Psi_\text{in}\right\rangle ,\label{eq:psiF}
\end{equation}
where the ``asymptotic-in'' ket 
\begin{equation}
  \left|\Psi_\text{in}\right\rangle \equiv\mathcal{U}_{0}(0,t_0)\left|\Psi(t_0)\right\rangle,
\end{equation}
is what an initial ket $|\Psi(t_0)\rangle$ would evolve to at $t=0$ were there no interaction $V$, with $t_0<t_I,t_\text{min}$ the initial time (see Eq. \eqref{eq: quantum absorption}). Note that since $t_0$ is long before any interaction occurs, $\left|\Psi_\text{in}\right\rangle$ is independent of $t_0$. From the definition (\ref{eq:Udef}) we can show, using (\ref{eq:UUnought}), that for $t_{a},t_{a}'\leq t_\text{min}$ we have $U(t,t_{a})=U(t,t_{a}'$), and thus both must be equal to $U(t,-\infty)$; if we put 
\begin{equation}
  U(t)\equiv U(t,-\infty),\label{eq:Utdef}
\end{equation}
we can write (\ref{eq:psiF}) as 
\begin{equation}
  \left|\Psi(t)\right\rangle =\mathcal{U}_{0}(t,0)U(t)\left|\Psi_\text{in}\right\rangle.
\end{equation}

We now move to an interaction picture, and begin by introducing the time evolution of each Schr\"{o}dinger operator $O_{\alpha}$ according to $H_0$, 
\begin{equation}
  \hat{O}_{\alpha}(t)\equiv\mathcal{U}_{0}(0,t)O_{\alpha}\mathcal{U}_{0}(t,0),\label{eq:OHdef}
\end{equation}
with an initial time of zero. For our operators of interest we will have, schematically, 
\begin{equation}
  \hat{O}_{\alpha}(t)=g_{\alpha}(t)O_{\alpha},\label{eq:OHexpression}
\end{equation}
where the $g_{\alpha}(t)$ are (classical) functions of time. For example, if $O_{\alpha}$ is $a_s(\boldsymbol{k})$
then
\begin{equation}
\begin{gathered}
  \hat{a}_{s}(\boldsymbol{k},t)=e^{-i\omega_{k}t}a_{s}(\boldsymbol{k}).
\end{gathered}
\end{equation}
With the expressions (\ref{eq:OHdef},\ref{eq:OHexpression}), and
the definitions of $\mathcal{U}_{0}(t_{a},t_{b})$, $\mathcal{U}(t_{a},t_{b})$,
and $U(t)$, it is easy to confirm that the expectation value of any operator $O_\alpha$ satisfying Eq. \eqref{eq:OHexpression} is
\begin{equation}
\label{eq:expectation value}
    \bra{\Psi(t)}O_\alpha\ket{\Psi(t)} = \bra{\Psi_\text{in}}\bar O_\alpha(t)\ket{\Psi_\text{in}},
\end{equation}
where
\begin{equation}
\begin{split}
  \bar{O}_{\alpha}(t)&=U^{\dagger}(t)\mathcal{U}_{0}(0,t)O_{\alpha}\mathcal{U}_{0}(t,0)U(t)\label{eq:bar_from_caret}\\
  &=g_{\alpha}(t)\check{O}_{\alpha}(t),
\end{split}
\end{equation}
and
\begin{equation}
  \check{O}_{\beta}(t)\equiv U^{\dagger}(t)O_{\beta}U(t),\label{eq:caret_def}
\end{equation}
which is the time evolution including the interaction with the start time $t = -\infty$. To relate everything back to the full Heisenberg picture we have
\begin{equation}
    \begin{split}
        O^H_\alpha(t) = \mathcal{U}_0(t_0,0)\bar{O}_\alpha(t)\mathcal{U}_0(0,t_0),
    \end{split}
\end{equation}
then
\begin{equation}
    \begin{split}
        \bra{\Psi(t)}O_\alpha\ket{\Psi(t)} &= \bra{\Psi_\text{in}}\bar O_\alpha(t)\ket{\Psi_\text{in}}\\
        & = \bra{\Psi(t_0)}O^H_\alpha(t)\ket{\Psi(t_0)}.
    \end{split}
\end{equation}

Since the interaction between the system and field only occurs for times $t$ within the ``interaction interval'' $t_\text{min}\le t \le t_\text{max}$, as long as $t_0<t_I<t_\text{min}$ and $t_F>t_\text{max}$, we can extend the limits of integration for the absorption, extinction, and scattered energy (Eq. \eqref{eq: quantum absorption}, \eqref{eq: quantum extinction}, \eqref{eq: quantum scattering}) to range from negative to positive infinity, because none of them acquire contributions from those additional times.

To solve for the expectation value of an operator, we could use Eq. \eqref{eq:expectation value} combined with Eq. \eqref{eq:bar_from_caret} and solve for $g_\alpha(t)$ due to the free evolution, and $\check O _\alpha(t)$ due to the interaction. Equivalently, we could also choose to work directly with the operators $\bar O _\alpha(t)$ and form the dynamical equations in terms of them. Since it is the latter that naturally appears in the equations, we choose to work with the operators $\bar O_\alpha(t)$.

To determine their dynamics we start with the complicated part due to their dependence on the $\check{O}_{\alpha}(t)$ (see Eq. \eqref{eq:bar_from_caret}), the evolution of which follows from the evolution of $U(t)$. From its definition (\ref{eq:Utdef}) the dynamical equation for $U(t)$ can be found,
\begin{equation}
  i\hbar\frac{d}{dt}U(t)=V\left(\left\{ \hat{O}_{\alpha}(t)\right\}\right)U(t),
\end{equation}
which leads to the dynamical equation for $\check{O}_{\alpha}(t),$
\begin{equation}
  i\hbar\frac{d}{dt}\check{O}_{\alpha}(t)=\left[\check{O}_{\alpha}(t),V\left(\left\{ \bar{O}_{\alpha}(t)\right\}\right)\right].\label{eq:dynamics Ocheck}
\end{equation}
Specifying now to the system operators we will consider, each $g_\alpha(t)$ that will arise is of the form
\begin{equation}
    \label{eq:g_alpha}
    g_\alpha(t) = e^{-i\Omega_\alpha t}
\end{equation}
for some constant $\Omega_\alpha$. Then the equation for the dynamics of $\check{O}_{\alpha}(t)$ (Eq. \eqref{eq:dynamics Ocheck}) is equivalent to
\begin{equation}
  i\hbar\left(\frac{d}{dt} + i\Omega_\alpha\right)\bar{O}_{\alpha}(t)=\left[\bar{O}_{\alpha}(t),V\left(\left\{ \bar{O}_{\alpha}(t)\right\}\right)\right],\label{eq:dynamics Obar}
\end{equation}
which has the same structure as the Heisenberg equations for $O^H_\alpha(t)$. However, while the ket relevant for expectation values of the $O^H_\alpha(t)$ is $\left|\Psi(t_0)\right\rangle$, the ket relevant for $\bar{O}_{\alpha}(t)$ is $\left|\Psi_\text{in}\right\rangle$. And while the $O^H_\alpha(t)$ is equal to the corresponding Schr\"{o}dinger operators $O_\alpha$ at $t_0$, it is the operators $\check{O}_{\alpha}(t)$ that equal the corresponding Schr\"{o}dinger operators $O_\alpha$ at $t=-\infty$. We return to this point below.

\section{Four level system \label{sec:simplesystem}}
\begin{figure}
    \centering
    \includegraphics[width = 0.25\linewidth]{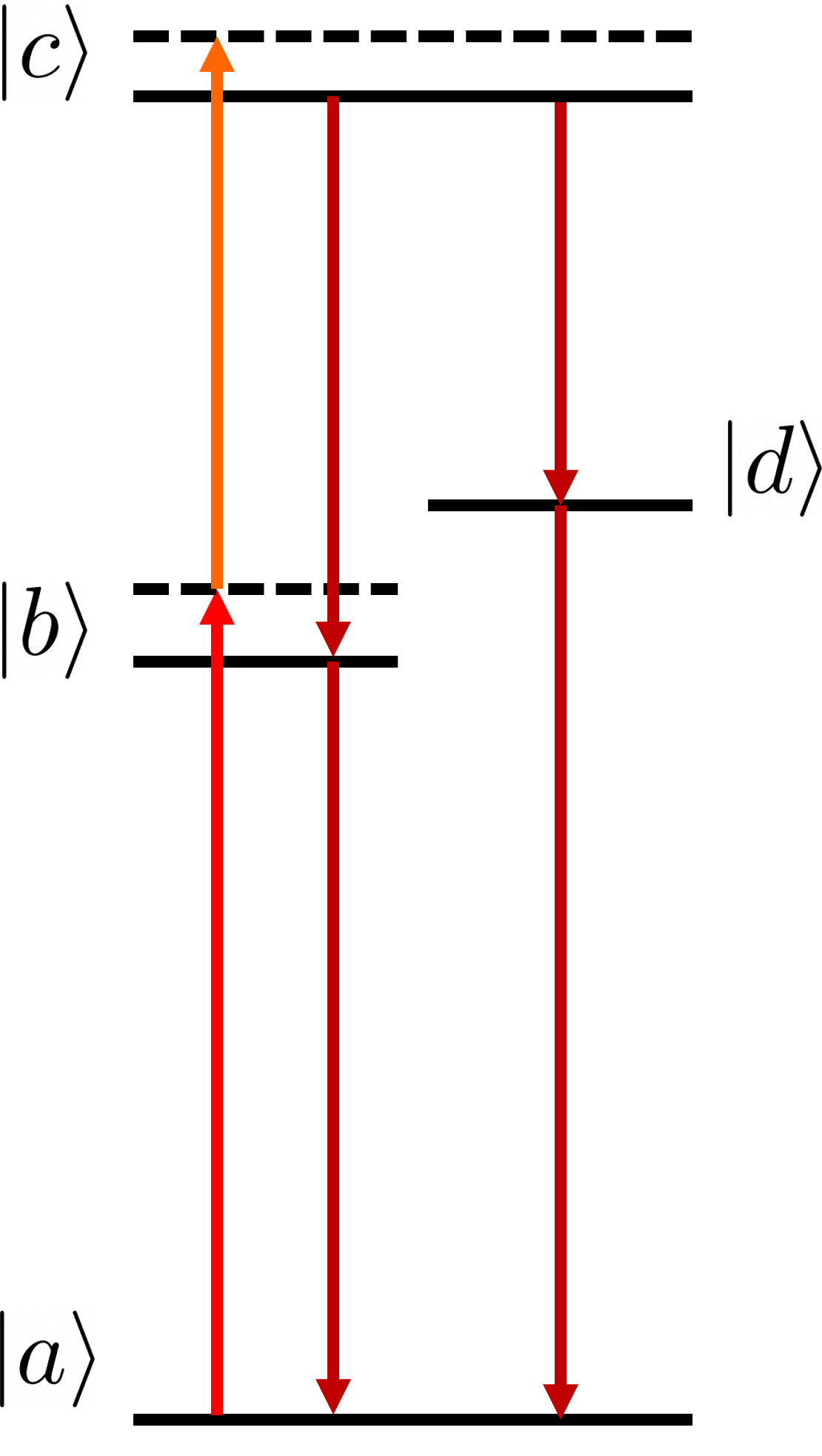}
    \caption{Schematic of the four level system we are considering. Upward arrows denote the pumping scheme we will consider below.}
    \label{fig:4lvlsystem}
\end{figure}

In this section, we specialize to a four level system. This serves the purpose of allowing us to derive further results in detail, but everything we consider here can be generalized to more complicated systems with many levels. For the four level system shown schematically in Fig. \ref{fig:4lvlsystem},  $\ket{a}$ is the ground state. While we will initially work generally and consider any pumping scheme, our focus will be on the specific situation where $\ket{c}$ is the final state in a two-photon excitation process (denoted by the upward pointing vertical lines in Fig. \ref{fig:4lvlsystem}) that involves an intermediate state $\ket{b}$, and after transition from $\ket{c}$ to another state $\ket{d}$ we envision fluorescence associated with the decay of the system from $\ket{d}$ to the ground state. We take the only dipole allowed transitions to be $\ket{a}\leftrightarrow\ket{b}$, $\ket{b}\leftrightarrow\ket{c}$, $\ket{c}\leftrightarrow\ket{d}$, and $\ket{d}\leftrightarrow\ket{a}$; in a later section we consider a concrete example and show that  for systems where parity is a good quantum number, this is often valid, at least to good approximation. Such a scenario is relevant to a wide variety of spectroscopic procedures where the fluorescence is measured during an experiment \cite{lichtman2005fluorescence}. 

Henceforth referring to this four-level scheme as the ``atom'' we have
\begin{equation}
    H_\text{S} = \hbar\omega_a \sigma_{aa} + \hbar\omega_b \sigma_{bb} + \hbar\omega_c \sigma_{cc} + \hbar\omega_d \sigma_{dd},
\end{equation}
where $\sigma_{pq} = \ket{p}\bra{q}$, and we assume these states can be taken to provide a complete set, 
\begin{equation}
    \label{eq:complete_relation}
    \hat 1 = \sigma_{aa} + \sigma_{bb} + \sigma_{cc} + \sigma_{dd}. 
\end{equation}
For the set of allowed dipole transitions, the dipole moment can be expanded as
\begin{equation}
\label{eq:muplus}
    \boldsymbol{\mu}_+ = \boldsymbol{\mu}_{ab}\sigma_{ab} + \boldsymbol{\mu}_{bc}\sigma_{bc} + \boldsymbol{\mu}_{dc}\sigma_{dc} + \boldsymbol{\mu}_{ad}\sigma_{ad},
\end{equation}
where each contribution to $\boldsymbol{\mu}_+$ corresponds to an allowed transition in Fig. \ref{fig:4lvlsystem}. 

Next we consider the situation where the energy level difference $\hbar\omega_{ij} \equiv \hbar\omega_i - \hbar\omega_j$ between each pair of states is sufficiently larger than the bandwidth(s) of the incident light that we can split the electric field operator $\boldsymbol{E}_\pm$ into a set of ``frequency bands'' 
\begin{equation}
    \label{eq: E freqband}
    \boldsymbol{E}_\pm = \boldsymbol{E}_\pm^{ba} + \boldsymbol{E}_\pm^{cb} + \boldsymbol{E}_\pm^{cd} + \boldsymbol{E}_\pm^{da},
\end{equation}
where, for example, $\boldsymbol{E}_\pm^{ba}$ is a field operator corresponding to the frequency band centered at $\omega_{ba}$, which is the transition frequency between $\ket{a}$ and $\ket{b}$, and contains all the frequency components that are important for that transition. This is often valid for atomic and molecular systems, and while this simplification is not necessary to calculate further results, we will see that when we include the radiation reaction terms this assumption leads to the very intuitive scenario that we discuss below. Since each field operator exists within its own frequency band they satisfy
\begin{equation}
    \left[\boldsymbol{E}_+^{pq},  \boldsymbol{E}_-^{rs}\right] = 0,
\end{equation}
for all transitions $pq$ and $rs$, and so we can treat them independently. 

Using the ``frequency-band'' expansion and the dipole moment expansion (Eq. \eqref{eq:muplus}), we expand the interaction $H_{\text{S-EM}}$ (Eq. \eqref{eq:H_M-EM v3}); within our assumptions, to good approximation it simplifies to
\begin{equation}
    \label{eq:H_M-EM v4}    
    \begin{split}
        H_{\text{S-EM}} =& -\boldsymbol{E}_-^{ba}\cdot \boldsymbol{\mu}_{ab}\sigma_{ab}-\boldsymbol{E}_-^{cb}\cdot \boldsymbol{\mu}_{bc}\sigma_{bc}\\
        &-\boldsymbol{E}_-^{da}\cdot \boldsymbol{\mu}_{ad}\sigma_{ad}-\boldsymbol{E}_-^{cd}\cdot \boldsymbol{\mu}_{dc}\sigma_{dc} + \text{H.c.},
    \end{split}
\end{equation}
since we are assuming that the RWA is applicable and so transitions of the system  between $\ket{p}$ and $\ket{q}$ ($p>q$) will only be mediated by fields contained in the corresponding operator $\boldsymbol{E}_\pm^{pq}$.

\section{quantum reservoir \label{sec:quantum reservoir}}
We now turn to the quantum reservoir and follow the treatment highlighted by  
Fisher \cite{fischer2018derivation} and used by others \cite{PhysRevA.106.023115, vernon2015spontaneous, vendromin2024highly, quesada2022beyond}, where the effects of a reservoir are described by treating it as a fictitious one-dimensional waveguide.  In this model, for a given transition the system is effectively always interacting with the vacuum state of the associated waveguide; this implements the physical assumption that the coherence generated between the system and reservoir decays quickly.

In our use of this approach, we assume there is an independent non-radiative decay pathway associated with each dipole allowed transition. Of course non-radiative decays are not limited by dipole selection rules, and one could include decay pathways involving all energy levels; however, since this will not lead to any new physics, and can easily be added at a later date, we do not include them at this time. So for each decay process depicted in Fig.  \ref{fig:4lvlsystem}, we introduce the waveguide ``reservoir operators" $\psi_{pq}(z)$ and $\psi^\dagger_{pq}(z)$, which will correspond to transitions between $\ket{p}$ and $\ket{q}$ ($p>q$) and we take the system ``position" of the atom to be at $z = 0$ \cite{PhysRevA.106.023115}. We assume each non-radiative decay is independent, which is equivalent to setting
\begin{equation}
    \left[\psi_{pq}(z),  \psi^\dagger_{rs}(z^\prime)\right] = \delta_{pr}\delta_{qs} \delta(z - z^\prime).
\end{equation}

Then the full reservoir Hamiltonian is
\begin{equation}
    H_\text{R} = H_\text{R}^{ba} + H_\text{R}^{cb} + H_\text{R}^{cd} + H_\text{R}^{da},
\end{equation}
where
\begin{equation}
    \begin{split}
        \label{eq:reservoir hamlitonian for ij}
        H_\text{R}^{pq} &= \hbar \omega_{pq}\int dz \psi^\dagger_{pq}(z)\psi_{pq}(z)+\frac{i\hbar v_{pq}}{2}\int dz \left(\frac{d\psi^\dagger_{pq}(z)}{dz}\psi_{pq}(z) - \text{H.c.}\right),
    \end{split}
\end{equation}
for ($p>q$) is the reservoir Hamiltonian corresponding to the transition between $\ket{p}$ and $\ket{q}$; $\omega_{pq}$ is taken as a center frequency of the waveguide modes considered in the indicated frequency band, and $v_{pq}$ is the group velocity of excitation propagation in the waveguide. 

For a given transition between $\ket{p}$ and $\ket{q}$ ($p>q$), the interaction Hamiltonian between the system and reservoir is
\begin{equation}
    H_\text{S-R}^{pq} = \hbar \eta_{pq}^*\sigma_{pq}\psi_{pq}(0) + \hbar \eta_{pq}\psi^\dagger_{pq}(0)\sigma_{qp},
\end{equation}
where we adopt a pointwise coupling and again work within the RWA; the total interaction Hamiltonian is then given by
\begin{equation}
\label{eq:reservoir interaction ij}
    H_\text{S-R} = H_\text{S-R}^{ba} + H_\text{S-R}^{cb} + H_\text{S-R}^{cd} + H_\text{S-R}^{da}.
\end{equation}
In equation \eqref{eq:reservoir hamlitonian for ij} we introduced a nominal reference frequency and group velocity, and in Eq. \eqref{eq:reservoir interaction ij} we introduced a nominal coupling constant $\eta_{pq}$. In fact, the group velocity $v_{pq}$ and coupling $\eta_{pq}$ will lead to a new constant that will identify the non-radiative decay widths associated with each transition $\ket{p}\to \ket{q}$ ($p>q$); it will be the physical parameter that can be identified by experiment, and there will be one of these physical parameters for each coupling of a transition to a waveguide reservoir. 

To simplify the calculations below, we define the operators 
\begin{equation}
\label{eq:Fij+}
    F_+^{pq} = \frac{\boldsymbol{\mu}_{pq}}{\hbar}\cdot\boldsymbol{E}_+^{pq} - \eta_{pq}^*\psi_{pq}(0),
\end{equation}
for $p>q$ and their adjoints,
\begin{equation}
    (F_+^{pq})^\dagger = \frac{\boldsymbol{\mu}_{qp}}{\hbar}\cdot\boldsymbol{E_-}^{pq} - \eta_{pq}\psi_{pq}^\dagger(0) \equiv F_-^{qp},
\end{equation}
which is a combined field-reservoir operator. With these definitions, the interaction Hamiltonian \eqref{eq:V(t)}, 
\begin{equation}
    \label{eq:interaction with F definition}
        V = -\hbar\sigma_{ba}F_+^{ba}-\hbar\sigma_{cb}F_+^{cb} -\hbar\sigma_{cd}F_+^{cd} -\hbar\sigma_{da}F_+^{da} + \text{H.c.},
\end{equation}
is now written in a form that allows a simple interpretation: The energy of the system is increased (decreased) by annihilating (creating) energy in the external field or reservoir, mediated by the operator $F_+^{pq}$ ($F_-^{qp}$).

\section{Dynamics \label{sec:effective Heisenberg equations}}
With the Hamiltonian of the system-field-reservoir now defined, we move to evaluating and formally solving the dynamical equations for the system operators. We begin with the free solution of the system operators given by 
\begin{equation}
    \hat\sigma_{pq}(t) = e^{-i\omega_{qp}t}\sigma_{pq},
\end{equation}
where we have employed $H_0$, which sets $\Omega_\alpha \to \omega_{qp}$ (see Eq. \eqref{eq:g_alpha}), with the dynamics of each $\sbar{pq}$ determined by Eq. \eqref{eq:dynamics Obar} and given in detail in Appendix A by Eq. \eqref{app:dynamics sbar v1}. In calculating each equation for $\sbar{pq}$ we defined the quantity
\begin{equation}
\label{eq:fbardef}
    \fbar{pq}{+} = \frac{\boldsymbol{\mu}_{pq}}{\hbar}\cdot \bar{\boldsymbol{E}}^{pq}_+(t) - \eta_{pq}^* \bar{\psi}_{pq}(0,t), 
\end{equation}
for ($p>q$) where we set $\bar{\boldsymbol{E}}^{pq}_+(t)\equiv \boldsymbol{E}^{pq}_+(\{\bar{a}_{s}(\boldsymbol{k},t)\})$ to denote  the time evolution of each individual frequency band field operator,  and $\bar{a}_{s}(\boldsymbol{k},t)$ and $\bar{\psi}_{pq}(0,t)$ are formally given by Eq. \eqref{eq:bar_from_caret}. 

The set of equations for each $\sbar{pq}$ given by Eq. \eqref{app:dynamics sbar v1} is a set of third order coupled nonlinear differential equations, where each $\fbar{pq}{\pm}$ is still an unknown quantity; the 
equations are third order because the dynamics of $\bar{\boldsymbol{E}}^{pq}_+(t)$ involve the radiation reaction contribution.  To eventually solve for the system operators perturbatively, we first work out the solution to each $\fbar{\pm}{pq}$, which depends on $\bar{\psi}_{pq}(0,t)$ and $ \bar{\boldsymbol{E}}^{pq}_+(t)$. The first contribution, $\bar{\psi}_{pq}(0,t)$, is exactly solvable and given by \cite{PhysRevA.106.023115}
\begin{equation}
\label{eq:psibar solution}
    \bar{\psi}_{pq}(0,t) = \hat\psi_{pq}(0,t) - i\frac{\eta_{pq}}{2v_{pq}}\sbar{qp},
\end{equation}
where
\begin{equation}
    \hat\psi_{pq}(z,t) = \psi_{pq}(z - v_{pq}t)e^{-i\omega_{pq}t}.
\end{equation}
Note that $\hat\psi_{pq}(0,t)$ is proportional to a Schr\"{o}dinger operator, and can be though of as an ``input'' of the reservoir field to the system; $\bar{\psi}_{pq}(0,t)$ is then taken to be the ``output'' which contains a contribution from the system proportional to $\sbar{qp}$, for which we have yet to solve. 

To solve for the dynamics of $\bar{\boldsymbol{E}}^{pq}_+(t)$ we start with the solution to the full field operator $\bar{\boldsymbol{E}}_+(t)$ given by Eq. \eqref{eq:E^H(t)}, which in our interaction picture is
\begin{equation}
    \label{eq:Ebar(t)}
    \bar{\boldsymbol{E}}_+(t) = \hat{\boldsymbol{E}}_+(t) + \frac{1}{6\pi\epsilon_0c^3}\frac{d^3\bar{\boldsymbol{\mu}}_+(t)}{dt^3},
\end{equation}
where
\begin{subequations}
    \begin{gather}
        \hat{\boldsymbol{E}}_+(t) \equiv \boldsymbol{E}_+(\{\hat{a}_{s}(\boldsymbol{k},t)\})\\ \hspace{18mm} = \boldsymbol{E}_+(\{a_{s}(\boldsymbol{k})e^{-i\omega_k t}\}),\nonumber\\
        \bar{\boldsymbol{\mu}}_+(t) \equiv \boldsymbol{\mu}_+(\{\sbar{pq} \}),
    \end{gather}
\end{subequations}
and again because $\hat{\boldsymbol{E}}_+(t)$ is a function of Schr\"{o}dinger operators it can be thought of as an ``input.'' The quantity $\bar{\boldsymbol{E}}_+(t)$ can then be thought of as a ``output'' which contains a contribution from the third time derivative of the dipole moment which we classically understand as the radiative contribution from the dipole at the position of the dipole \cite{dalibard1982vacuum, jackson2021classical}. Now for each allowed transition given by the dipole moment (Eq. \eqref{eq:muplus}) there is a corresponding radiation term in Eq. \eqref{eq:Ebar(t)}. Here we come to the the main implication of our frequency band approximation: If each transition is sufficiently far away in energy then we can input the dipole moment (Eq. \eqref{eq:muplus}) and the frequency band decomposition of the incident field operator (Eq. \eqref{eq: E freqband}) into Eq. \eqref{eq:Ebar(t)}, and extract that the time evolution of each frequency band is given by
\begin{equation}
    \label{eq:Ebar_xy(t)sol}
    \bar{\boldsymbol{E}}^{pq}_+(t) = \hat{\boldsymbol{E}}^{pq}_+(t) + \frac{\boldsymbol{\mu}_{qp}}{6\pi\epsilon_0c^3}\frac{d^3\bar{\sigma}_{qp}(t)}{dt^3}.
\end{equation}
That is, each frequency band field operator only ``sees'' the field from the system due to the corresponding system transition. 

At this point we can use the formal expressions \eqref{eq:psibar solution} and \eqref{eq:Ebar_xy(t)sol} for $\bar{\psi}_{pq}(z,t)$ and $\bar{\boldsymbol{E}}^{pq}_+(t)$ respectively in the equation \eqref{eq:fbardef} for $\fbar{pq}{+}$, which then goes into each dynamical equation for $\sbar{rs}$ given by Eq. \eqref{app:dynamics sbar v1}. However, before implementing a perturbative solution we must consider the third order time derivatives associated with the radiation reaction term in $\bar{\boldsymbol{E}}^{pq}_+(t)$. For even a well-behaved lowest order solution cannot be identified unless one forces ``by-hand'' that all population is \emph{initially} in the ground state; this is analogous to the runaway solutions that occur in the classical treatment of radiation reaction \cite{jackson2021classical}. Here we follow a standard approach \cite{milonni2004influence}, and in calculating the radiation reaction term in \eqref{eq:Ebar_xy(t)sol} we take
\begin{equation}
    \label{eq:sigmabarapprox}
    \frac{d}{dt} \bar{\sigma}_{rs}(t)= \frac{d}{dt}\left( e^{-i\omega_{sr}t}\check\sigma_{rs}(t)\right)\approx -i\omega_{sr}\bar{\sigma}_{rs}(t).
\end{equation}
This relies on the RWA that we have adopted from the start, and therefore on near-resonant excitation, for it is only under such an assumption that the time evolution of $\bar{\sigma}_{rs}(t)$ is close to an oscillation at frequency $\omega_{sr}$.

Using this approximation, the expression for $\fbar{pq}{+}$ identified at the start of the preceding paragraph is evaluated and simplifies to
\begin{equation}
    \label{eq:fbarsol2}
    \fbar{pq}{+} = \fhat{pq}{+}+i\frac{\Gamma_{pq}}{2}\sbar{qp}.
\end{equation}
Here we have defined $\fhat{pq}{+}$ as
\begin{equation}
\label{eq:fhatdef}
    \fhat{pq}{+} = \frac{\boldsymbol{\mu}_{pq}}{\hbar}\cdot \hat{\boldsymbol{E}}^{pq}_+(t) - \eta_{pq}^* \hat{\psi}_{pq}(0,t), 
\end{equation}
and set
\begin{equation}
    \Gamma_{pq} = \Gamma_{pq}^\mathrm{r} + \Gamma_{pq}^\text{nr},
\end{equation}
which is the total decay rate from $\ket{p}\to\ket{q}$ ($p>q$) and includes the radiative ($\Gamma_{pq}^\mathrm{r}$) and non-radiative ($\Gamma_{pq}^\mathrm{nr}$) contribution. The radiative contribution is given by
\begin{equation}
\label{eq:gammarad}
    \Gamma^\text{r}_{pq} = \frac{\omega_{pq}^3}{3\pi\epsilon_0c^3\hbar}|\boldsymbol{\mu}_{pq}|^2,
\end{equation}
and has the usual form of the ``natural'' linewidth, which leads to the 
``Einstein A coefficient'' \cite{loudon2000quantum}.  In solving for the reservoir response we combined the reservoir parameters to form the constant
 \begin{equation}
  \Gamma^{\text{nr}}_{pq}=\frac{\left|\eta_{pq}\right|^{2}}{v_{pq}},
\end{equation}
which we can identify as the \emph{non-radiative} decay rate. Finally, it will be useful in what follows to define a total decay rate of state $p$ by
\begin{equation}
\label{eq:total_decay_rate}
    \Gamma_p = \sum_q \Gamma_{pq},
\end{equation}
where $1/\Gamma_p$ is the lifetime of state $\ket{p}$. We see that the quantity $\fbar{pq}{+}$ (Eq. \eqref{eq:fbarsol2}) can be understood as a sum of two contributions: The first, given by $\fhat{pq}{+}$, is an ``input'' from the field and reservoir, and the second is a contribution from the system response. That system response, characterized by the $\sbar{rs}$, still must be determined.

To do that, we use the expressions $\fbar{pq}{+}$ (Eq. \eqref{eq:fbarsol2}) in each equation for $\sbar{rs}$ (Eq.\eqref{app:dynamics sbar v1}) to identify a first order differential equation for each  $\sbar{rs}$. The resulting equations for each $\sbar{rs}$ will be normally ordered and will remain normally ordered after each perturbative iteration. Because of this, we are able to drop all contributions to the dynamical equations that involve a $\fhat{pq}{+}$ for which there is no incident light near the frequency $\omega_{pq}$, because both the electromagnetic field operator $\hat{\boldsymbol{E}}_+^{pq}(t)$ and reservoir operator $\hat\psi_{pq}(0,t)$ will then annihilate the initial state $\ket{\Psi_\text{in}}$.

To this point the pumping scheme has been kept general, but we now focus on the situation where incident light drives a two-photon transition from $\ket{a}$ to $\ket{c}$ via $\ket{b}$. Then we can drop each term that contains either $\fhat{da}{+}$ or $\fhat{cd}{+}$ or their adjoints. The resulting equations \eqref{app:dynamics sbar v2} form a set of linear coupled differential equations involving the $\sbar{pq}$ that contain the effects of both radiative and non-radiative decay.  We will solve them perturbatively, but we first consider expressions for the scattering and absorbed energy. 

\section{Scattered and Absorbed energy}\label{sec:Scattered and Absorbed energy}
In this section we revisit the equations for the scattered and absorbed energy (Eq. \eqref{eq: quantum scattering} and \eqref{eq: quantum absorption}) for the specified four level system.

We begin with the scattered energy (Eq. \eqref{eq: quantum scattering}) and show in Appendix \ref{sec:app:Scattered and absorbed energy} that if we move to the interaction picture, consider the interaction Hamiltonian given by Eq. \eqref{eq:H_M-EM v4}, and apply the slowly varying approximation of Eq. \eqref{eq:sigmabarapprox}, the total scattered energy is given by
\begin{equation}
    \begin{split}
    \label{eq:scattenergyfinal}
        \mathcal{S}& = \hbar\omega_{ba}\Gamma_{ba}^\text{r}\int\limits_{-\infty}^{\infty}\langle\bar{\sigma}_{bb}(t)\rangle dt 
        + \hbar\omega_{da}\Gamma_{da}^\text{r}\int\limits_{-\infty}^{\infty}\langle\bar{\sigma}_{dd}(t)\rangle dt \\
        &+ \hbar\omega_{cb}\Gamma_{cb}^\text{r}\int\limits_{-\infty}^{\infty}\langle\bar{\sigma}_{cc}(t)\rangle dt
        + \hbar\omega_{cd}\Gamma_{cd}^\text{r}\int\limits_{-\infty}^{\infty}\langle\bar{\sigma}_{cc}(t)\rangle dt.
    \end{split}
\end{equation}
The physical meaning of the terms contributing to the scattered energy is then easily identified: For example, population in the intermediate state ($\langle \bar\sigma_{bb}(t)\rangle$) can radiate energy ($\hbar\omega_{ba}$) and decay to the ground state at a rate given by $\Gamma_{ba}^\text{r}$. At any instant, the amount of scattered energy is then $\hbar\omega_{ba}\Gamma_{ba}^\text{r}\langle \bar\sigma_{bb}(t)\rangle dt$; to determine the total scattered energy we integrate over all time. For each allowed radiative transition (shown in Fig. \ref{fig:4lvlsystem}) there is a corresponding term in Eq. \eqref{eq:scattenergyfinal}. 

In this context, the slowly varying approximation of Eq. \eqref{eq:sigmabarapprox} manifests itself as the assumption that the radiative linewidth with which the system is decaying is sufficiently narrow such that to good approximation the radiation is emitted at the transition frequency. For frequencies in the optical range this is typically valid for atomic and molecular systems.

Continuing in Appendix \ref{sec:app:Scattered and absorbed energy} and starting with the general form of the absorbed energy (Eq. \eqref{eq: quantum absorption}), then applying the same steps as the derivation of the scattered energy, using the equation for $\fbar{pq}{+}$ (Eq. \eqref{eq:fbardef}) and the dynamics of each $\sbar{rs}$ (Eq. \eqref{app:dynamics sbar v1}), we show that the total absorbed energy is given by
\begin{equation}
    \begin{split}
    \label{eq:absorbedenergyfinal}
        \mathcal{A}& = \hbar\omega_{ba}\Gamma_{ba}^\text{nr}\int\limits_{-\infty}^{\infty}\langle\bar{\sigma}_{bb}(t)\rangle dt 
        + \hbar\omega_{da}\Gamma_{da}^\text{nr}\int\limits_{-\infty}^{\infty}\langle\bar{\sigma}_{dd}(t)\rangle dt \\
        &+ \hbar\omega_{cb}\Gamma_{cb}^\text{nr}\int\limits_{-\infty}^{\infty}\langle\bar{\sigma}_{cc}(t)\rangle dt
        + \hbar\omega_{cd}\Gamma_{cd}^\text{nr}\int\limits_{-\infty}^{\infty}\langle\bar{\sigma}_{cc}(t)\rangle dt,
    \end{split}
\end{equation}
which corresponds to the expression for the scattered energy $\mathcal{S}$ (Eq. \eqref{eq:scattenergyfinal}) but with the radiative decay rates $\Gamma_{pq}^\text{r}$ replaced with the non-radiative decay rates $\Gamma_{pq}^\text{nr}$; the physical meaning of the contributing terms here can likewise be easily identified.

As in the calculation for the scattered energy, the slowly varying approximation again implies that the linewidth for non-radiative decay is sufficiently narrow so that the energy transfer is given by the transition frequency. In Appendix \ref{app:Absorption check}, we show that this is equivalent to assuming that we are working within the RWA.

Finally, we immediately calculate the extinction
$\mathcal{E} = \mathcal{A} + \mathcal{S}$, which has the same form as $\mathcal{S}$ and $\mathcal{A}$ but with the \emph{total} decay rate $\Gamma_{pq}$ appearing, which includes both the radiative and non-radiative decay rate. We now have a clear physical model that
reflects our intuition of the interaction: The system responds to the incident electromagnetic field, which drives population into excited states that then decay in two ways, \emph{radiatively} back into the electromagnetic field, which we classify as scattered light, or \emph{non-radiatively} into the reservoirs, which we classify as absorption. 

To confirm the consistency of this approach, we directly calculate the change in energy of the reservoir $\Delta E_\text{R}$ in Appendix \ref{sec:app:Scattered and absorbed energy}, and show that it is identical to the calculated absorption $\mathcal{A}$. So despite the approximation made using Eq.\eqref{eq:sigmabarapprox}, the calculated quantities $\mathcal{A}$ and $\mathcal{S}$ are consistent. This should be expected, since the way we model the reservoir mimics the Markov approximation \cite{fischer2018derivation}.

\section{Fluorescence emission count}
\label{sec:solving the equations of motion}
We begin solving the equations of motion by noting that the general form of each $\sbar{rs}$ (see Eq. \eqref{app:dynamics sbar v2}) is 
\begin{equation}
  \left(\frac{d}{dt}+\frac{\Gamma_{r}}{2}+\frac{\Gamma_{s}}{2}+i\omega_{sr}\right)\bar{\sigma}_{rs}(t)=iK_{sr}(t),\label{eq:first}
\end{equation}
for the appropriate right-hand side $K_{sr}(t)$. Using the relation between $\sbar{rs}$ and $\check\sigma_{rs}(t)$ (Eq. \ref{eq:bar_from_caret}) this is equivalent to 
\begin{equation}
  \left(\frac{d}{dt}+\frac{\Gamma_{r}}{2}+\frac{\Gamma_{s}}{2}\right)\check{\sigma}_{rs}(t)=e^{i\omega_{sr}t}iK_{sr}(t),
\end{equation}
which has a formal solution that includes a homogeneous solution satisfying the initial condition at $t=-\infty$, where each
$\check{\sigma}_{rs}(t)$ is equal to the corresponding Schr\"{o}dinger
operator $\sigma_{rs}$. Because of the damping terms $\Gamma_{r},\Gamma_{s}$,
that homogeneous solution will vanish for finite times. Correspondingly,
in a formal solution of (\ref{eq:first}) the homogeneous solution
will vanish as well, and introducing a Green function
\begin{equation}
    \label{eq:Green function}
    G_{sr}(t) \equiv i e^{-\left(\frac{\Gamma_{s}}{2} + \frac{\Gamma_{r}}{2} + i\omega_{sr}\right)t},
\end{equation}
the exact solution to Eq. \eqref{eq:first} and each operator in Eq. \eqref{app:dynamics sbar v2} is
\begin{equation}
    \label{eq:exact solution to sigma_ij}
    \bar\sigma_{rs}(t) = \int\limits_{-\infty}^{t}dt_1G_{sr}(t - t_1)K_{sr}(t_1),
\end{equation}
for the appropriate $K_{sr}(t)$ on the right-hand side. 

In this work, we are interested in the emission count from fluorescence associated with the transition from $\ket{d}$ to $\ket{a}$; it is given by
\begin{equation}
    n = \Gamma_{da}^\text{r}\int\limits_{-\infty}^{\infty}\langle\bar{\sigma}_{dd}(t)\rangle dt,
\end{equation}
where we have divided the scattered energy by the energy $\hbar\omega_{da}$ (see Eq. \eqref{eq:scattenergyfinal}) associated with that transition to obtain the number of photons emitted in this transition. Since there is no incident light pumping the system from $\ket{a}\to \ket{d}$, the only way for there to be a nonzero population in state $\ket{d}$ is via decay from state $\ket{c}$ (radiatively or non-radiatively), and so
\begin{equation}
    \label{eq:sigma_dd exact solution}
    \sbar{dd} = \Gamma_{cd}\int\limits_{-\infty}^t dt_1 e^{-\Gamma_{d}(t - t_1)}\bar{\sigma}_{cc}(t_1).
\end{equation}
In what follows it will be useful to write the fluorescence count as
\begin{equation}
\label{eq:total count}
    n = \frac{\Gamma_{da}^\text{r}}{\Gamma_d}\frac{\Gamma_{cd}}{\Gamma_c}p,
\end{equation}
where we set
\begin{equation}
    \label{eq:pexcitation}
    p = \Gamma_c\Gamma_d \int\limits_{-\infty}^{\infty} dt \int\limits_{0}^{\infty}d\tau e^{-\Gamma_d\tau} \langle \bar\sigma_{cc}(t - \tau)\rangle,
\end{equation}
which is the \textit{total} system excitation probability to state $\ket{c}$ (henceforth referred to as the \textit{excitation probability}); it will be our primary interest in the remaining sections. The fluorescence count \eqref{eq:total count} is then given by a product of three terms: 1) The total probability of excitation to state $\ket{c}$ ($p$); 2) the branching ratio for the decay \emph{either} radiatively or nonradiatevly from $\ket{c}\to\ket{d}$ ($\Gamma_{cd}/\Gamma_c$); 3) and the branching ratio for $\ket{d}$ to decay to $\ket{a}$ \emph{radiatively} ($\Gamma_{da}^\text{nr}/\Gamma_d$). Since the calculation of $n$ from $p$ is trivial, in the next section we focus on the excitation probability and explicitly calculate it using perturbation theory. 

\section{perturbative solution \label{sec:perturbative solution}}
To solve the coupled system of equations \eqref{app:dynamics sbar v2} we expand each $\bar{\sigma}_{rs}(t)$ in a perturbative expansion and iterate order-by-order. For the zeroth order solution when there is no external electromagnetic field, there is no coherence generated; no population is moved out of the ground state, since the initial ket of the reservoir is the vacuum state and the equations are normal ordered. Then using the completeness relation \eqref{eq:complete_relation} -- which is also satisfied in the interaction picture in which we work -- we solve for the last unknown system operator, and for finite times we have
\begin{equation}
\label{eq:ggnought}
    \bar\sigma_{aa}^{(0)}(t) = \hat 1.
\end{equation}
This occurs because the interaction between the system and reservoir is still operative, and since we take the initial ket of the reservoir to be vacuum, the system can only lose energy to the reservoir. So for $\ket{\Psi_\text{in}}$ describing the system in any state, for finite times  $t>-\infty$ the system will decay to the ground state, and Eq. \eqref{eq:ggnought} necessarily results. 

To generate each higher order term we input the zeroth order term into the right-hand side of each formal solution and then continue order-by-order. At each order the solution will be given in terms of a time integral, which we evaluate by moving to frequency space. For any Hermitian operator $O(t)$, such as the electric field operator at each site of the system, we write 
\begin{equation}
    O(t) = \int\limits_{-\infty}^{\infty} \dbarw{} O(\omega)e^{-i\omega t},
\end{equation}
where we define $\dbarw{} = d\omega/\sqrt{2\pi}$. Introducing positive and negative frequency components in the usual way, we then write all frequency integrals to range from $0$ to $\infty$, and leave these limits implicit in the expressions below. Since we are only interested in the probability of excitation $p$  (Eq. \eqref{eq:pexcitation}) we need to calculate $\sbar{cc}$ to the lowest nonzero order, which is fourth order. 

In Appendix \ref{app:Perturbation theory} we start at zeroth order and calculate all terms that contribute to  $\sbar{cc}$, and with the $\sbar{cc}$ that results the probability of excitation of an atom is given by
\begin{equation}
\label{eq:pwork}
    \begin{split}
        p =& 2\pi\int\dbarw{\mathrm{II}}^\prime\dbarw{\mathrm{I}}^\prime\dbarw{\mathrm{I}}\dbarw{\mathrm{II}}G_{ba}^*(\omega_\mathrm{I}^\prime)G_{ba}(\omega_\mathrm{I}) \\
        & \times2\text{Im}\left[G_{ca}(\omega_\mathrm{I} + \omega_\mathrm{II})\right]2\pi\delta(\omega_\mathrm{I} + \omega_\mathrm{II} -\omega_\mathrm{I}^\prime - \omega_\mathrm{II}^\prime)\\
        &\times \langle \hat{F}_{-}^{bc}(-\omega_\mathrm{II}^\prime)\hat{F}_{-}^{ab}(-\omega_\mathrm{I}^\prime)\hat{F}_{+}^{ba}(\omega_\mathrm{I})\hat{F}_{+}^{cb}(\omega_\mathrm{II})\rangle,
    \end{split}
\end{equation}
where 
\begin{equation}
        G_{pq}(\omega) = \frac{1}{\omega_{pq} - \omega -i\frac{\Gamma_{p}}{2} - i\frac{\Gamma_{q}}{2}}.
\end{equation}
The resulting probability of excitation depends on the intermediate state linewidth $G_{ba}(\omega_\mathrm{I})$, the final state linewidth $G_{ca}(\omega_\mathrm{I} + \omega_\mathrm{II})$, and the four-point correlation function; the latter depends on the operators $\hat{F}_{+}^{ba}(\omega_\mathrm{I})$ and $\hat{F}_{+}^{cb} (\omega_\mathrm{II})$, which annihilate a photon centered at the first and second transition respectively. We choose to label the incident frequencies by $\omega_\mathrm{I}$ ($\omega_\mathrm{I}^\prime$) and $\omega_\mathrm{II}$ ($\omega_\mathrm{II}^\prime$) to be clear that the incident light we will be considering is centered at two different frequencies. 

For now we have left the final result for the probability of excitation in terms of each $\hat{F}_+^{pq}(\omega)$ operator because it is notationally more compact. But since the initial ket of the reservoir is the vacuum state, 
\begin{equation}
    \hat F_{+}^{pq}(\omega)\ket{\Psi_\text{in}} \to \frac{\boldsymbol{\mu}_{pq}}{\hbar}\cdot \hat{\boldsymbol{E}}^{pq}_{+}(\omega)\ket{\Psi_\text{in}},
\end{equation} 
and because the correlation function is normally ordered, transitions are only made by the external field, as expected. The expression for the probability of excitation of the system  (Eq. \eqref{eq:pexcitation}), with possible resonant excitation to the intermediate state, and with both radiative and non-radiative decay, is a main result of this paper. It is valid near resonance, and for any non-degenerate pulsed or CW quantum state of light that is propagating in three dimensions with any polarization. 

To provide a sample calculation, we consider light incident on the system propagating in the $\hat{\boldsymbol{z}}$ direction. As we consider pumping the system at two different frequencies, there are field operators corresponding to each of the frequency bands $\omega_{ba}$ and $\omega_{cb}$. If we assume a single polarization for each frequency band and neglect dispersive effects, the positive frequency components for each frequency band are
\begin{subequations}
\label{eq:Efieldoperatorv1}
    \begin{gather}
    \hat{\boldsymbol{E}}^{ba}_{\boldsymbol{+}}(\boldsymbol{r},\omega_\mathrm{I}) =\boldsymbol{e}_\mathrm{I}l_\mathrm{I}(x,y,z)\sqrt{\frac{\hbar \omega_\mathrm{I}}{2\epsilon_0c}}a_\mathrm{I}(\omega_\mathrm{I})e^{i\omega_\mathrm{I} z/c},\\
    \hat{\boldsymbol{E}}^{cb}_{\boldsymbol{+}}(\boldsymbol{r},\omega_\mathrm{II}) =\boldsymbol{e}_\mathrm{II}l_\mathrm{II}(x,y,z)\sqrt{\frac{\hbar \omega_\mathrm{II}}{2\epsilon_0c}}a_\mathrm{II}(\omega_\mathrm{II})e^{i\omega_\mathrm{II} z/c},
    \end{gather}
\end{subequations}
where $\boldsymbol{r} = (x,y,z)$, $a_J(\omega)$ for $J = \mathrm{I}, \mathrm{II}$ is an annihilation operator centered at the frequency $\bar\omega_{J}$ within the respective frequency band, and $\boldsymbol{e}_J$ is the polarization. In writing Eq. \eqref{eq:Efieldoperatorv1} for the field operators in each frequency band, we have expanded the use of the subscript on the annihilation operator to label not only polarization, as in Eq. \eqref{eq:HEM}, but also the frequency band of interest. To include the effects of different beam spatial profiles we include the spatial dependent function $l_J(\boldsymbol{r})$ normalized according to
\begin{equation}
    \int dxdy|l_J(x,y,z)|^2 = 1,
\end{equation}
which ensures that $|l_J(x,y,z_0)|^2\langle a_J^\dagger(\omega)a_J(\omega)\rangle$ is the spectral density per unit area of light passing through the plane located at $z = z_0$. In this treatment we model the beam profile in a phenomenological way; however, in the simplest limit of a field propagating with a uniform area $A$, the normalization condition of the spatial profile enforces $l_J(x,y,z) = 1/\sqrt{A}$ and we recover the result of Blow et al. \cite{blow1990continuum}. More sophisticated treatments of quantization within different regimes have been considered, and are useful in properly characterizing higher order spatial modes of incident light; see for example Garrison and Chiao \cite{garrison2008quantum}. Since our goal here is to model the effects of a simple experiment with a Gaussian pulse, this phenomenological treatment will suffice. Finally, we make the usual approximation that the bandwidths we will be considering are much smaller than the center frequency; the positive frequency component of each electric field operator then simplifies to
\begin{subequations}
\label{eq:Efieldoperatorv2}
    \begin{gather}
        \hat{\boldsymbol{E}}_+^{ba}(\boldsymbol{r},\omega_\mathrm{I}) = \boldsymbol{e}_\mathrm{I}l_\mathrm{I}(x,y,z) E_\mathrm{I} a_\mathrm{I}(\omega_\mathrm{I})e^{i\omega_\mathrm{I} z/c}, \\
        \hat{\boldsymbol{E}}_+^{cb}(\boldsymbol{r},\omega_\mathrm{II}) = \boldsymbol{e}_\mathrm{II}l_\mathrm{II}(x,y,z)E_\mathrm{II} a_\mathrm{II}(\omega_\mathrm{II})e^{i\omega_\mathrm{II} z/c},
    \end{gather}
\end{subequations}
where 
\begin{equation}
    \label{eq:script E}
    E_J \equiv \sqrt{\frac{\hbar\bar{\omega}_J}{2\epsilon_0 c}}, 
\end{equation}
with $J = \mathrm{I},\mathrm{II}$ used for the remainder of this paper. Note that in our treatment of absorption we have already made this assumption; see the discussion surrounding Eq. \eqref{eq:sigmabarapprox}. 

Using the form of the field operator (Eq. \eqref{eq:Efieldoperatorv2}), the probability of excitation \eqref{eq:pwork} of an atom or molecule at position $\boldsymbol{r}$ is
\begin{equation}
\begin{split}
    p(\boldsymbol{r}) &= \eta \hspace{-1mm}\int\dbarw{\mathrm{II}}^\prime\dbarw{\mathrm{I}}^\prime\dbarw{\mathrm{I}}\dbarw{\mathrm{II}}G_{ba}^*(\omega_\mathrm{I}^\prime)G_{ba}(\omega_\mathrm{I})L(\omega_\mathrm{I} + \omega_\mathrm{II}) \\
    & \times{\langle a_\mathrm{II}^\dagger(\omega_\mathrm{II}^\prime)a_\mathrm{I}^\dagger(\omega_\mathrm{I}^\prime)a_\mathrm{I}(\omega_\mathrm{I})a_\mathrm{II}(\omega_\mathrm{II})\rangle}|l_\mathrm{I}(\boldsymbol{r})|^2|l_\mathrm{II}(\boldsymbol{r})|^2\\
    & \times 2\pi\delta(\omega_\mathrm{I} + \omega_\mathrm{II} -\omega_\mathrm{I}^\prime - \omega_\mathrm{II}^\prime)
\end{split}
\end{equation}
where we set 
\begin{equation}
\label{eq:eta}
    \eta = 2\pi \frac{\bar\omega_\mathrm{I}\bar\omega_\mathrm{II}}{(2\epsilon_0 c \hbar)^2}|\boldsymbol{e}_\mathrm{II}\cdot \boldsymbol{\mu}_{cb}|^2|\boldsymbol{e}_\mathrm{I}\cdot \boldsymbol{\mu}_{ba}|^2,
\end{equation}
and put
\begin{equation}
    L(\omega) = \frac{1}{2\pi}\frac{\Gamma_c}{(\omega_{ca} - \omega)^2 + \frac{\Gamma_c^2}{4}},
\end{equation}
which is normalized to unity. Note here we explicitly include the spatial dependence of the excitation probability on the left-hand side. 

Taking $\rho(x,y,z)$ to be the number density of atoms,
\begin{equation}
    \int dxdydz \rho(x,y,z) = N_\mathrm{atoms},
\end{equation}
where $N_\mathrm{atoms}$ is the total number of atoms, the total fluorescence emission count is then 
\begin{equation}
    \begin{split}
        \overline n& = \int\rho(\boldsymbol{r})n(\boldsymbol{r})d\boldsymbol{r}\\
        & = \overline p\frac{\Gamma_{da}^\text{r}}{\Gamma_d}\frac{\Gamma_{cd}}{\Gamma_c} N_\mathrm{atoms},
    \end{split}
\end{equation}
where $\overline p$ is given by
\begin{equation}\label{eq:pexcfinal}
    \begin{split}
        \overline p &= \eta \hspace{-1mm}\int\dbarw{\mathrm{II}}^\prime\dbarw{\mathrm{I}}^\prime\dbarw{\mathrm{I}}\dbarw{\mathrm{II}}G_{ba}^*(\omega_\mathrm{I}^\prime)G_{ba}(\omega_\mathrm{I})L(\omega_\mathrm{I} + \omega_\mathrm{II}) \\
    & \times\frac{\langle a_\mathrm{II}^\dagger(\omega_\mathrm{II}^\prime)a_\mathrm{I}^\dagger(\omega_\mathrm{I}^\prime)a_\mathrm{I}(\omega_\mathrm{I})a_\mathrm{II}(\omega_\mathrm{II})\rangle}{A_\mathrm{eff}^2}\\
    & \times 2\pi\delta(\omega_\mathrm{I} + \omega_\mathrm{II} -\omega_\mathrm{I}^\prime - \omega_\mathrm{II}^\prime),
    \end{split}
\end{equation}
and is the excitation probability of a single atom averaged over all atoms in the sample; the relevant effective area $A_\mathrm{eff}$ is given by 
\begin{equation}
    \frac{1}{A_\mathrm{eff}^2} = \frac{1}{N_\mathrm{atoms}}\int |l_\mathrm{I}(\boldsymbol{r})|^2|l_\mathrm{II}(\boldsymbol{r})|^2\rho(\boldsymbol{r})d\boldsymbol{r} .
\end{equation}

It is illustrative to change variables and evaluate the delta function, which we do in Appendix \ref{app:changingvariables}; the excitation probability is then
\begin{equation}\label{eq:pexcfinalfinal}
    \overline p=\eta\int L(\omega) \left|\int\hspace{-1mm} d\1G_{ba}(\1)\frac{a_\mathrm{II}(\omega - \1)a_\mathrm{I}(\1)\ket{\Psi_\mathrm{in}}}{A_\mathrm{eff}}\right|^2 \hspace{-1mm}\frac{d\omega}{2\pi},
\end{equation}
which has a simple description: The probability of excitation is the sum over all the ways in which two photons can be annihilated, weighted by the intermediate state linewidth function $G_{ba}(\1)$, so that the sum of their frequencies equals $\omega$; then we sum over all possible sums of the two-photon terms weighted by the two-photon linewidth function $L(\omega)$.

The probability $\overline p$ of the excitation of an atom is now in its final form; considering its evaluation for different states of light will be the focus of the remaining sections. We point out that this result is what is to be expected from a simpler calculation involving just the dipole moment and a Fermi's Golden Rule perturbative calculation. Here the advantage is that we include a quantum reservoir and radiation reaction, which has two consequences: 1) Radiative and non-radiative lifetimes are included in the linewidth functions, unlike in the Fermi's Golden Rule calculation; 2) we take into account the dynamics of the reservoir, allowing us to calculate the absorbed and scattered energy in a consistent way to maintain energy conservation.  

\section{Classical light\label{sec:CW Coherent state}}
We begin with two classical light sources modeled by coherent states
\begin{equation}
    \label{eq:coherent state v1}
    \ket{\psi_\text{cl}}  =
    D_\mathrm{I}(\alpha_\mathrm{I})D_\mathrm{II}(\alpha_\mathrm{II}) \ket{\text{vac}},
\end{equation}
where $D_J(\alpha_J)$ is the usual displacement operator
\begin{equation}
    D_J(\alpha_J) \equiv e^{\alpha_J\int d\omega \varphi_J(\omega)a
    ^\dagger(\omega) - \text{H.c.}},
\end{equation}
and $\alpha_J$ is a complex coefficient. The spectral amplitudes of the states are given by $\varphi_J(\omega)$ and are normalized to satisfy
\begin{equation}
   \int d\omega |\varphi_J(\omega)|^2 = 1.
\end{equation}

Using the usual transformation property for coherent states \cite{loudon2000quantum}
\begin{equation}
\label{eq:D opertor properties}
        D_J^\dagger a_{J^\prime}(\omega)D_J = (a_J(\omega) + \alpha_J \varphi_J(\omega))\delta_{JJ^\prime},
\end{equation}
and the unitarity of the displacement operator, we evaluate
\begin{equation}
\label{eq:classical1pa}
    \bra{\psi_\text{cl}}a_J^\dagger(\omega)a_J(\omega)\ket{\psi_\text{cl}} = |\alpha_J|^2|\phi_J(\omega)|^2,
\end{equation}
and 
\begin{equation}
\label{eq:classical2pa}
    \bra{\psi_\text{cl}}a\one^\dagger(\1^\prime) a\two^\dagger(\2^\prime)   a\two(\omega\two)a\one(\omega\one)\ket{\psi_\text{cl}} = |\alpha\two|^2|\alpha\one|^2\varphi\one^*(\1^\prime) \varphi\two^*(\2^\prime)   \varphi\two(\omega\two)\varphi\one(\omega\one),
\end{equation}
with
\begin{equation}
    N_J^\text{cl} = \int d\omega \bra{\psi_\text{cl}} a_J^\dagger(\omega)a_J(\omega)\ket{\psi_\text{cl}} = |\alpha_J|^2
\end{equation}
being the average number of photons in each frequency band. Finally, with  Eq. \eqref{eq:classical2pa} the average excitation probability is worked out to be
\begin{equation}
\label{eq:pulsed_classical_excitation_prob}
    \begin{split}
        \overline p^\mathrm{cl} = \eta \frac{|\alpha\one|^2}{A_\mathrm{eff}} \frac{|\alpha\two|^2}{A_\mathrm{eff}} \int L(\omega) \left|\int\limits G_{ba}(\omega_\mathrm{I})\varphi_\mathrm{I}(\omega_\mathrm{I})\varphi_\mathrm{II}(\omega - \omega_\mathrm{I})\dbarw{\mathrm{I}}\right|^2d\omega.
    \end{split}
\end{equation}
As expected for classical light modeled by coherent states, the excitation probability is ``coherent" in the sense that we integrate over $\omega\one$, with the phase dependence of $G_{ba}(\1)$ allowing for the system to be coherently controlled \cite{stowe2008control}. The fluorescence emission count for a pulsed source of classical light is then given by
\begin{equation}
    \overline n_\mathrm{pulse}^\mathrm{cl} = \overline p^\text{cl}_\text{pulse}\frac{\Gamma_{da}^\text{r}}{\Gamma_d}\frac{\Gamma_{cd}}{\Gamma_c} N_\mathrm{atoms},
\end{equation}
where $\overline p^\text{cl}_\text{pulse}$ is given by Eq. \eqref{eq:pulsed_classical_excitation_prob}.

To consider the continuous wave limit, we consider fields oscillating at the definite frequencies $\bar\1$ and $\bar\2$ for a duration $T$. Then as $T\to \infty$ each spectral amplitude function is strongly peaked at its center frequency, and the excitation probability \emph{rate} is
\begin{equation}
\label{eq:classical_exciattionrate_cw_limit}
    \overline r^\text{cl}_\text{cw} = \frac{\overline p^\mathrm{cl}_\text{cw}}{T} = F\one F\two \sigma(\bar\1, \bar\2),
\end{equation}
where 
\begin{equation}
    \sigma(\1, \2) = \eta L(\1 + \2)|G_{ba}(\1)|^2
\end{equation}
is the two-photon cross-section, in agreement with Boyd \cite{boyd2020nonlinear} in the appropriate limit, with $F_J = |\alpha_J|^2 / (A_\text{eff}T)$ the effective incident photon flux at frequency $\bar{\omega}_J$; see Appendix \ref{sec:app:Classical light: The CW limit}. The fluorescence emission rate in the continuous wave limit is then given by
\begin{equation}
    \overline R_\mathrm{cw}^\mathrm{cl} = \frac{\overline n^\mathrm{cl}_\mathrm{cw}}{T} = \overline r^\text{cl}_\text{cw}\frac{\Gamma_{da}^\text{r}}{\Gamma_d}\frac{\Gamma_{cd}}{\Gamma_c} N_\mathrm{atoms}.
\end{equation}

\section{Squeezed Light \label{sec:Entangled States of Light}}
We now turn to non-degenerate squeezed light, which we model with the state
\begin{equation}
    \label{eq:squeezed state}
    \ket{\psi_\text{sq}} = S \ket{\text{vac}},
\end{equation}
with
\begin{equation}
    \label{eq:squeezing operator}
    S \equiv e^{\beta\int d\omega_\mathrm{I} d\omega_\mathrm{II}\gamma(\omega_\mathrm{I},\omega_\mathrm{II})a_\mathrm{I}^\dagger(\omega_\mathrm{I})a_\mathrm{II}^\dagger(\omega_\mathrm{II}) - \text{H.c.} },
\end{equation}
the unitary squeezing operator and $\beta = |\beta|e^{i\theta}$ the squeezing parameter. The function $\gamma(\omega_\mathrm{I},\omega_\mathrm{II})\neq\gamma(\omega_\mathrm{II},\omega_\mathrm{I})$ is the joint spectral amplitude (JSA), which satisfies
\begin{equation}
    \int d\omega_\mathrm{I} d\omega_\mathrm{II} |\gamma(\omega_\mathrm{I},\omega_\mathrm{II})|^2 = 1,
\end{equation}
and is centered at $\omega\one=\bar\omega\one$ and $\omega\two= \bar\omega\two$.

To evaluate the quantities needed for the excitation probability using the squeezed state given by Eq. \eqref{eq:squeezed state} is not as simple as for classical light, and generally one needs to make a distinction between pulsed and CW sources. For pulsed sources, one typically performs a Schmidt decomposition of the joint spectral amplitude, decomposing it into an orthonormal and complete set of Schmidt modes; the excitation probability will then involve sums over the different Schmidt mode contributions. When considering the CW limit, this method is not convenient because the number of Schmidt modes tends to infinity. Instead, one can work directly with the squeezing operator, and process a squeezing operator transform similar to that in the single mode limit. Since in the CW limit we can derive closed form expressions for the excitation probability rate, we will consider it first to build intuition, and then move to the more general case of pulsed sources.  

\subsection{Continuous wave limit}
\label{sec:squeezedlight_cw}
For a typical spontaneous parametric downconversion process \cite{grice1997spectral}, the form of the JSA is
\begin{subequations}
    \begin{gather}
        \label{eq:jsa def}
        \gamma(\omega_\mathrm{I},\omega_\mathrm{II}) = \alpha(\omega_\mathrm{I} +\omega_\mathrm{II})\phi(\omega_\mathrm{I}, \omega_\mathrm{II}).
    \end{gather}
\end{subequations}
The function $\alpha(\omega)$ is related to the spectral amplitude of the pump used to generate the squeezed light, and $\phi(\omega_\mathrm{I}, \omega_\mathrm{II})$ is the phase-matching function describing the dispersion properties of the medium mediating the generation \cite{lerch2013tuning}. 

To model CW excitation, we set the pump amplitude to be
\begin{subequations}
    \label{eq:squeezed state pump}
    \begin{gather}
    \tilde\alpha(t) = \frac{e^{-i\bar\omega_p t}}{\sqrt{T_p}}, \hspace{5mm} -\frac{T_p}{2}\le t\le \frac{T_p}{2},\\
    \alpha(\omega) = \frac{1}{\sqrt{\Omega_p}}\sinc\left(\frac{(\omega - \bar\omega_p)\pi}{\Omega_p}\right),
    \end{gather}
\end{subequations}
which satisfies
\begin{equation}
\label{eq:pumpnorm}
    \int d\omega |\alpha(\omega)|^2 = \int dt |\tilde\alpha(t)|^2 =  1,
\end{equation}
where $\bar\omega_p = \bar{\omega}_\mathrm{I} + \bar\omega_\mathrm{II}$ is the pump center frequency, $T_p$ is the duration of the pump, and $\Omega_p = 2\pi/T_p$ identifies the pump bandwidth. In the CW limit, $\alpha(\omega)$ will be strongly peaked near $\omega \approx \bar\omega_p$ and each photon pair generated will be strongly correlated so that $\omega_\mathrm{I} + \omega_\mathrm{II} \approx \bar\omega_p$. We use this to approximate the JSA as
\begin{equation}
\label{eq:jsapprox}
    \begin{split}
    \gamma(\omega_\mathrm{I},\omega_\mathrm{II}) &\approx \alpha(\omega_\mathrm{I} + \omega_\mathrm{II})\phi(\omega_\mathrm{I}, \bar\omega_p - \omega_\mathrm{I})\\
    &\approx \alpha(\omega_\mathrm{I} +\omega_\mathrm{II})\phi(\bar\omega_p - \omega_\mathrm{II}, \omega_\mathrm{II}),
    \end{split}
\end{equation}
where in the first (second) line we set, to good approximation, $\omega_\mathrm{II}\to \bar\omega_p - \1$ ($\omega_\mathrm{I}\to \bar\omega_p - \2$); the phase-matching function then varies over as single variable $\omega_\mathrm{I}$ ($\omega_\mathrm{II}$). The approximations are on the same footing because the range in the frequency $\omega_\mathrm{I}$ from its center value must always be approximately equal and opposite to the range in frequency $\omega_\mathrm{II}$ from its center value. 
We include both expressions \eqref{eq:jsapprox} because they arise naturally below. In fact, it will be useful to introduce two new functions given by this approximation
\begin{equation}
\label{eq:phi_J(w_J)}
    \begin{split}
    \phi_\mathrm{I}(\omega_\mathrm{I})& = \phi(\omega_\mathrm{I}, \bar\omega_p - \omega_\mathrm{I}),\\
    \phi_\mathrm{II}(\omega_\mathrm{II})& =\phi(\bar\omega_p - \omega_\mathrm{II}, \omega_\mathrm{II}),
    \end{split}
\end{equation}
where $\phi_J(\omega_J)$ is a function centered at $\bar\omega_J$ which satisfies $\phi_\mathrm{I}(\bar\omega_\mathrm{I}) = \phi_\mathrm{II}(\bar\omega_\mathrm{II})$ and $\phi_\mathrm{I}(\bar\omega_p - \omega_\mathrm{II}) = \phi_\mathrm{II}(\omega_\mathrm{II})$ (or equivalently $\phi_\mathrm{I}(\1) = \phi_\mathrm{II}(\bar\omega_p - \omega_\mathrm{I})$).
Given this approximation and the normalization of the pump function (Eq. \eqref{eq:pumpnorm}), the JSA remains normalized when
\begin{equation}
\label{eq:pmfnorm}
        \int d\omega |\phi_J(\omega)|^2 = 1,
\end{equation}
which we will also use in our calculations below. This is useful because $\phi_J(\omega)$ sets the range of frequencies for which the pair generation is effective: Photons are generated in pairs with their frequency components centered at $\bar\omega_\mathrm{I}$ and $\bar\omega_\mathrm{II}$, but which range over a width set by $\phi_J(\omega)$. To quantify this range, for square normalized functions \cite{landes2021quantifying} we set
\begin{equation}
\label{eq:bandwidth}
    \Omega_c = \frac{1}{|\phi_\mathrm{I}(\bar\omega_\mathrm{I})|^2} = \frac{1}{|\phi_\mathrm{II}(\bar\omega_\mathrm{II})|^2},
\end{equation}
where we take the maximum value of $\phi_J(\omega_J)$ to occur at at $\bar\omega_J$, the center frequency of the generation. The range of frequencies of the generated photon pairs is then $\bar\omega_\mathrm{I} - \frac{\Omega_c}{2}\lesssim \omega_\mathrm{I} \lesssim \bar\omega_\mathrm{I} + \frac{\Omega_c}{2}$ and $\bar\omega_\mathrm{II} - \frac{\Omega_c}{2}\lesssim \omega_\mathrm{II} \lesssim \bar\omega_\mathrm{II} + \frac{\Omega_c}{2}$, such that $\omega_\mathrm{I} + \omega_\mathrm{II} = \bar\omega_p$ is satisfied, with $\Omega_c$ setting the effective bandwidth.

\begin{figure}
    \centering
    \includegraphics[width = 0.5\linewidth]{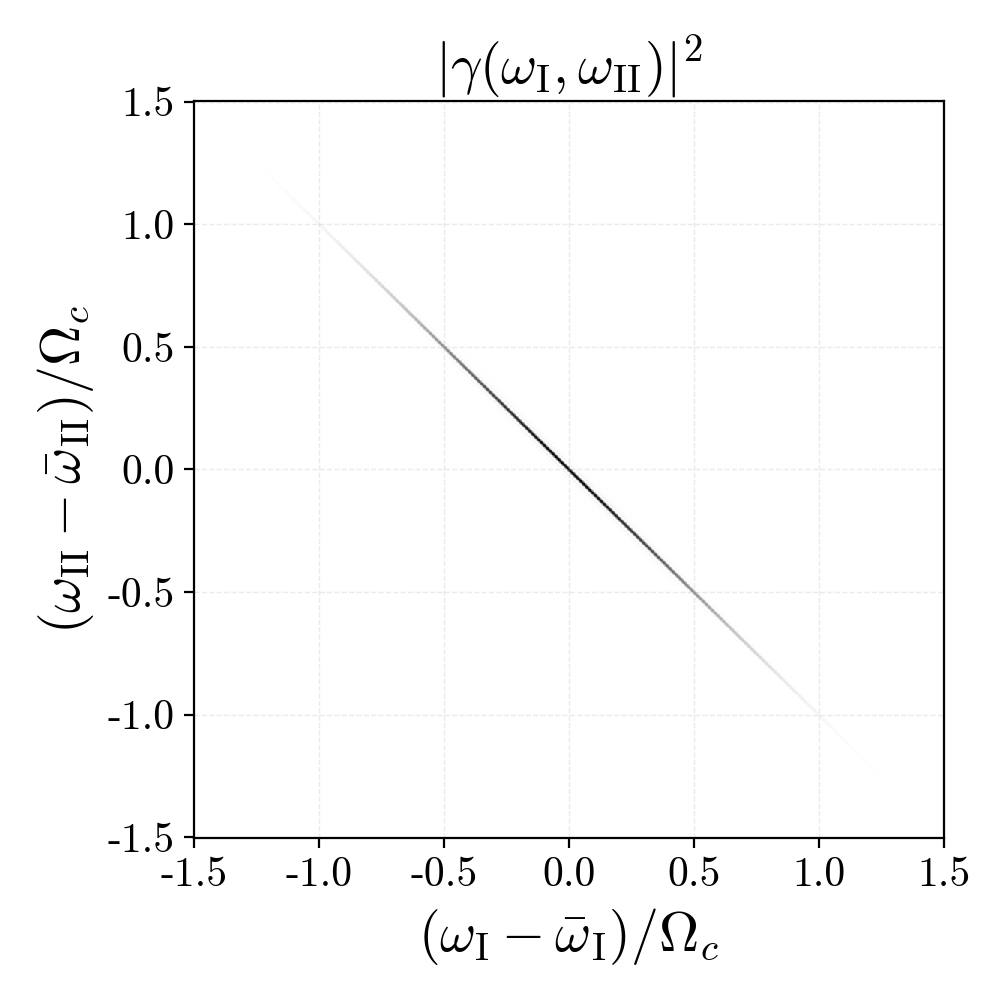}
    \caption{Schematic of the joint spectral intensity in the CW limit corresponding to the phase-matching function given by Eq. \eqref{eq:gaussianpmf}.}
    \label{fig:JSA_schematic}
\end{figure}

Using the effective bandwidth $\Omega_c$, we define a related time parameter $T_c = 2\pi/\Omega_c$ which is often referred to as the ``entanglement time'' \cite{lee2006entangled,schlawin2017entangled,szoke2020entangled,villabona2018two,PhysRevApplied.15.044012,corona2022experimental,raymer2020two,raymer2021large,schlawin2018entangled,fei1997entanglement} or ``coherence time'' of photon pairs; it is the time scale over which one expects to detect two photons if one neglects loss. As an example, consider the phase-matching function modeled by a Gaussian,
\begin{equation}
\label{eq:gaussianpmf}
    \phi(\omega_\mathrm{I}, \omega_\mathrm{II}) = \left(\frac{1}{\pi\bar\sigma_c^2}\right)^{\frac{1}{4}}e^{-\frac{[(\omega_\mathrm{I} - \bar\omega_\mathrm{I}) - (\omega_\mathrm{II} - \bar\omega_\mathrm{II})]^2}{4\bar\sigma_{c}^2}},
\end{equation}
which leads to
\begin{equation}
    \phi_J(\omega) = \left(\frac{1}{\pi\bar\sigma_c^2}\right)^{\frac{1}{4}}e^{-\frac{(\omega - \bar\omega_J)^2}{2\bar\sigma_c^2}},
\end{equation}
which is properly square normalized. We choose a Gaussian because it has the qualitative behavior of the phase-matching function generated in either a resonant or non-resonant nonlinear material \cite{quesada2022beyond}; it has been used in the past to model nonlinear processes with great success \cite{fedorov2009gaussian}. In the Gaussian given by Eq. \eqref{eq:gaussianpmf} the width is identified by $\bar\sigma_c$, and the effective bandwidth defined by Eq. \eqref{eq:bandwidth} is given by $\Omega_c^2 = \pi\bar\sigma_c^2$. In Fig. \ref{fig:JSA_schematic} we plot the corresponding joint spectral intensity ($|\gamma(\1, \2)|^2$), but we stress that this is only a schematic; in the CW limit the narrow width would be infinitely narrow. 

\subsubsection{Correlation functions}
To calculate the excitation probability rate, we now turn our attention to the correlation functions. Since the Schmidt decomposition is not practical in the CW limit, we directly work with the squeezing operator. Alternatively, one could solve directly for the mode operators for the nonlinear process \cite{raymer2022theory}, where time-ordering effects that describe the power broadening of the JSA are included. However, with that strategy one does not extract an analytical JSA as we do here. With the method we use we have the best of both worlds: 
We can choose a JSA that includes a physical model of the phase-matching function -- which could well include time-ordering corrections, although admittedly these are often negligible in squeezed light generation -- and as well we can generate closed form solutions for the correlation functions, unlike what would result from using the Schmidt decomposition. 

We begin in Appendix \ref{app:squeezing operator} by calculating the squeezing operator transformation for non-degenerate squeezed light and an arbitrary phase-matching function (see Eq. \eqref{eq:sqtransform}), which for each $a_J(\omega)$ is given by
\begin{equation}
    \begin{split}
        &\hspace{1mm}S^\dagger a_J(\omega)S \\
        &=  a_J(\omega)+ \frac{(c_J(\omega) - 1)}{\sqrt{\Omega_p}}\int d\omega^\prime\alpha^*(\omega^\prime + \bar\omega_p - \omega)a_J(\omega^\prime)+ \frac{e^{i\theta_J(\omega)}s_J(\omega)}{\sqrt{\Omega_p}}\int d\omega\alpha(\omega + \omega^\prime)a_{\bar{J}}^\dagger(\omega^\prime),
    \end{split}
\end{equation}
where we set
\begin{subequations}
    \begin{gather}
        s_J(\omega) = \text{sinh}(|\beta\phi_J(\omega)\sqrt{\Omega_p}|),\\
        c_J(\omega) = \text{cosh}(|\beta\phi_J(\omega)\sqrt{\Omega_p}|),\\
        e^{i\theta_J(\omega)} = \frac{\beta \phi_J(\omega)}{|\beta \phi_J(\omega)|},
    \end{gather}
\end{subequations}
and use ${\bar{J}}$ to denote the \emph{opposite} of $J$, that is, if $J = \mathrm{I}$ then ${\bar{J}} = \mathrm{II}$, and vice-versa. 

It will be useful to introduce the quantity
\begin{equation}
    r_J(\omega) = \frac{\phi_J(\omega)}{|\phi_J(\bar\omega_J)|},
\end{equation}
which has a maximum magnitude of one; the range of frequencies over which its magnitude is significant sets the effective range of frequencies that are relevant. Then the two functions $s_J(\omega)$ and $c_J(\omega)$ can be written as
\begin{subequations}
    \begin{gather}
        s_J(\omega) = \text{sinh}(|\bar\beta r_J(\omega)|),\\
        c_J(\omega) = \text{cosh}(|\bar\beta r_J(\omega)|).
    \end{gather}
\end{subequations}
where we set 
\begin{equation}
\label{eq:barbeta}
\begin{split}
    \bar\beta &= \beta \sqrt{\Omega_p}|\phi_J(\bar\omega_J)|\\
    &=\beta \sqrt{\frac{T_c}{T_p}}.
\end{split}
\end{equation}
To gain insight into the definition of $\bar\beta$, we calculate the first-order correlation function 
\begin{equation}
\label{eq: G1CWsqueezed}
    \begin{split}
        G^{(1)}_J(\omega, \omega^\prime)& \equiv \langle\psi_\text{sq}| a_J^\dagger(\omega) a_J(\omega^\prime)\ket{\psi_\text{sq}}\\
        &= s_J^2(\omega) \times \frac{1}{\Omega_p} \text{sinc}\left(\frac{(\omega - \omega^\prime)\pi}{\Omega_p}\right),
    \end{split}
\end{equation}
where we have used the property that $\alpha(\omega)$ is strongly peaked. Taking the CW limit when $\Omega_p \to 0$
\begin{equation}
    G^{(1)}_J(\omega, \omega^\prime) \to s_J^2(\omega)\delta(\omega - \omega^\prime),
\end{equation}
the expected result for squeezed light in the CW limit with $s_J^2(\omega)$ being the photon spectral density \cite{loudon1987squeezed}. From this result, it is straightforward to evaluate the photon rate
\begin{equation}
\label{eq:CWrate}
    \begin{split}
        \frac{N_J^\text{sq}}{T_p} = \int \frac{d\omega}{2\pi}\text{sinh}^2(|\bar\beta r_J(\omega)|),
    \end{split}
\end{equation}
and we interpret $\bar\beta$ as the ``normalized'' squeezing parameter. Indeed, in the low photon number rate limit, which we take to be when $|\bar\beta|\ll1$
\begin{equation}
\label{eq:photon rate low beta bar}
    \begin{split}
        \frac{N_J^\text{sq}}{T_p}& \approx \int \frac{d\omega}{2\pi}|\bar\beta r_J(\omega)|^2\\
        & = \frac{|\bar\beta|^2}{T_c}\\
        &=\frac{|\beta|^2}{T_p},
    \end{split}
\end{equation}
as expected. We now see the usefulness of defining $\bar\beta$: While $|\beta|^2$ diverges in the CW limit, because the number of photons diverges, $|\bar\beta|^2$ remains finite, and is further interpreted -- at least in the low photon rate limit -- as the number of photons within the coherence time $T_c$.

The incident rate of photons depends on two parameters, $\Omega_c$ (or $T_c$) and $|\bar\beta|$. For a fixed $|\bar\beta|$, as the bandwidth $\Omega_c$ increases, so does the photon rate. Mathematically, this follows from Eq. (\ref{eq:CWrate}) because $r_J(\omega)$ has a maximum of unity while it is nonzero over a wider frequency range as $\Omega_c$ increases; physically, it arises because the coherence time $T_c$ is decreasing and so there are more photons per unit time. We will treat $\bar\beta$ as an independent parameter, which then changes the photon rate for a fixed $\Omega_c$, and so plays the role of a ``gain'' parameter \cite{PhysRevA.106.013717}.

Next, we evaluate the second-order correlation function
\begin{equation}
\label{eq:sqtwo-photoncorr}
    \begin{split}
        &\bra{\psi_\text{sq}} a_\mathrm{II}^\dagger(\omega_\mathrm{II}^\prime)a_\mathrm{I}^\dagger(\omega_\mathrm{I}^\prime)a_\mathrm{I}(\omega_\mathrm{I})a_\mathrm{II}(\omega_\mathrm{II})\ket{\psi_\text{sq}} \\
        &=\frac{T_p}{2\pi}s_\mathrm{I}(\omega_\mathrm{I}^\prime)c_\mathrm{I}(\omega_\mathrm{I}^\prime)c_\mathrm{I}(\omega_\mathrm{I})s_\mathrm{I}(\omega_\mathrm{I})e^{i(\theta_\mathrm{I}(\omega_\mathrm{I}) - \theta_\mathrm{I}(\omega_\mathrm{I}^\prime))} \alpha(\omega_\mathrm{I} + \omega_\mathrm{II})\alpha^*(\omega_\mathrm{I}^\prime + \omega_\mathrm{II}^\prime)+G_\mathrm{I}^{(1)}{(\omega\one, \omega\one^\prime)}G_\mathrm{II}^{(1)}{(\omega\two, \omega\two^\prime)},
    \end{split}
\end{equation}
where we have used the properties of $\phi_J(\omega)$ and that $\alpha(\omega)$ is strongly peaked. The two-photon correlation function has two contributions: 1) The first, previously referred to as the ``coherent'' contribution \cite{dayan2007theory}, is due to photons that are anti-correlated in frequency, such that $\omega_\mathrm{I} + \omega_\mathrm{II} \approx \bar\omega_p$; 2) the second, previously referred to as the ``incoherent'' contribution \cite{dayan2007theory}, is the product of first-order correlation functions, where the two frequencies $\omega\one$ and $\omega\two$ ($\omega\one^\prime$ and $\omega\two^\prime$) are independent. In contrast to the coherent contribution, the range of nonzero frequency components in the incoherent contribution is determined by the functions $s_J(\omega)$, and is typically broadband.

\subsubsection{Excitation probability rate}
With the second-order correlation function in the CW limit (Eq. \eqref{eq:sqtwo-photoncorr}), the excitation probability \emph{rate} is
\begin{equation}
\label{eq:squeezed_exciattionrate_cw_limit}
    \overline r^\text{sq}_\text{cw} = \overline  r^\text{sq,c}_\text{cw} + \overline  r^\text{sq,ic}_\text{cw}
\end{equation}
where
\begin{equation}
\label{eq:coherentcont_CW}
    \overline  r^\text{sq,c}_\text{cw} = \frac{\overline p^\text{sq,c}_\text{cw}}{T_p} = \eta L(\bar\omega_p)\left|\int G_{ba}(\omega_\mathrm{I})e^{i\theta_\mathrm{I}(\omega_\mathrm{I})}\frac{s_\mathrm{I}(\omega_\mathrm{I})c_\mathrm{I}(\omega_\mathrm{I})}{A_\text{eff}}\frac{d\omega\one}{2\pi}\right|^2\hspace{-2mm},
\end{equation}
follows from the \emph{coherent} contribution to the correlation function and
\begin{equation}
\label{eq:incoherentcont_CW}
    \overline  r^\text{sq,ic}_\text{cw} = \frac{\overline p^\text{sq,ic}_\text{cw}}{T_p} = \eta\int L(\omega) \int\left|G_{ba}(\omega_\mathrm{I})\frac{s_\mathrm{II}(\omega - \omega_\mathrm{I})s_\mathrm{I}(\omega_\mathrm{I})}{A_\text{eff}}\right|^2   \frac{d\omega d\omega\one}{(2\pi)^2},
\end{equation}
follows from the \emph{incoherent} contribution. The two-photon linewidth function $L(\omega)$ plays a central role in determining the behavior of both of these contributions to the excitation probability rate, which we now discuss. The fluorescence emission rate in the continuous wave limit is then given by
\begin{align}
    \overline  R_\mathrm{cw}^\mathrm{sq} = \frac{\overline n^\mathrm{sq}_\mathrm{cw}}{T_p} = \overline r^\text{sq}_\text{cw}\frac{\Gamma_{da}^\text{r}}{\Gamma_d}\frac{\Gamma_{cd}}{\Gamma_c} N_\mathrm{atoms},
\end{align}
or given by the sum of the coherent and incoherent contributions
\begin{equation}
\overline  R_\mathrm{cw}^\mathrm{sq} = \overline  R_\mathrm{cw}^\mathrm{sq,c} + \overline  R_\mathrm{cw}^\mathrm{sq,ic} 
\end{equation}
where the terms on the right-hand side are defined in the obvious way.

Starting with the coherent contribution, we note that it is proportional to the two-photon linewidth function evaluated at the pump frequency; this is a result of the spectral correlations in the CW limit where photons are anti-correlated such that $\omega_\mathrm{I} + \omega_\mathrm{II} \approx \bar\omega_p$. Despite each individual photon having a potentially large bandwidth, the spectral correlations are such that the photon pairs that contribute can always be resonant with the two-photon transition. 

The coherent contribution is also proportional to the norm of a single integral where we integrate over the intermediate state linewidth function, the frequency-dependent phase term $\theta_\mathrm{I}(\omega_\mathrm{I})$, as well as the terms $s_\mathrm{I}(\omega_\mathrm{I})$ and $c_\mathrm{I}(\omega_\mathrm{I})$ that set the bandwidth of the incident squeezed light. Since we integrate first and then take the norm, by tuning the phase of the JSA one can ensure that all the frequency components 
add constructively (or destructively) increasing (or decreasing) the probability of excitation; this was discussed and experimentally verified by Dayan et al. \cite{dayan2004two}, where a spatial light modulator was used to manipulate the phase and coherently control the two-photon transition driven by squeezed light. 

Aside from the role spectral features of squeezed light play on the coherent contribution to two-photon excitation rate, the overall scaling differs from that for classical light. While the classical result scales quadratically with the incident photon rate (see Eq. \eqref{eq:classical_exciattionrate_cw_limit}), the scaling for squeezed light depends on the function $s_\mathrm{I}(\omega_\mathrm{I})$ and $c_\mathrm{I}(\omega_\mathrm{I})$.

In the low-photon rate limit, when $|\bar\beta|\ll 1$ and $c_\mathrm{I}(\omega_\mathrm{I})\approx 1$, the integrand is proportional to $s_\mathrm{I}(\omega_\mathrm{I})$, and there is a \emph{linear} dependence on the photon spectral density. This could, \emph{in principle}, enhance the excitation rate compared to classical light in the same limit and \emph{may} be useful for spectroscopic applications \cite{schlawin2018entangled, schlawin2017entangled, dorfman2016nonlinear, dayan2007theory, fei1997entanglement, schlawin2013photon, oka2018two, lee2006entangled}. We stress \emph{in principle}, because low photon rates are required, and the resulting signal may then not be intense enough to be useful. Whether or not it is has been a subject of debate \cite{PhysRevApplied.15.044012, mikhaylov2022hot, hickam2022single, landes2024limitations, Raymer:21}, highlighting the importance of calculating fluorescence emission rates for real systems. On the other hand, in the high photon rate limit, when $|\bar\beta| \gg 1$ and $c_\mathrm{I}(\omega_\mathrm{I})\approx s_\mathrm{I}(\omega_\mathrm{I})$, we recover the classical-like quadratic scaling with the photon density.

The incoherent contribution (see Eq. \eqref{eq:incoherentcont_CW}) is similar to the classical result in that it depends on two integrals, where $s_J(\omega_J)$ plays a role analogous to that of $\varphi_J(\omega_J)$ in the excitation of classical light. This analogy is expected because the frequencies involved in the incoherent contribution are not anti-correlated. However, the structure of the two integrals is distinct from the classical result, as it depends on the integral of the \emph{norm} of the intermediate linewidth function and photon densities.
Contributions from photons of different frequencies, $\omega_\mathrm{I}$, always add positively, with no phase dependence from the linewidth function or the incident light; this is the source of the label ``incoherent contribution.''

This behavior is a direct result of the dependence of the second-order correlation function on the first-order correlation function (see Eq. \eqref{eq:sqtwo-photoncorr}). In the CW limit, the first-order correlation function (see Eq. \eqref{eq: G1CWsqueezed}) is largest when its arguments are approximately equal. As a result, the incoherent contribution to the second-order correlation function is largest when $\omega_\mathrm{I}\approx \omega_\mathrm{I}^\prime$ and $\omega_\mathrm{II}\approx \omega_\mathrm{II}^\prime$. This is analogous to the photon statistics of a thermal (or chaotic) state; indeed, the reduced density operator of a non-degenerate squeezed state \emph{is} a thermal state \cite{quesada2019broadband}. And since the incoherent contribution involves photons for which the frequencies $\omega_\mathrm{I}$ and $\omega_\mathrm{II}$ are uncorrelated, the system is effectively ``seeing'' a thermal state.

In Section \ref{sec:CW comparison} we compare the fluorescence rates of classical and squeezed light in the CW limit. Before that, we consider the excitation probability driven by squeezed light in the pulsed regime. 

\subsection{Pulsed regime}
To consider the pulsed regime, we begin with an arbitrary JSA $\gamma(\1,\2)$, where in general $\gamma(\1,\2)\neq \gamma(\2,\1)$. We perform a Schmidt decomposition of the JSA 
\begin{equation}
    \gamma(\1,\2) = \sum_n\sqrt{p_n}f_{\mathrm{I}n}(\1)f_{\mathrm{II}n}(\1),
\end{equation}
where each set of functions $\{f_J(\omega)\}$ ($J = \mathrm{I},\mathrm{II}$) forms an orthonormal basis, and the positive numbers $p_n$ satisfy $\sum_n p_n = 1$. Then the squeezing operator is written as
\begin{equation}
    S = e^{\sum\limits_n \beta_n A_{\mathrm{I}n}^\dagger A_{\mathrm{II}n}^\dagger - \text{H.c.}},
\end{equation}
where we put $\beta_n = \beta\sqrt{p_n}$, and
\begin{equation}
\label{eq:Ad}
    A^\dagger_{Jn} = \int d\omega f_{Jn}(\omega)a_J^\dagger(\omega)
\end{equation}
are the creation operators associated with a set of ``super-modes''; the operators satisfy
\begin{equation}
    \left [A_{Jn}, A^\dagger_{J^\prime m} \right] = \delta_{JJ^\prime}\delta_{nm}.
\end{equation}
Then from the standard squeezing operator transform \cite{loudon2000quantum}
\begin{equation}
    \begin{split}
        S^\dagger A_{Jn}S = c_nA_{Jn} + e^{i\theta}s_nA_{\bar{J}n}^\dagger\\
        S^\dagger A_{Jn}^\dagger S = c_nA_{Jn}^\dagger + e^{-i\theta}s_nA_{\bar{J}n},
    \end{split}
\end{equation}
where we put $c_n = \mathrm{cosh}(|\beta_n|)$, $s_n = \mathrm{sinh}(|\beta_n|)$ and again set $\bar J$ to denote the opposite of $J$.

\begin{figure}
    \centering
    \includegraphics[width = \linewidth]{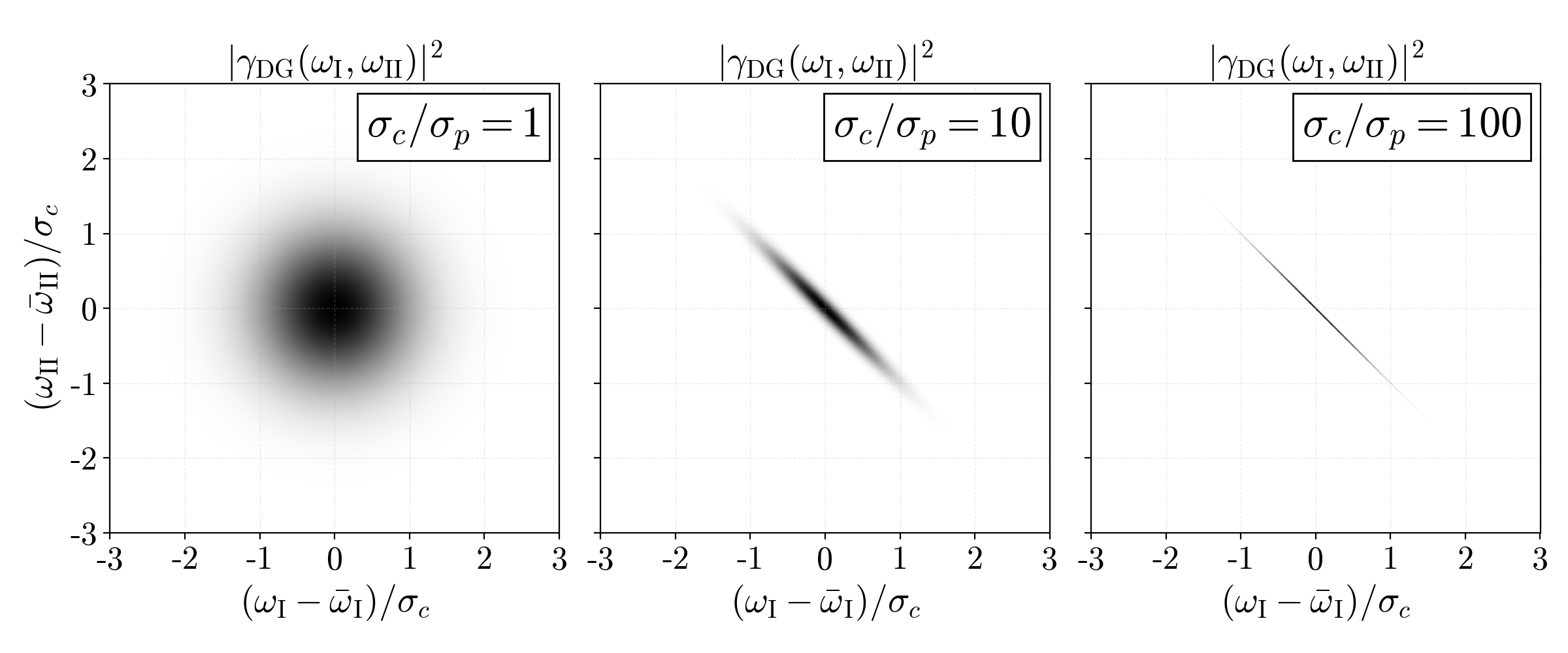}
    \caption{Schematic of the double-Gaussian (Eq. \eqref{eq:double_gaussian}) joint spectral intensity in the pulsed regime for $\sigma_c / \sigma_p = 1, 10, 100$.}
    \label{fig:pulsed_JSI_schematic}
\end{figure}

As an example of a pulsed source of squeezed light, we will consider the ``double-Gaussian'' JSA defined by
\begin{equation}
\label{eq:double_gaussian}
    \gamma_\text{DG}(\1, \2) = \sqrt{\frac{1}{\pi\sigma_p\sigma_c}}e^{-\frac{(\1 + \2 - \bar\omega_c)^2}{4\sigma_p^2}}e^{-\frac{[(\1 -\bar\1) - (\2 - \bar\2)]^2}{4\sigma_c^2}},
\end{equation}
where $\sigma_p$ and $\sigma_c$ are two parameters. We plot the JSI in Fig. \ref{fig:pulsed_JSI_schematic} for $\sigma_c / \sigma_p = 1, 10, 100$. In each plot, we identify $\sigma_p$ as the effective width along the diagonal direction and $\sigma_c$ as the effective width along the anti-diagonal direction. Similar to the CW limit, where the effective bandwidth was set by $\Omega_c$ (or the Gaussian width $\bar\sigma_c$) and we identified a corresponding coherence time (see Section \eqref{sec:squeezedlight_cw}), $\sigma_c$ sets the effective bandwidth of each photon, and its inverse approximately sets the coherence time of the photon pairs that make up the pulse of squeezed light.  In contrast to the CW limit, where the photon pairs have a strong anti-correlation such that $\1 + \2 \approx \bar\omega_p$, in the pulsed regime the degree of this correlation is set by the ratio of $\sigma_c/\sigma_p$ and can be uncorrelated for a ``separable'' JSA, shown in the leftmost plot of Fig. \ref{fig:pulsed_JSI_schematic}. We discuss how different degrees of correlation influence the excitation probability and fluorescence emission below.

\subsubsection{Correlation functions}
As before, we calculate the first-order correlation function
\begin{equation}
\label{eq:G1pulsed}
    G_J^{(1)}(\omega, \omega^\prime) = \sum_nf_{Jn}^*(\omega)f_{Jn}(\omega^\prime) s_n^2,
\end{equation}
with the total number of photons given by
\begin{equation}
\label{eq:pulsesqueezedphotons}
    N_J^\text{sq} = \sum_n \text{sinh}^2(|\beta_n|),
\end{equation}
and second-order correlation function
\begin{equation}
\label{eq:pulsedG2}
\begin{split}
    &\bra{\beta} a^\dagger_\mathrm{II}(\omega_\mathrm{II}^\prime)a^\dagger_\mathrm{I}(\omega_\mathrm{I}^\prime)a_\mathrm{I}(\omega_\mathrm{I})a_\mathrm{II}(\omega_\mathrm{II})\ket{\beta}\\ 
    &=\sum_{n,m} f_{\mathrm{II}n}^*(\2^\prime)f_{\mathrm{I}n}^*(\1^\prime)f_{\mathrm{I}m}(\1)f_{\mathrm{II}m}(\2)s_nc_ns_mc_m+G\one^{(1)}(\omega\one, \omega\one^\prime)G\two^{(1)}(\omega\two, \omega\two^\prime).
\end{split}    
\end{equation}
Similar to the result for the CW limit, here the second-order correlation function also involves two contributions. The first involves a sum over $m$ where the frequencies involved are $\omega\one$ and $\omega\two$. In other words, we \emph{coherently} add the contributions from those two frequencies; this will lead to the \emph{coherent} contribution to the excitation probability discussed below. The second contribution is written in terms of the first order correlation function in the same way as the CW limit. But while in the CW limit it led to an \emph{incoherent} contribution, we will see below that here it will lead to a contribution to the excitation probability that can be best thought of as \emph{partially incoherent}.  

\subsubsection{Excitation probability}
In the pulsed regime, the excitation probability is
\begin{equation}
\label{eq:squeezed_pulsed_Excitation_probability}
    \overline p^\text{sq}_\mathrm{pulse} = \overline p^\text{sq,c}_\text{pulse} + \overline p^\text{sq,ic}_\text{pulse},
\end{equation}
where
\begin{equation}
\label{eq:coherentcont_PULSED}
    \overline p^\mathrm{sq,c}_\mathrm{pulse} = \eta \hspace{-1mm}\int \hspace{-1mm} L(\omega)\left|\sum_n\int G_{ba}(\omega_\mathrm{I})f_\mathrm{IIn}(\omega - \omega\one)f_\mathrm{In}(\omega\one)\frac{s_n c_n}{A_\mathrm{eff}}\dbarw{\mathrm{I}}\right|^2\hspace{-2mm}d\omega,
\end{equation}
is derived from the coherent contribution to the correlation function (Eq. \eqref{eq:sqtwo-photoncorr}) and
\begin{equation}
\label{eq:incoherentcont_PULSED}
    \overline p^\text{sq,ic}_\text{pulse} = \eta\int L(\omega) \sum_{n,m}\left|\int G_{ba}(\omega_\mathrm{I})f_\mathrm{IIn}(\omega - \omega\one)f_\mathrm{Im}(\omega\one)\frac{s_ns_m}{A_\text{eff}}\dbarw{\mathrm{I}}\right|^2d\omega,
\end{equation}
is derived from its incoherent contribution. The fluorescence emission count in the pulsed regime is then given by
\begin{align}
    \overline  n_\mathrm{pulse}^\mathrm{sq} = \overline n^\text{sq}_\text{pulse}\frac{\Gamma_{da}^\text{r}}{\Gamma_d}\frac{\Gamma_{cd}}{\Gamma_c} N_\mathrm{atoms},
\end{align}
or given by the sum of the coherent and incoherent contributions
\begin{equation}
\overline  n_\mathrm{pulse}^\mathrm{sq} = \overline  n_\mathrm{pulse}^\mathrm{sq,c} + \overline  n_\mathrm{pulse}^\mathrm{sq,ic} 
\end{equation}
where the terms on the right-hand side are defined in the obvious way.

Similar to the excitation probability rate in the CW limit, the excitation probability here has two unique contributions that can be labeled ``coherent" and ``incoherent," but in the pulsed regime the terms ``coherent'' and ``incoherent'' acquire a double meaning. The ``coherent" term involves an integral over $\omega_I$ of each product $f_\mathrm{IIn}(\omega - \omega\one)f_\mathrm{In}(\omega\one)$ of two Schmidt modes, followed by the addition of the contributions from all such pairs of Schmidt modes. This contribution is coherent both in the sense that each frequency component of each product of Schmidt mode adds coherently, \emph{and} in the sense that the net contributions from each product of Schmidt modes add coherently. From Eq. \eqref{eq:pulsesqueezedphotons} the number of photons within each Schmidt mode is $s_n^2$, and we can identify two limits: 1) When $|\beta_n|\ll 1$ $s_n \to |\beta_n|$, $c_n \to 1$ and the excitation probability scales linearly with the incident photon number; 2) if instead $|\beta_n|\gg 1$, $c_n \to s_n$ and the probability of excitation scales quadratically with the incident photon number. 

The ``incoherent" contribution to the excitation probability here also has a similar structure to the corresponding contribution to the excitation probability rate in the CW limit, but with some distinctions. Here we have an integral over $f_\mathrm{IIn}(\omega - \omega\one)f_\mathrm{Im}(\omega\one)$, where now $n$ can be different from $m$; the integral is then squared and we sum over all contributions of $n$ and $m$. We refer to this contribution as \emph{partially incoherent} because the terms with $n=m$ match the coherent contribution -- aside from the scaling with $s_nc_n$ -- in that each frequency contribution adds coherently. This is in contrast to the incoherent contribution in the CW limit, where the frequency contributions do not add coherently. The reason is that in the CW limit the field is stationary and so $G^{(1)}(\omega, \omega^\prime)\propto\delta(\omega - \omega^\prime)$, while in the pulsed regime (see Eq. \eqref{eq:G1pulsed}) correlations between different frequency arguments survive \cite{quesada2019broadband}. Despite the frequency coherence when $n=m$, the contributions from the different pairs of Schmidt modes add \emph{incoherently} because we first square the integral and then perform the double sum over the contributions. 

\section{Example system}
\label{sec:example system}
\begin{figure}
    \centering
    \includegraphics[width = 0.25\linewidth]{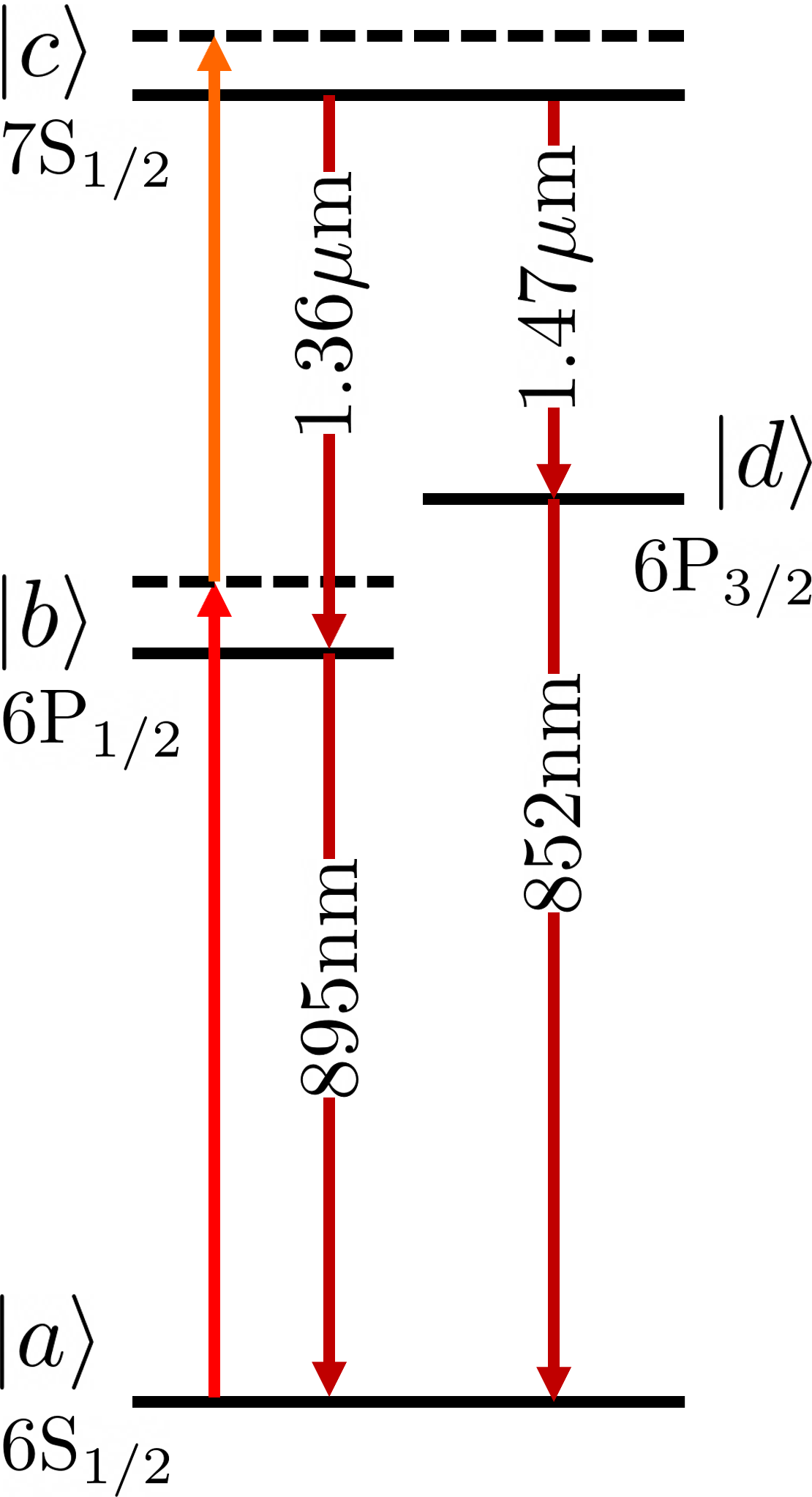}
    \caption{Schematic of the relevant energy levels and their dipole allowed decay pathways (downward arrows) for Cesium. The pumping scheme we consider is shown with the upward arrows.}
    \label{fig:Csleveldiagram}
\end{figure}

To provide a realistic calculation for expected fluorescence rates we use cesium (Cs) atoms as an example system. We choose cesium because of its simple atomic structure, with the relevant energy levels shown in Fig. \ref{fig:Csleveldiagram}. For our purpose, we are interested in the $6\mathrm{S}_{1/2}$, $6\mathrm{P}_{1/2}$, $6 \mathrm{P}_{3/2}$ and $7\mathrm{S}_{1/2}$ energy levels. Cesium also has a $5\mathrm{D}$ manifold between $6 \mathrm{P}_{3/2}$ and $6 \mathrm{S}_{1/2}$, which we neglect because it will be very far detuned from any incident light. Also, the energy level $7\mathrm{S}_{1/2}$ is dipole forbidden from decaying to the $5\mathrm{D}$ energy levels, so it is not relevant for our fluorescence calculations and we can safely neglect it.

Associated with each energy level are a set of degenerate states due to the magnetic quantum number of the atom, which we label as $m_j$, where $j$ labels the total angular momentum \cite{sakurai2020modern}. One could go beyond this description and include the hyperfine structure as well; however, for our purposes a fine structure model is sufficient, and we leave the inclusion of the hyperfine structure to future work. Including the degenerate levels, there are two possible ground states associated with $6 \mathrm{S}_{1/2}$, two (three) intermediate states associated with $6\mathrm{P}_{1/2}$ ($6\mathrm{P}_{3/2}$) and two final states associated with $7\mathrm{S}_{1/2}$. Since the intermediate states are both p-orbitals, transitions between them are dipole forbidden and thus significantly suppressed. Further, the energy levels are well separated, so we treat them as independent for the bandwidths we will be considering. In this example, we are pumping near 
$895 \mathrm{nm}$ ($6 \mathrm{S}_{1/2}\to 6 \mathrm{P}_{1/2}$) and $1.36 \mu\mathrm{m}$ ($6 \mathrm{P}_{1/2}\to7 \mathrm{S}_{1/2}$), a non-degenerate two-photon transition process.

In the two-photon excitation process, the magnetic quantum number of the intermediate and final states that are excited depend on the initial ground state, incident polarizations, and angular momentum selection rules. Therefore, we must consider both ground states as our starting points, since they are degenerate and equally likely. In Appendix \ref{app:Magnetic quantum number}, we demonstrate that for linearly polarized light both ground states lead to distinct excitation pathways, and can be treated as independent. The associated matrix elements are evaluated in terms of the total decay rates set by experiment. For the excitation pathway of Cs that we are considering, the matrix elements of both pathways are the same, and so we can effectively drop the magnetic quantum number and map the states in our model to $\ket{a} = \ket{6\mathrm{S}_{1/2}}$, $\ket{b} = \ket{6\mathrm{P}_{1/2}}$, $\ket{c} = \ket{7\mathrm{S}_{1/2}}$, and $\ket{d} = \ket{7\mathrm{S}_{1/2}}$. The relevant matrix elements for linearly polarized light are (see Appendix \ref{app:Magnetic quantum number})
\begin{equation}
\begin{split}
    &|\boldsymbol{e}_\mathrm{I}\cdot\boldsymbol{\mu}_{ba}|^2  = \frac{1}{3} \left( \frac{\Gamma_{ba}^\mathrm{r} 3\pi\epsilon_0c^2\hbar}{\omega_{ba}^3}\right),\\
    &|\boldsymbol{e}_\mathrm{II}\cdot\boldsymbol{\mu}_{cb}|^2  = \frac{1}{3} \left( \frac{\Gamma_{cb}^\mathrm{r} 3\pi\epsilon_0c^2\hbar}{\omega_{cb}^3}\right),
\end{split}
\end{equation}
where $\boldsymbol{e}_\mathrm{I}=\boldsymbol{e}_\mathrm{II}=\boldsymbol{\hat{z}}$.

Since generally the enhancement squeezed light provides -- if any -- is predominantly in the low photon number limit, it can be very sensitive to line broadening effects. For example, in the CW limit, Doppler broadening can reduce the coherent contribution while leaving the incoherent contribution unaffected \cite{drago2023two}. For this reason we have chosen to consider the Cs atoms in a MOT, effectively removing the effects of Doppler broadening. Following this theme, we also assume that the temperature of the MOT is sufficiently small that other broadening effects, such as the non-radiative decay of excited states as well as dephasing -- both due to the interaction of Cs atoms with each other and the environment -- are sufficiently small that they can be neglected. That is, we consider the detection of fluorescence from Cs atoms pumped with squeezed light under the most optimal conditions. While this may seem unrealistic if the goal is to compare to actual experiments, we will show that even in such situations the enhancement that squeezed light provides is still generally hard to detect. 

The Cs distribution inside the MOT is modeled by setting $\rho(\boldsymbol{r}) \to N_\mathrm{atoms}g(x)g(y)g(z)$, where
\begin{equation}
    g(x) = \frac{1}{\sqrt{2\pi\sigma^2}}e^{-\frac{x^2}{2\sigma^2}},
\end{equation}
is a Gaussian distribution normalized to unity. This assumes the width of the distribution is the same in all directions, which is experimentally accessible, but also easily generalized. We model the spatial distribution of the incident field as a circular Gaussian beam and set
\begin{equation}
    l_J(\boldsymbol{r}) = \sqrt{\frac{2}{\pi w^2(z)}}e^{-\frac{(x^2 + y^2)}{w^2(z)}},
\end{equation}
where
\begin{equation}
    w^2(z) = w_0^2 \left( 1 + \left(\frac{z}{z_R}\right)^2\right),
\end{equation}
with $w_0$ the beam waist and $z_R$ the Rayleigh length. With the density of Cs atoms and the field profile, the effective is area is
\begin{equation}
    \frac{1}{A_\text{eff}^2} = \frac{1}{l_0}\int \frac{dz}{A_0(z)}\frac{e^{-\pi\frac{z^2}{l_0^2}}}{2l_0^2 + A_0(z)},
\end{equation}
where we set $l_0 = \sqrt{2\pi\sigma^2}$, $A_0(z) = \pi w^2(z)/2$; the expression for $A_\text{eff}$ must be evaluated numerically. For the calculations below, we set the number of atoms inside the MOT to be $N_\mathrm{atoms}=10^6$ and set the FWHM along each axis to be $\Delta l = 0.1$mm, both achievable with current technology.  The beam waist FWHM in the x-y plane is taken to be $\Delta r = 0.1 $mm. Combining these values, the effective area is $A_\text{eff} \approx 220 \mu \text{m}^2$.

In the following section we consider incident classical and squeezed light, pulsed and CW, for different pulse widths, incident photon rates, and photon numbers; to make the calculations we rely on the equations derived from perturbation theory. To ensure that the equations we derive are justified, we only consider incident energies such that the predicted maximum population of the intermediate state $\bar\sigma^{(2)}_{bb}(t)$ remains less than 0.1 for all time.

\section{Classical and squeezed light fluorescence \label{sec:Classical and squeezed light fluorescence}}
Our goal in this section is to determine if excitation by squeezed light leads to a measurable advantage in the fluorescence generated over excitation by classical light. However, there are many ways the two sources of light could be compared. The first is to take both to be CW. Then the coherence time of classical light tends to infinity, while for squeezed light it is set by $T_c$ (see the text in the paragraph after Eq. \eqref{eq:bandwidth}). This is the comparison most often considered in the literature for degenerate light, and is typically what is most easily accessible experimentally. In this situation, the parameter space is reduced to the incident photon fluxes -- which should always be set equal for the two sources -- the detuning of the fields from resonance, and the coherence time of squeezed light; the resonant or nonresonant contributions for different coherence times are then considered. 

Here we take a different approach, since we are working in the non-degenerate regime and we can independently set the incident center frequencies to be resonant with both transitions; for classical light in the CW, limit this is the ``optimal condition'' for generating fluorescence, and we use this scenario as our reference. In comparing excitation with classical light to excitation with squeezed light, the only parameter to be varied is then the coherence time of the squeezed light, and we explore the trade-off between confinement in frequency or time by varying it.

A second scenario to consider is when both sources are pulsed; here the situation is more complicated because there are more independent parameters to consider. For pulsed sources of classical light we can identify a coherence time of the photons, effectively set by the inverse of the bandwidth, and there are two classical pulses to consider. For squeezed light in the pulsed regime we also have two parameters to identify, the coherence time of the photon pairs (inverse bandwidth) and the width of the photon anti-correlations; for the double-Gaussian JSA, these are identified by $\sigma_c$ and $\sigma_p$ respectively. In a very general analysis one could consider optimizing the classical bandwidths given the intermediate and final linewidths $\Gamma_b$ and $\Gamma_c$; however, for Cs $\Gamma_c$ is on the order of $\Gamma_b$, and so we set the frequency width of both classical pulses to be the same. Then the most natural way to compare the effectiveness of classical and squeezed light for pulsed excitation is to set the frequency width of the excitations to be the same. That is, we set the coherence time of the classical pulses to be the same as that of the squeezed pulse. And, as in comparing the CW excitations, we set each center frequency to be on resonance and only compare the two sources at equal photon numbers per pulse. 

The comparisons outlined above for CW light and for pulsed light are what we present below. Other comparisons are certainly also possible.  For example, one could consider a pulsed source of classical light where the coherence time matches that for CW squeezed light. However, care must be taken in appropriately setting the pulse energy of the classical light and the time over which the CW excitation is considered so that comparable incident energies are involved. We leave this analysis for future work.

\subsection{CW comparison}
\label{sec:CW comparison}
\begin{figure}
    \centering
    \includegraphics[width = \linewidth]{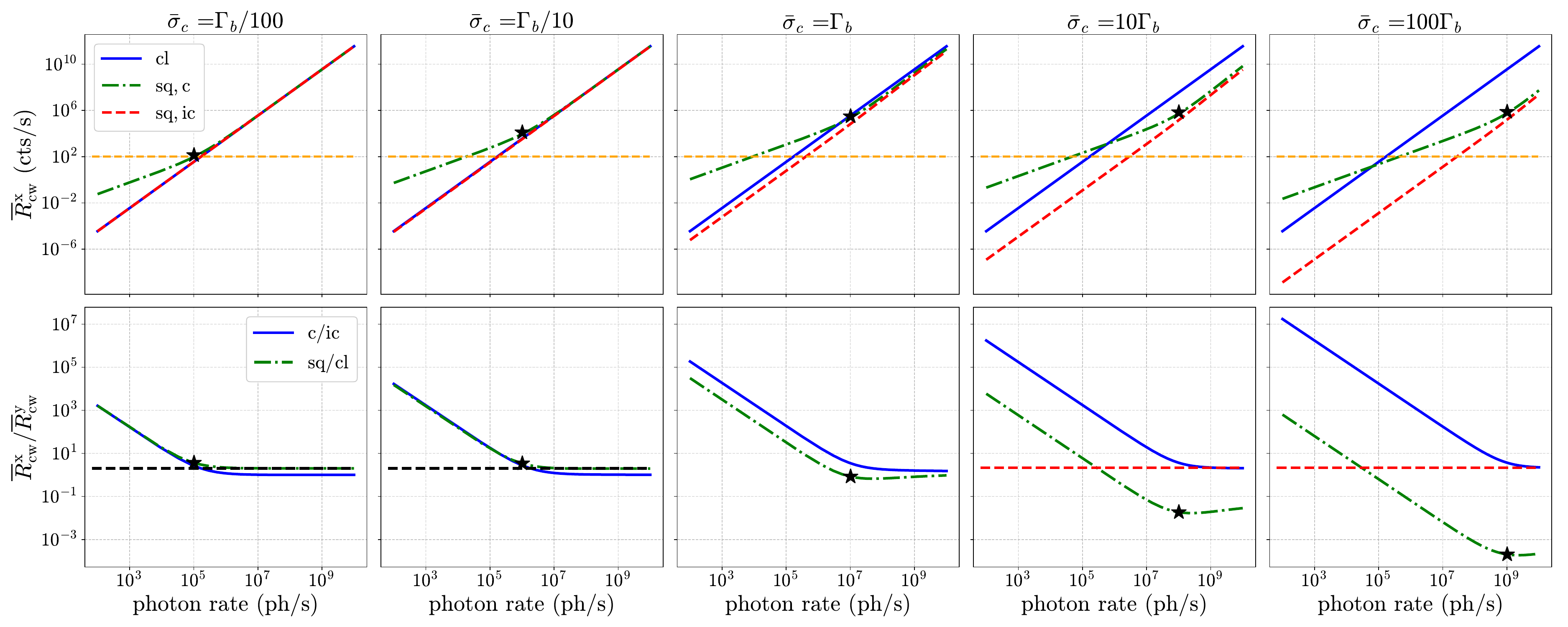}
    \caption{Fluorescence emission rates for squeezed and classical light (top row) and the ratio between the coherent/incoherent and squeezed/classical contributions (bottom row). From left to right $\sigma_c$ is increasing starting from the narrowband ($\bar\sigma_c/\Gamma_b = 0.01$)
    to the broadband limit ($\bar\sigma_c/\Gamma_b = 100$).  
    In the top row the horizontal orange dashed line is set to $100$ cts/s. In each plot the star corresponds to the ``cross-over'' point when $|\bar\beta| = 1$ and denotes where the scaling of the coherent contribution changes from linear to quadratic. The horizontal black dashed lines in the lower row correspond to the limiting case when $\overline R^\mathrm{sq}_\mathrm{cw} / \overline R^\mathrm{cl}_\mathrm{cw}\to 2$, and the red dashed lines correspond to $\bar R^\mathrm{sq,c}_\mathrm{cw} / \bar R^\mathrm{sq,ic}_\mathrm{cw} \to \Gamma_b / \Gamma_c$.}
    \label{fig:CWplot}
\end{figure}

We begin with the comparison of fluorescence emission rates for CW excitation. For classical light we directly apply Eq. \eqref{eq:classical_exciattionrate_cw_limit}, and for squeezed light we use the result of Eq. \eqref{eq:squeezed_exciattionrate_cw_limit} with the Gaussian model for the phase-matching function given in Eq. \eqref{eq:gaussianpmf}. With the photon rates set to be equal for classical and squeezed light excitation, the only free parameter is $\bar\sigma_c$, the effective bandwidth of the squeezed light. We consider a range of values $\bar\sigma_c / \Gamma_b = 0.01,,  
0.1,  1,  10,  100$, which covers the narrowband, intermediate and broadband limits. In Fig. \ref{fig:CWplot} we plot the calculated fluorescence emission rate for excitation with classical light, and the coherent and incoherent contributions to the fluorescence emission rate for excitation with squeezed light (top row) for increasing $\bar\sigma_c$ (left-to-right). In the bottom set of plots we include the ratio of the fluorescence emission rates for squeezed light excitation to those of classical light excitation, as well as the ratio between the coherent and incoherent contributions to the fluorescence emission rates for squeezed light excitation. Each plot is on a log-log scale, and the range of $|\bar\beta|$ is chosen so that the photon rate varies over the same range for each column. In the top row of plots, the horizontal orange dashed line is set to $100$ cts/s as a reasonable limit of what can be experimentally detected before other sources of loss, such as a limited collection efficiency, are included. In each plot the star corresponds to the ``cross-over'' point when $|\bar\beta| = 1$ and denotes where the scaling of the coherent contribution changes from linear to quadratic. The horizontal black dashed lines in the second row of plots correspond to the limiting case when $\overline R^\mathrm{sq}_\mathrm{cw} / \overline R^\mathrm{cl}_\mathrm{cw}\to 2$, and the red dashed lines there correspond to $\bar R^\mathrm{sq,c}_\mathrm{cw} / \bar R^\mathrm{sq,ic}_\mathrm{cw} \to \Gamma_b / \Gamma_c$; each is discussed below. Note that in all columns, the fluorescence emission rate driven by classical excitation is the same, since we are always setting the bandwidth of each photon to be sufficiently narrow and on resonance. We now discuss the scaling of the squeezed light emission rate in each plot, and how it compares to the classical result. In doing so we refer to the ``low'' and ``high'' photon rate limit as being before ($|\bar\beta| < 1$) or after ($|\bar\beta| > 1$) the cross-over point respectively.

Consider the first column in Fig. \ref{fig:CWplot} where $\bar\sigma_c = \Gamma_c/100$, corresponding to the narrowband limit where each squeezed photon is sharply distributed about the center frequency values. In the high photon rate limit, the coherent and incoherent contribution to squeezed light are identical and equal to the classical result. This is expected because all photons are narrowband and the classical and squeezed light contributions all scale quadratically. In the low photon rate limit, the ratio of the coherent contribution to squeezed light and the classical result becomes very large, which would usually be associated with a large enhancement due to squeezed light. However, this region of parameter space corresponds to extremely low fluorescent count rates that are below detection capabilities.

Increasing the effective bandwidth of each squeezed photon, set by $\bar\sigma_c$, by an order of magnitude, we focus on the second column in Fig. \ref{fig:CWplot} and observe qualitatively the same behavior. The cross-over point, however, has moved to a higher incident photon rate. Physically this arises because we have decreased the coherence time, leading to  more photons per unit time; see the discussion in the paragraph after Eq. \eqref{eq:photon rate low beta bar}. Since we always want to compare the squeezed results to the classical result at equal incident photon rates, we must decrease the value of $|\bar\beta|$ appropriately. Consequently, the linear scaling that is set by the coherent contribution persists to high incident photon rates. Despite the region being narrow and the enhancement minimal, to the left of the cross-over point there is a region where the coherent contribution to the squeezed light fluorescence emission rate is approximately an order of magnitude larger that the classical result \textit{and} is within an experimentally detectable regime.

As a result of the coherent and incoherent contributions to the fluorescence rate following excitation by squeezed light being identical and equal to that following excitation by classical light in the high photon rate limit, the ratio $\overline R^\mathrm{sq}_\mathrm{cw} / \overline R^\mathrm{cl}_\mathrm{cw}\to 2$, and there is an admittedly undramatic factor of two enhancement of the result for squeezed light over that for classical light; the ratio is denoted by the horizontal black dashed line in the first and second column. This is expected and equivalent to the ratio of the second order correlation functions for the states of light in this regime \cite{loudon2000quantum}.

Next, we consider the intermediate regime where the effective bandwidth of the squeezed light is on the order of the intermediate state linewidth. The results for this regime are plotted in the third column of Fig. \ref{fig:CWplot}. Despite the increase of the spectral-temporal correlations due to the larger bandwidth, in the high photon rate limit both the coherent and incoherent contributions to the squeezed light fluorescence rate are less efficient compared to the CW classical result. On the other hand, in the low photon rate limit, the coherent contribution is enhanced by spectral-temporal correlations and provides a factor of $\sim$ 100 times the classical result \textit{and} is in an experimentally detectable range.

In the fourth and fifth column of Fig. \ref{fig:CWplot} the effective bandwidth of squeezed photons is successively increased by an order of magnitude. Similar to the intermediate regime, the coherent and incoherent contribution to the squeezed light fluorescence emission rates continue to decrease relative to the classical result since the bandwidth is now much larger than the intermediate state linewidth.

Notice an interesting behavior in columns four and five of Fig. \ref{fig:CWplot} (broadband regime) where the ratio of the coherent and incoherent contribution of the squeezed light emission rate approach a fixed limit denoted by the horizontal red dashed line corresponding to $\Gamma_b / \Gamma_c$. To understand this, consider the broadband limits
\begin{equation}
    \overline r^\text{sq,c}_\mathrm{cw} \to \eta L(\bar\omega_p)\frac{s_\mathrm{I}^2(\omega_{ba})c_\mathrm{I}^2(\omega_{ba})}{A_\text{eff}^2}\left|\int G_{ba}(\omega_\mathrm{I})\frac{d\omega\one}{2\pi}\right|^2\hspace{-2mm},
\end{equation}
\begin{equation}
    \overline r^\text{sq,ic}_\mathrm{cw} \to \eta \frac{s^2_\mathrm{II}(\omega_{ca} - \omega_{ba})s^2_\mathrm{I}(\omega_{ba})}{A_\text{eff}^2}
    \int\left|G_{ba}(\omega_\mathrm{I})\right|^2   \frac{ d\omega\one}{(2\pi)^2},
\end{equation}
where to good approximation the only photons within the incident bandwidth that participate in the excitation are the ones directly on resonance; i.e., we set $\omega_\mathrm{I}\approx \omega_{ba}$ and $L(\omega)\approx \delta(\omega - \omega_{ca})$. Evaluating both integrals, setting the incident frequency to be resonant with the two-photon transition ($\bar\omega_p = \omega_{ca}$) and using the property that $s^2_\mathrm{II}(\bar\omega_{p} - \omega_{ba}) = s^2_\mathrm{I}(\omega_{ba})$, the coherent and incoherent excitation rates simplify to
\begin{equation}
\label{eq:squeezed_CW_broadband_coherent}
    \overline r^\text{sq,c}_\mathrm{cw} \to  \frac{\eta}{\Gamma_c}\frac{s_\mathrm{I}^2(\omega_{ba})c_\mathrm{I}^2(\omega_{ba})}{2\pi A_\text{eff}^2},
\end{equation}
\begin{equation}
    \overline r^\text{sq,ic}_\mathrm{cw} \to \frac{\eta}{\Gamma_b} \frac{s^4_\mathrm{I}(\omega_{ba})}{2\pi A_\text{eff}^2}. 
\end{equation}
Taking the ratio of the two contributions 
\begin{equation}
\label{eq:cw_coh/incoh}
    \frac{\overline r^\text{sq,c}_\mathrm{cw}}{\overline r^\text{sq,ic}_\mathrm{cw}} = \frac{\Gamma_b}{\Gamma_c} \left(1 + \frac{1}{s^2_\mathrm{I}(\omega_{ba})}\right) = \frac{\overline R^\text{sq,c}_\mathrm{cw}}{\overline R^\text{sq,ic}_\mathrm{cw}},
\end{equation}
which approaches $\Gamma_b/\Gamma_c$ when $|\bar\beta|\gg 1$ and $1/s^2_\mathrm{I}(\omega_{ba})\ll1$. 

\begin{figure}
    \centering
    \includegraphics[width = 0.25\linewidth]{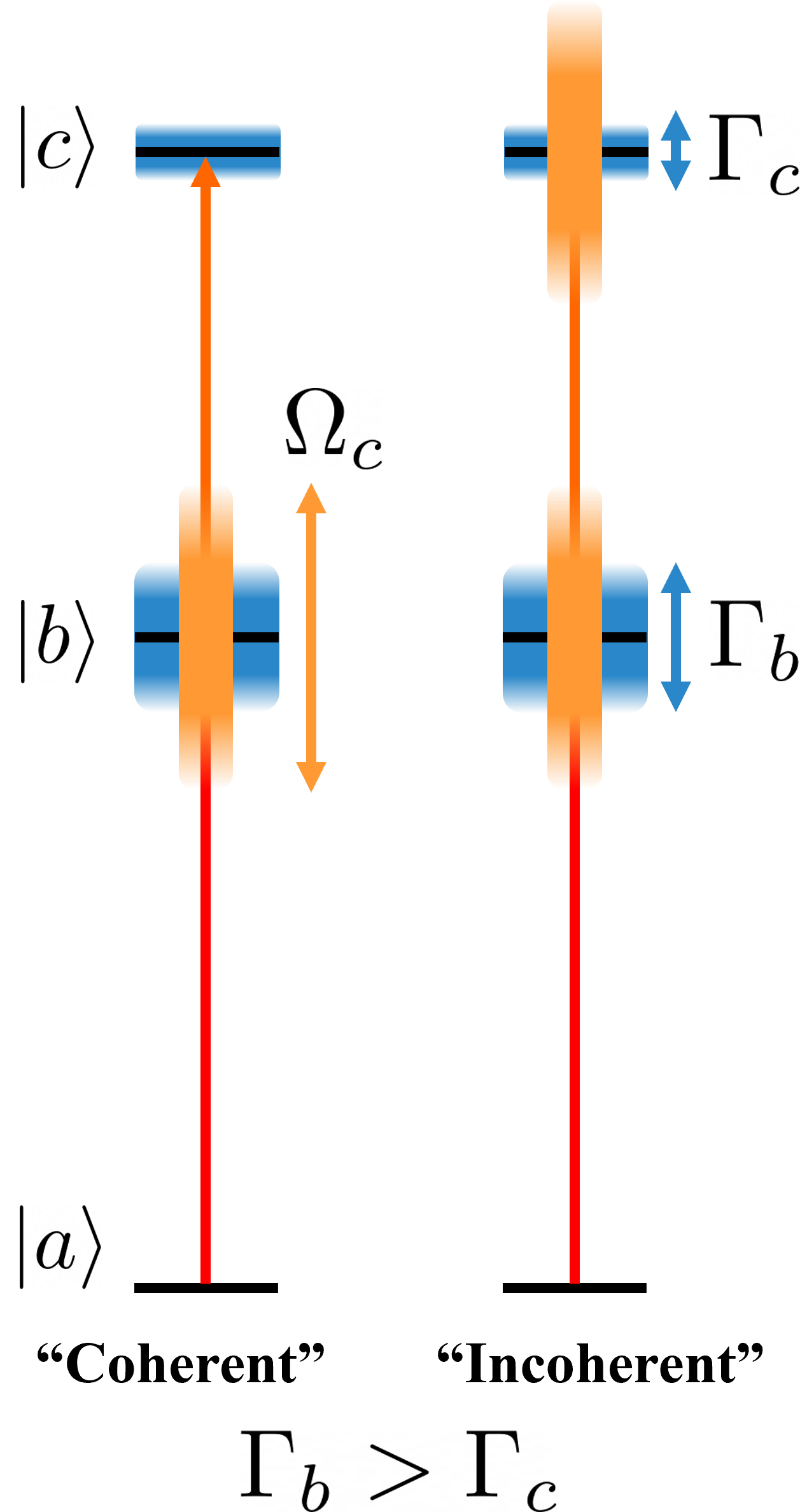}
    \caption{Schematic of two-photon excitation driven by squeezed light in the broadband limit and CW limit when $\Gamma_b>\Gamma_c$. Here we set the incident center frequencies to be resonant and so the coherent contribution is resonant with the two-photon transition. The blue box at each energy level represent that linewidth of that state being either $\Gamma_b$ or $\Gamma_c$. The orange box denotes the bandwidth of squeezed $\Omega_c$.}
    \label{fig:3lvldiag}
\end{figure}

From Eq. \eqref{eq:cw_coh/incoh}, we see very clearly that the difference between the coherent and incoherent contribution in the broadband limit is very sensitive to the properties of the system  
being considered. In Fig. \ref{fig:3lvldiag} we show a schematic of the limit when $\Gamma_b > \Gamma_c$. For the coherent contribution in the broadband limit, when the center frequencies are set to be on resonance with the two-photon transition, all intermediate state photons participate to the excitation and so the limiting factor to the excitation is the two-photon excited state linewidth. On the other hand, for the incoherent contribution in the broadband limit, it is the intermediate state linewidth that is the limiting factor to the excitation rate. Therefore, the coherent contribution is larger (smaller) when $\Gamma_b > \Gamma_c$ ($\Gamma_b < \Gamma_c$) compared to the incoherent contribution. For isolated Cs atoms, $\Gamma_b / \Gamma_c \approx 2.11$ and so their contributions are approximately equal in the broadband limit. However, we stress that this may be very different for other systems such as fluorescent molecules and demonstrates the importance of having a well classified system before calculations are made \cite{PhysRevA.106.023115}. 

In summary, within the CW limit of both classical and squeezed light, we identify narrow regions where squeezed light offers an enhancement. Our calculations indicate that these 
regions should be within the experimentally detectable range. Further, given that this enhancement first increases and then decreases with increasing bandwidth (see second to third column in Fig. \ref{fig:CWplot}), one could find the optimal effective squeezed bandwidth for a given system. However, we stress that our predictions of fluorescence enhancement due to excitation by squeezed light are on the cusp of what could be detected, given that we have not included: 1) dispersion, which would broaden the coherence time of the photon pairs and decrease the enhancement, 2) finite collection efficiencies; 3) other broadening mechanisms such as non-radiative decay, dephasing, or residual Doppler broadening. 

\subsection{Pulsed comparison}
Next we consider the pulsed regime. For classical light, we consider two pulsed sources modeled by the square normalized Gaussian amplitudes: 
\begin{equation}
    \varphi_J(\omega) = \left(\frac{1}{\pi\sigma_J^2}\right)^\frac{1}{4} e^{-\frac{(\omega - \bar\omega_J)^2}{2\sigma_J^2}}.
\end{equation}
To model a pulsed source of squeezed light, we use the double-Gaussian JSA given by Eq. \eqref{eq:double_gaussian} and shown in Fig. \ref{fig:pulsed_JSI_schematic}. As in the CW limit, we set the center frequencies of both incident fields to be resonant with the respective transitions. In general, the number of photons per pulse of squeezed light depends on the two parameters $(\sigma_c/\sigma_p,|\beta|)$. Similar to the CW limit, for the values of $\sigma_c/\sigma_p$ that we consider we adjust the range of $|\beta|$ appropriately so we are always comparing equal incident photon numbers of squeezed and classical light. 

After setting the above restrictions, there are four independent parameters: The two effective bandwidths of the classical sources $\sigma_\mathrm{I}$, $\sigma_\mathrm{II}$, and the widths of the JSA, set by $\sigma_c$ and $\sigma_p$. For our current analysis, we are only interested in frequency anti-correlations, where one expects squeezed light to be advantageous over the use of classical light; thus, we limit the parameter space to the regime where $\sigma_p \le \sigma_c$. To investigate both narrowband and broadband behavior, we consider $\sigma_p/\Gamma_b = 0.1, 1, 10$, and for each of these values we explore $\sigma_c/\sigma_p = 1, 10, 100$. In Fig. \ref{fig:pulsed_JSI_schematic}, we plot the JSA for the three values of $\sigma_c/\sigma_p$, from which we can see it ranges from uncorrelated to strongly correlated.  For pulsed classical light, the coherence time of each individual photon is approximately set by $\sim 1 / \sigma_\mathrm{I}, 1 / \sigma_\mathrm{II}$ for each pulse. On the other hand, for pulsed squeezed light, the photon \emph{pairs} have a coherence time approximately set by $\sim 1/\sigma_c$, similar to the CW limit. To provide a fair comparison to classical light, we explore the regime where the classical pulses have the same coherence time as the squeezed pulse of light; that is, we set $\sigma_\mathrm{I} = \sigma_\mathrm{II} = \sigma_c$. In setting this restriction, each individual photon has the same effective bandwidth, so when we vary $\sigma_c/\sigma_p$, we are effectively probing the influence that the squeezed light spectral-temporal correlations have on the fluorescence. 

\begin{figure}
    \centering
    \includegraphics[width = \linewidth]{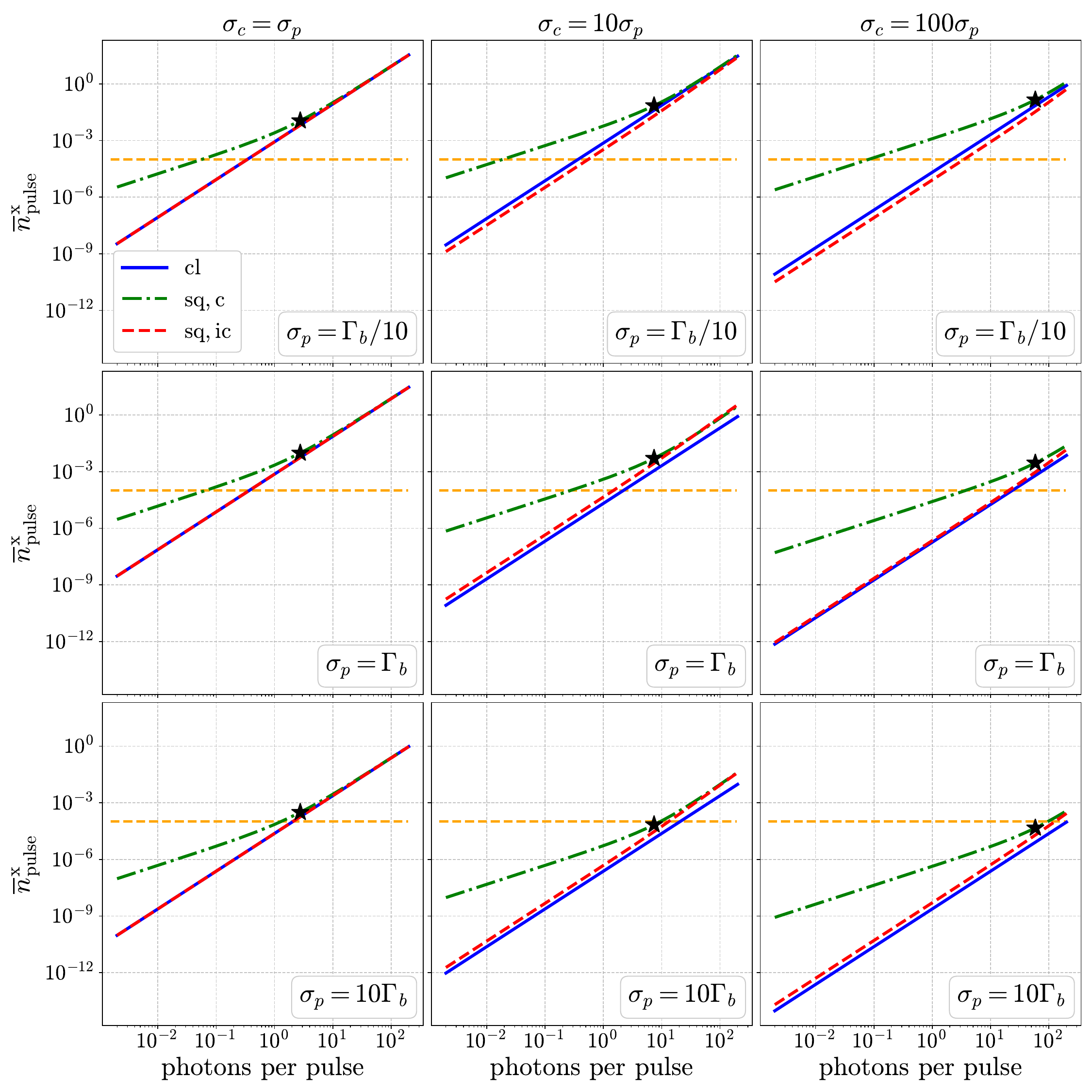}
    \caption{Fluorescence emission count per pulse for classical and the coherent and incoherent contributions to squeezed light. The orange dashed horizontal line at $100/10^{6}$ corresponds to 100 counts/second for a modest repetition rate of 1 MHz.}
    \label{fig:pulsed_plot_1}
\end{figure}

In Fig. \ref{fig:pulsed_plot_1} we plot the fluorescence count for a pulse of classical light (see Eq. \eqref{eq:pulsed_classical_excitation_prob}) and the coherent and incoherent contributions for squeezed light (see Eqs. \eqref{eq:coherentcont_PULSED} and \eqref{eq:incoherentcont_PULSED}) for each pair of parameters discussed above. Each plot is a function of the incident photon number per pulse. As we go down a column and across a row, $\sigma_p$ and $\sigma_c$ respectively increase. Since we are plotting the emission count \emph{per pulse}, we include the orange dashed horizontal line at $100/10^{6}$, which for a modest repetition rate of 1 MHz identifies an effective fluorescence count rate of 100 counts/second. We take this to identify the minimum count rate for detection, in agreement with the detectable regime set in the CW limit. In each plot, the star represents the approximate ``cross-over'' point from a linear to quadratic scaling. This is plotted by calculating the photon number and fluorescence count when $|\beta|\sqrt{p_0} = 1$, analogous to the CW limit (see Fig. \ref{fig:CWplot}). As in the CW limit,  the cross-over point generally moves from left to right in each row due to the change in the incident photon number with $\sigma_c$. The identification of the cross-over point for pulsed excitation is in principle more subtle than for CW excitation, because for pulsed excitation there are many Schmidt modes that contribute to the scaling; nonetheless, by taking the largest contribution due to $p_0$ we find that we get a good approximation of the cross-over point. We refer to ``low'' and ``high'' photon number limit as being before or after the cross-over point respectively.

Consider the first column of Fig. \ref{fig:pulsed_plot_1}, which corresponds to $\sigma_c = \sigma_p$ and $\sigma_p/\Gamma_b = 1/10,1,10$. As we move down a column, the classical result for the fluorescence, the coherent contribution to the fluorescence from the excitation by squeezed light, and the incoherent contribution to fluorescence from the excitation by squeezed light all decrease. This is due to the bandwidths increasing, yielding pulse energies more spread out over the atomic linewidths. For this set of parameters, the incoherent contribution is always equal to the classical result and is equal to the coherent contribution in the high photon number limit. This behavior is as result of the JSA being separable and equal to
\begin{equation}
    \gamma(\1,\2)  \to \varphi_\mathrm{I}(\1)\varphi_\mathrm{II}(\2), 
\end{equation}
where the right-hand side involves the classical pulse spectral amplitudes; this holds since we set $\sigma_\mathrm{I}=\sigma_\mathrm{II} = \sigma_c$. In this limit, the photons within each pair have the same effective bandwidth, set to be equal to the bandwidth of each photon in the classical pulse. It then follows that $\bar n^\mathrm{sq,c}_\mathrm{pulse} / \bar n^\mathrm{sq,ic}_\mathrm{pulse} = (1 + 1 / N_\mathrm{I})$ and $\bar n^\mathrm{sq}_\mathrm{pulse}= (2 + 1 / N_\mathrm{I})\bar n^\mathrm{cl}_\mathrm{pulse}$, where $N_\mathrm{I} = N_\mathrm{II}$ is the number of photons in each classical pulse. Thus, in the separable limit for pulsed sources, we recover the $ 1 / N_\mathrm{I}$ dependence, which shows that as $N_\mathrm{I}$ decreases the ratio of squeezed to classical light increases, but the overall fluorescence decreases. Finally, in the narrowband regime, we find a small region where the coherent contribution is larger than the classical result, and should be experimentally detectable. 

Next consider the first row, where $\sigma_p = \Gamma_b/10$ and $\sigma_c$ increases from left to right. Notice that the classical result, squeezed coherent contribution, and squeezing incoherent contribution to the fluorescence all generally decrease with increasing bandwidth.  However, similar to the CW limit (see the discussion surrounding Fig. \ref{fig:CWplot}), the cross-over point moves to the right, extending the region where the coherent contribution is larger than the incoherent contribution and the classical result. It is in this region of parameter space that we find the largest benefit of using squeezed light over classical light. The fluorescence count from using squeezed light, due to the coherent contribution, is detectable and approximately three orders of magnitude larger than the classical result. 

Consider now the bottom row in Fig. \ref{fig:pulsed_plot_1}, where $\sigma_p = 10\Gamma_b$, and $\sigma_c$ is increasing from left to right. In this regime, photons from classical and squeezed light are broadband. Further, the coherent contribution to the fluorescence from excitation with squeezed light involves excitation by photons that are anti-correlated, with a bandwidth larger than the two-photon excited state linewidth. Moving from the left to rightmost column, the classical result and the coherent and incoherent contributions to fluorescence from the excitation by squeezed light all decrease, but not by equal amounts. In the middle and right columns, the classical result decreases, but the coherent and incoherent contributions remain approximately equal in the high photon number limit. Na\"{i}vely, one would expect the incoherent contribution to decrease as does the classical result, due to the lack of the photon pair correlations and larger bandwidth.  

The unique relationship between the coherent and incoherent contributions to the fluorescence in the broadband and high photon number limit is subtle but straightforward. The photon number in each Schmidt mode is equal to $\mathrm{sinh}^2(|\beta_n|)$ (see \eqref{eq:pulsesqueezedphotons}) and scales nonlinearly with $|\beta_n|$. This nonlinear scaling effectively ``washes out'' the spectral correlations of the squeezed light in the high photon number limit. To see this effect in detail, we take the broadband limit of the coherent and incoherent contributions to the fluorescence following the excitation by squeezed light. To good approximation, Eq. \eqref{eq:coherentcont_PULSED} and \eqref{eq:incoherentcont_PULSED} are given by
\begin{align}
        \overline p^\text{sq,c}_\text{pulse} \to \frac{\eta}{2A_\text{eff}^2} \left|\sum_n f_\mathrm{IIn}(\omega_{cb})f_\mathrm{In}(\omega_{ba})s_n c_n\right|^2\equiv \frac{\eta}{2A_\text{eff}^2} G^{(2)}_\text{c}(\omega_{ba},\omega_{cb}),\\
        \overline p^\text{sq,ic}_\text{pulse} \to \frac{\eta}{2A_\text{eff}^2} G^{(1)}_{\mathrm{I}}(\omega_{ba})G^{(1)}_{\mathrm{II}}(\omega_{cb}) \equiv \frac{\eta}{2A_\text{eff}^2}G^{(2)}_\text{ic}(\omega_{ba},\omega_{cb}),
\end{align}
where $G^{(1)}_J(\omega)$ is the first-order correlation given by Eq. \eqref{eq:G1pulsed} evaluated at the same frequencies, and $G^{(2)}_\mathrm{c}(\1,\2)$ and $G^{(2)}_\mathrm{ic}(\1,\2)$ are the coherent (first) and incoherent (second) contributions to the second-order correlation function, given by Eq. \eqref{eq:pulsedG2} evaluated at $\1 = \1^\prime$ and $\2 = \2^\prime$. In this limit, comparing the coherent and incoherent contributions to the fluorescence following the excitation by squeezed light is equivalent to comparing the coherent and incoherent contributions to the second-order correlation function of the squeezed light itself.

\begin{figure}
    \centering
    \includegraphics[width = 0.5\linewidth]{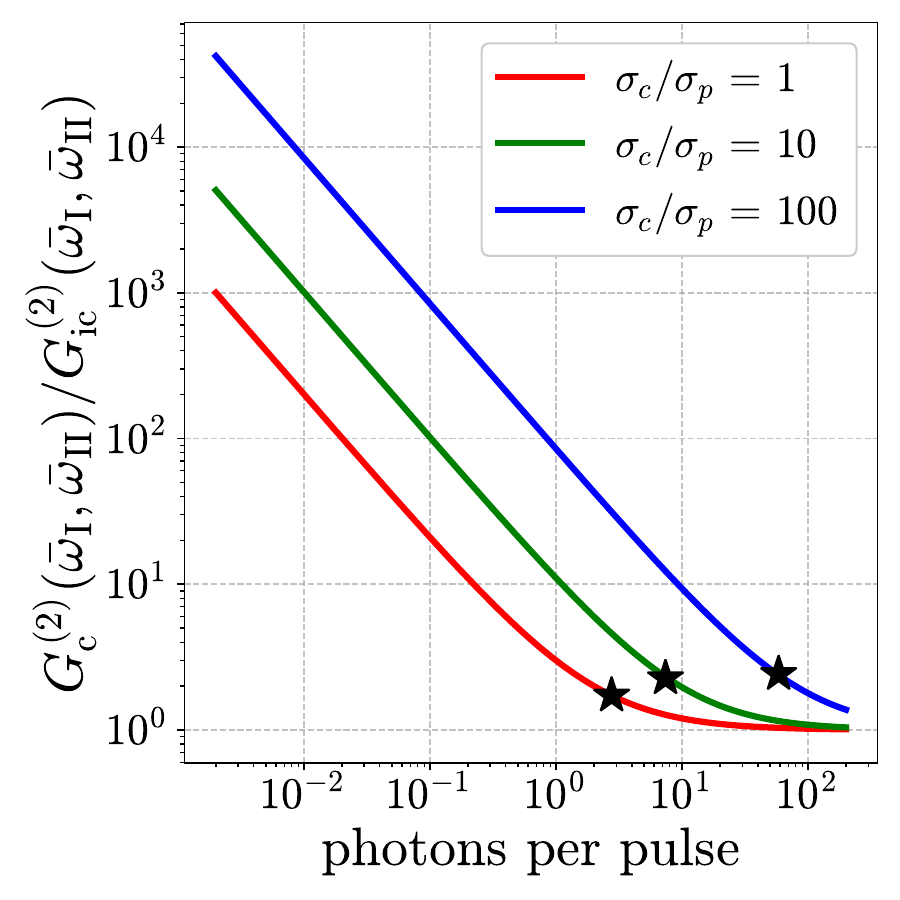}
    \caption{Ratio of the coherent to incoherent contribution to the second-order correlation function from squeezed light.}
    \label{fig:pulsed_plot_3}
\end{figure}

\begin{figure}
    \centering
    \includegraphics[width = \linewidth]{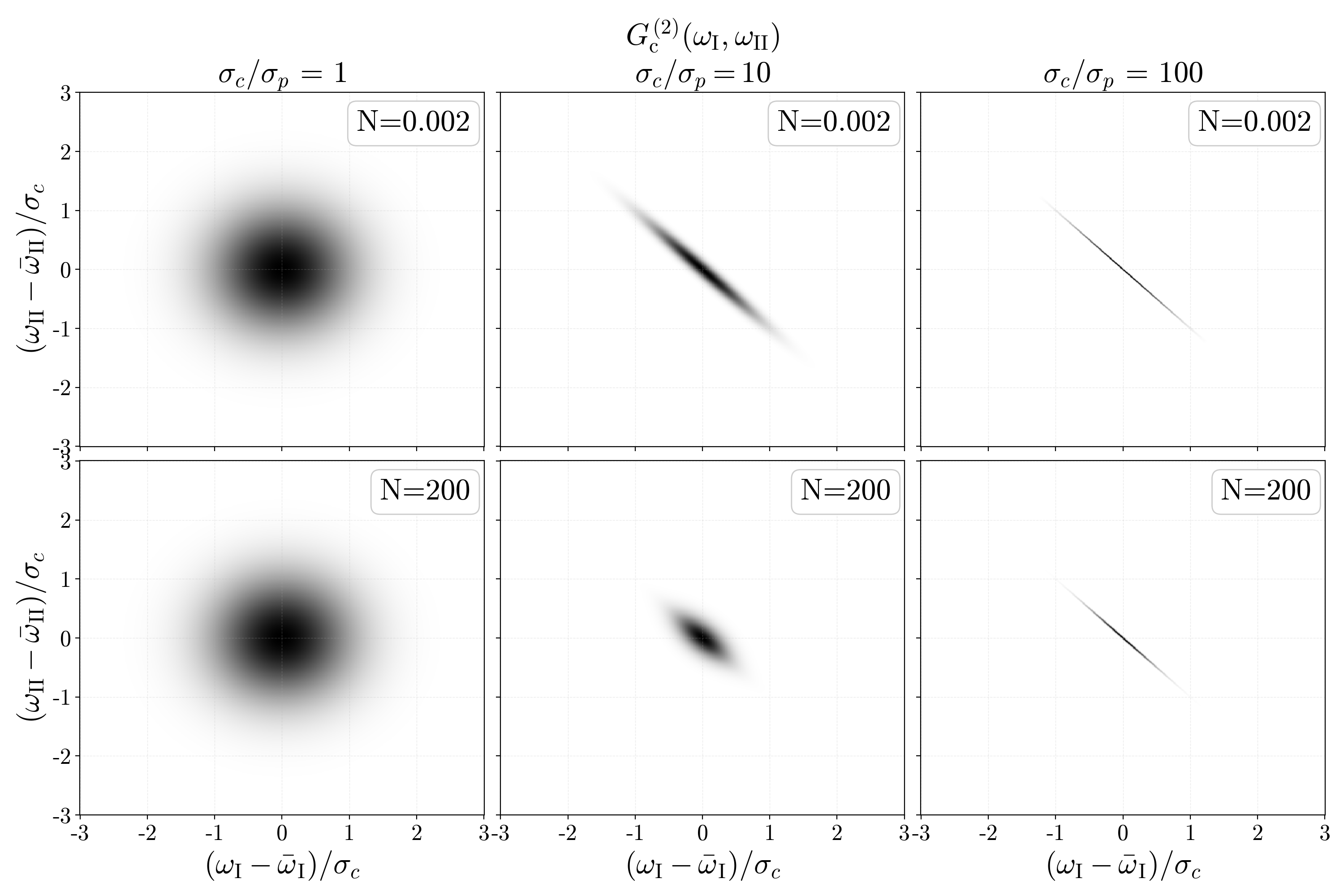}
    \caption{Coherent contribution to the second-order correlation function for squeezed light at difference degrees of spectral-temporal correlation and incident photon numbers. Each plot is scaled by its maximum value.}
    \label{fig:pulsed_plot_4}
\end{figure}

In Fig. \ref{fig:pulsed_plot_3} we plot the ratio of these contributions to the second-order correlation function -- evaluated at the center frequencies of the light -- as a function of photon number for the three values of $\sigma_c/\sigma_p$; the black stars represent the same cross-over points as in Fig. \ref{fig:pulsed_plot_1}. The cross-over points exhibit the same trend of moving to right. Notice that at sufficiently high photon numbers, all three curves approach unity. This is evident for $\sigma_c/\sigma_p = 1$ since the state is separable and there is only one Schmidt mode. For the other curves, this occurs because $s_n = \sinh(|\beta_n|)$, and for large enough $|\beta_n|$ we have $s_0\gg s_1$, so the correlation function is dominated by the first Schmidt mode contribution. In Fig. \ref{fig:pulsed_plot_4} we plot the coherent contribution to the second-order correlation function (normalized by its maximum value) for the three values of $\sigma_c/\sigma_p$, and the minimum and maximum values of the photon number in Fig. \ref{fig:pulsed_plot_3}; the corresponding JSA is shown in Fig. \ref{fig:pulsed_JSI_schematic}. The first column is separable, so only one Schmidt mode contributes to the correlation function, and it remains unchanged when the photon number is increased. However, for the second and third columns, the JSA is correlated, and its decomposition involves many Schmidt modes. As a result, we see that the coherent contribution to the correlation function narrows in the anti-diagonal direction, despite the fact that the JSA itself is still very correlated. Thus for large enough $|\beta_n|$ 
\begin{equation}
     \begin{split}
         G^{(2)}_\text{c}(\1,\2) \equiv \left|\sum_n f_\mathrm{IIn}(\1)f_\mathrm{In}(\2)s_n c_n\right|^2 \approx |f_\mathrm{II0}(\1)|^2|f_\mathrm{I0}(\2)|^2s_0^4\\
         G^{(2)}_\text{ic}(\1,\2) \equiv G^{(1)}_{\mathrm{I}}(\1)G^{(1)}_{\mathrm{II}}(\2) \approx |f_\mathrm{II0}(\1)|^2|f_\mathrm{I0}(\2)|^2s_0^4,
     \end{split}
\end{equation}
and so $G^{(2)}_\text{c}(\1,\2) / G^{(2)}_\text{ic}(\1,\2)  \approx 1$, and it follows that $n^\text{sq,c}_\text{pulse} / n^\text{sq,ic}_\text{pulse} \approx 1$ for excitation by pulsed light in the high photon number, broadband limit.

Finally, from Fig. \ref{fig:pulsed_plot_1} we note that there are regions where the coherent contribution to the fluorescence from the excitation by squeezed light is larger than the classical fluorescence within what we deem a detectable regime. Unsurprisingly, the best scenarios are in the first row of Fig. \ref{fig:pulsed_plot_1}, where $\sigma_p = \Gamma_b/10$ and $\sigma_c = \Gamma_b, 10\Gamma_b$. Beyond this regime, larger $\sigma_c$ and $\sigma_p$ begin to yield diminishing returns due to the larger bandwidths. For these parameters, we are achieving the best of both worlds of the spectral-temporal correlation. That is, the photon pairs are anti-correlated to sum to the two-photon transition within a width set by $\sigma_p = \Gamma_b/10$, and each individual photon (with an effective bandwidth on the order of $\sigma_c = \Gamma_b, 10\Gamma_b$) has a bandwidth comparable to the intermediate state linewidth. However, we again stress that our calculated fluorescence counts are for particular beam parameters and neglect other broadening effects, such as non-radiative decay, dephasing and Doppler broadening, as well as various collection efficiencies present in a real experiment.  

\section{Conclusion \label{sec:conculsion}}
Squeezed light is reported to significantly enhance the two-photon excitation of fluorescent molecules relative to the use of classical light \cite{tabakaev2021energy, tabakaev2022spatial, villabona2018two, li2020squeezed, villabona2017entangled, varnavski2017entangled, varnavski2020two, varnavski2022quantum}. If true, squeezed light could be a useful resource for microscopy applications. Given that the enhancement that squeezed light provides is only present when the incident photon rate (or photon number for a pulsed source) is very low, it has been the subject of much debate \cite{schlawin2018entangled, schlawin2017entangled, dorfman2016nonlinear, dayan2007theory, fei1997entanglement, schlawin2013photon, oka2018two, lee2006entangled, tabakaev2021energy, tabakaev2022spatial, villabona2018two, li2020squeezed, villabona2017entangled, varnavski2017entangled, PhysRevApplied.15.044012, mikhaylov2022hot, hickam2022single, landes2024limitations, Raymer:21, drago2023two, raymer2022theory, landes2021quantifying}. In this study, our goal was to make careful calculations of the two-photon excitation of a system driven by classical and squeezed light. Our attention was not limited to whether squeezed light provides an enhancement compared to classical light in a particular regime of parameter space, but crucially included the question of whether an enhancement -- if any -- could be detected in an ideal experiment.

To determine an answer, we developed a model of a system interacting with the quantized electromagnetic field. Our model includes the effects of radiation reaction and a quantum reservoir, which respectively lead to the radiative and non-radiative broadening of excited states. With these two contributions to the dynamics, we derived general equations for the total scattered and absorbed energy in a consistent way that respects energy conservation. From our model, we are able to make calculations that simulate either absorption or fluorescence based detection schemes of squeezed light two-photon excitation. In the present work, our focus was on the emitted fluorescence of a system when driven by classical and squeezed light.

To optimize our chances of simulating a detectable enhancement, we considered atomic cesium. Atomic systems provide a useful testbed over the use of fluorescent molecules due to their narrow linewidths, relatively simple structure, and well known properties. These features lead to higher excitation rates, and allow a simpler and more accurate model that does not need to take into account other physical effects \cite{mikhaylov2022hot, rocha2024reviewing}. Further, we consider the Cs atoms to be in a MOT. Cooling and confining the atoms at low temperature allows one to focus the incident beams tightly in an experiment, and decreases other broadening effects such as Doppler and collisional broadening. This enhances the excitation probability and improves detection rates in an experiment.

Using our model, we then calculated the fluorescence emission of Cs atoms driven by non-degenerate classical and squeezed light. In both cases we set the incident light to be resonant with the two atomic transitions. We consider this pumping scheme due to its effectiveness for two-photon excitation. We compared classical and squeezed light in the CW and pulsed regimes. In the CW limit, we explored the the transition from narrowband to broadband excitation, and calculated the fluorescence emission \textit{rate} as a function of the incident photon rate. In the pulsed regime, we similarly examined the narrowband to broadband as well as the correlated to uncorrelated regimes, and calculated the fluorescence emission \textit{count} as a function fo the incident photon number.

In the CW limit, we were able to find regions of parameter space where squeezed light provides a \textit{moderate} enhancement. This enhancement is in what we deem to be a detectable regime before other factors due to experimental conditions are considered. We demonstrated that there is a trade-off between the spectral and temporal correlation of the photon pairs provided by squeezed light when there is a resonant intermediate state. Finally, we discussed that -- contrary to usual assumptions -- the incoherent contribution to the squeezed light excitation cannot not be neglected, and is very dependent on the system parameters in the broadband limit.  

In the pulsed regime, we came to similar conclusions on the feasibility of measuring an enhancement due to squeezed light. Again, we found that the incoherent contribution cannot be neglected. Further, in the high photon number and broadband limit, the incoherent contribution to squeezed light always approaches the coherent contribution. It was shown that this behavior is a result of the nonlinear scaling of the second-order correlation function with photon number, and leads to a degradation of the correlations present within the coherent contribution.  

For CW and pulsed excitation of an atomic system, under ideal pumping and system conditions, squeezed light, in theory, does provide an enhancement when compared to classical light. The calculated enhancement is expected to be experimentally measurable. However, we stress that the enhancement is moderate compared to what has been previously reported \cite{schlawin2018entangled, schlawin2017entangled, dorfman2016nonlinear, dayan2007theory, fei1997entanglement, schlawin2013photon, oka2018two, lee2006entangled, tabakaev2021energy, tabakaev2022spatial, villabona2018two, li2020squeezed, villabona2017entangled, varnavski2017entangled}. Further, any other broadening effects that are not included in this study would only hinder detection capabilities.

To optimize an experiment for detecting an enhancement from squeezed light, it is best to start with a system that has narrow linewidths, and to minimize other sources of broadening. The bandwidth of the two-photon excitation should be less than or comparable to the two-photon linewidth, and is naturally met in the CW limit, but requires careful tuning in the pulsed regime.

When pumping near an intermediate state resonance, the bandwidth of each photon -- set by the inverse of the coherence time -- should be chosen to maximize the spectral overlap with the intermediate state, thereby enhancing excitation. Conversely, when pumping far from resonance, increasing the photon bandwidth can improve excitation. Ideally, the bandwidth should be broadened just enough for some frequency components to contribute resonantly, but not beyond that, again ensuring optimal overlap with the intermediate state.

\begin{acknowledgments}
The authors thank Alan McLean and Ralph Jimenez for valuable discussions. This work was supported by the Natural Sciences and Engineering Research Council of Canada (NSERC). C. D. acknowledges an Ontario Graduate Scholarship. 
\end{acknowledgments}

\onecolumngrid
\newpage
\appendix

\section{Equations of motion \label{app:Equations of motion}}
Using Eq. \eqref{eq:dynamics Obar} to determine the dynamics of each $\bar O_\alpha(t)$, we plug in each $\sbar{pq}$ and evaluate the commutator; the resulting set of coupled differential equations are given by
\begin{subequations}
\label{app:dynamics sbar v1}
    \begin{gather}
        \left(\frac{d}{dt} + i\omega_{ba}\right)\sbar{ab} = i\left[\sbar{aa}\fbar{ba}{+} - \sbar{bb}\fbar{ba}{+} + \fbar{bc}{-}\sbar{ac}-\sbar{db}\fbar{da}{+}\right],\\
        \left(\frac{d}{dt} + i\omega_{da}\right)\sbar{ad} = i\left[\sbar{aa}\fbar{da}{+} - \sbar{dd}\fbar{da}{+} + \fbar{dc}{-}\sbar{ac}-\sbar{bd}\fbar{ba}{+}\right],\\
        \left(\frac{d}{dt} + i\omega_{ca}\right)\sbar{ac} = i\left[ \sbar{ab}\fbar{cb}{+}+\sbar{ad}\fbar{cd}{+} - \sbar{bc}\fbar{ba}{+} - \sbar{dc}\fbar{da}{+}     \right],\\
        \left(\frac{d}{dt} + i\omega_{cb}\right)\sbar{bc} = i\left[\sbar{bb}\fbar{cb}{+} - \sbar{cc}\fbar{cb}{+} + \sbar{bd}\fbar{cd}{+} - \fbar{ab}{-}\sbar{ac} \right],\\
        \left(\frac{d}{dt} + i\omega_{cd}\right)\sbar{dc} = i\left[\sbar{dd}\fbar{cd}{+} - \sbar{cc}\fbar{cd}{+} + \sbar{db}\fbar{cb}{+} - \fbar{ad}{-}\sbar{ac} \right],\\
        \left(\frac{d}{dt} + i\omega_{db}\right)\sbar{bd} = i\left[\sbar{da}\fbar{ba}{+} + \sbar{dc}\fbar{cb}{+} - \fbar{ad}{-}\sbar{ab} - \sbar{cb}\fbar{cd}{+} \right],\\
        \frac{d}{dt}\sbar{aa} = i\left[ \fbar{ab}{-}\sbar{ab} - \sbar{ba}\fbar{ba}{+} + \fbar{ad}{-}\sbar{ad} - \sbar{ad}\fbar{da}{+} \right],\\
        \frac{d}{dt}\sbar{bb} = i\left[\sbar{ba}\fbar{ba}{+} - \fbar{ab}{-}\sbar{ab} + \fbar{bc}{-}\sbar{bc} - \sbar{bc}\fbar{cb}{+}\right],\\
        \frac{d}{dt}\sbar{dd} = i\left[\sbar{da}\fbar{da}{+} - \fbar{ad}{-}\sbar{ad} + \fbar{dc}{-}\sbar{dc} - \sbar{rc}\fbar{cd}{+}\right],\\
        \frac{d}{dt}\sbar{cc} = i\left[\sbar{cb}\fbar{cb}{+} - \fbar{bc}{-}\sbar{bc} + \sbar{cd}\fbar{cd}{+} - \fbar{dc}{-}\sbar{dc}   \right],
    \end{gather}
\end{subequations}
where $\fbar{ij}{+}$ is defined in Eq. \eqref{eq:fbardef}.

Taking the solution to each $\fbar{pq}{+}$ (Eq. \eqref{eq:fbarsol2}), using the definitions of $\Gamma_{p}$ for $p = b,d,c$ (Eq. \eqref{eq:total_decay_rate}) and setting the terms that do not contribute to zero because there is no incident light driving the transition (see discussion in the paragraph before Section \ref{sec:Scattered and Absorbed energy}), the equations for each $\sbar{pq}$ is given by 
\begin{subequations}
\label{app:dynamics sbar v2}
    \begin{gather}
        \left(\frac{d}{dt} + \frac{\Gamma_{b}}{2} + i\omega_{ba}\right)\sbar{ab} = i\left[\sbar{aa}\fhat{ba}{+} - \sbar{bb}\fhat{ba}{+} + \fhat{bc}{-}\sbar{ac}\right],\\
        \left(\frac{d}{dt} + \frac{\Gamma_{d}}{2} +i\omega_{da}\right)\sbar{ad} = -i\sbar{bd}\fhat{ba}{+},\\
        \left(\frac{d}{dt} + \frac{\Gamma_{c}}{2} +i\omega_{ca}\right)\sbar{ac} = i\left[ \sbar{ab}\fhat{cb}{+} - \sbar{bc}\fhat{ba}{+}\right],\\
        \left(\frac{d}{dt} + \frac{\Gamma_{b}}{2} + \frac{\Gamma_{c}}{2} +i\omega_{cb}\right)\sbar{bc} = i\left[\sbar{bb}\fhat{cb}{+} - \sbar{cc}\fhat{cb}{+} - \fhat{ab}{-}\sbar{ac} \right],\\
        \left(\frac{d}{dt} + \frac{\Gamma_{d}}{2} +\frac{\Gamma_{c}}{2} +i\omega_{cd}\right)\sbar{dc} = i\sbar{db}\fhat{cb}{+},\\
        \left(\frac{d}{dt} + \frac{\Gamma_{d}}{2} +\frac{\Gamma_{b}}{2}+i\omega_{db}\right)\sbar{bd} = i\left[\sbar{da}\fhat{ba}{+} + \sbar{dc}\fhat{cb}{+}\right],\\
        \frac{d}{dt}\sbar{aa} = i\left[ \fhat{ab}{-}\sbar{ab} - \sbar{ba}\fhat{ba}{+}\right] + \Gamma_{ba}\sbar{bb} + \Gamma_{da}\sbar{dd},\\
        \left(\frac{d}{dt} +\Gamma_{b}\right)\sbar{bb} = i\left[\sbar{ba}\fhat{ba}{+} - \fhat{ab}{-}\sbar{ab} + \fhat{bc}{-}\sbar{bc} - \sbar{bc}\fhat{cb}{+}\right] + \Gamma_{cb}\sbar{cc},\\
        \left(\frac{d}{dt} +\Gamma_{d}\right)\sbar{dd} = \Gamma_{cd}\sbar{cc},\\
        \left(\frac{d}{dt} +\Gamma_{c}\right)\sbar{cc} = i\left[\sbar{cb}\fhat{cb}{+} - \fhat{bc}{-}\sbar{bc} \right].
    \end{gather}
\end{subequations}

\section{Scattered and absorbed energy \label{sec:app:Scattered and absorbed energy}}
We start with the general form for the scattered energy in Eq. \eqref{eq: quantum scattering}. Moving to the interaction picture in which we work, the scattered energy is given by
\begin{equation}
\label{eq:B1}
    \mathcal{S} = \frac{1}{6\pi\epsilon_0 c^3}\int\limits_{-\infty}^{\infty}\langle \Psi_\text{in}|\frac{d^2\boldsymbol{\bar{\mu}}_{-}(t)}{dt^2}\cdot\frac{d^2\boldsymbol{\bar{\mu}}_{+}(t)}{dt^2}|\Psi_\text{in}\rangle dt +c.c.,
\end{equation}
where we can freely extend the limits of integration because the interaction only occurs between the ``interaction interval'' $t_\text{min}\le t\le t_\text{max}$, but we take $t_0< t_I < t_\text{min}$ and $t_F> t_\text{max}$. The derivation of scattered energy (see Eq. \eqref{eq: quantum scattering} and \eqref{eq:B1}) was carried out with the interaction Hamiltonian still in its general, but normally ordered form (Eq. \eqref{eq:H_M-EM v3}). In Section \ref{sec:simplesystem}, we specialized to the system field Hamiltonian (Eq. \eqref{eq:H_M-EM v4}), where we treat each frequency band separately. Since each frequency band is independent, the derivation in that limit follows in the 
same way for each term in Eq. \eqref{eq:H_M-EM v4}. Then for the set of frequency bands we are considering, the scattered energy is given by
\begin{equation}
    \begin{split}
        \mathcal{S} =& \frac{|\boldsymbol{\mu}_{ba}|^2}{6\pi\epsilon_0 c^3}\hspace{-1.5mm}\int\limits_{-\infty}^{\infty}\hspace{-1mm}\langle\ddot{\bar{\sigma}}_{ba}(t)\ddot{\bar{\sigma}}_{ab}(t)\rangle dt + \frac{|\boldsymbol{\mu}_{cb}|^2}{6\pi\epsilon_0 c^3}\hspace{-1.5mm}\int\limits_{-\infty}^{\infty}\hspace{-1mm}\langle\ddot{\bar{\sigma}}_{cb}(t)\ddot{\bar{\sigma}}_{bc}(t)\rangle dt\\
        &+\frac{|\boldsymbol{\mu}_{da}|^2}{6\pi\epsilon_0 c^3}\hspace{-1.5mm}\int\limits_{-\infty}^{\infty}\hspace{-1mm}\langle\ddot{\bar{\sigma}}_{da}(t)\ddot{\bar{\sigma}}_{ad}(t)\rangle dt + \frac{|\boldsymbol{\mu}_{cd}|^2}{6\pi\epsilon_0 c^3}\hspace{-1.5mm}\int\limits_{-\infty}^{\infty}\hspace{-1mm}\langle\ddot{\bar{\sigma}}_{cd}(t)\ddot{\bar{\sigma}}_{dc}(t)\rangle dt+ \text{c.c.},
    \end{split}
\end{equation}
where the ``\hspace{1mm}$\dot{}$\hspace{1mm}'' denotes a time derivative. Finally, we apply the slowly varying approximation (Eq. \eqref{eq:sigmabarapprox}) to each term. Together with the definition of $\Gamma_{pq}^\text{r}$ (Eq. \eqref{eq:gammarad}), and combining the complex conjugate terms, the total scattered energy is given by
\begin{equation}
    \begin{split}
    \label{app:eq:scattenergyfinal}
        \mathcal{S} = \hbar\omega_{ba}\Gamma_{ba}^\text{r}\int\limits_{-\infty}^{\infty}\langle\bar{\sigma}_{bb}(t)\rangle dt 
        + \hbar\omega_{da}\Gamma_{da}^\text{r}\int\limits_{-\infty}^{\infty}\langle\bar{\sigma}_{dd}(t)\rangle dt + \hbar\omega_{cb}\Gamma_{cb}^\text{r}\int\limits_{-\infty}^{\infty}\langle\bar{\sigma}_{cc}(t)\rangle dt
        + \hbar\omega_{cd}\Gamma_{cd}^\text{r}\int\limits_{-\infty}^{\infty}\langle\bar{\sigma}_{cc}(t)\rangle dt.
    \end{split}
\end{equation}

Consider the general form of the absorbed energy (Eq. \eqref{eq: quantum absorption}). Moving to our interaction picture and extending the limits of integration, the absorption is given by
\begin{equation}
    \mathcal{A} = \int\limits_{-\infty}^\infty\bra{\Psi_\text{in}}\bar{\boldsymbol{E}}_+(t)\cdot \frac{d\bar{\boldsymbol{\mu}}_+(t)}{dt}\ket{\Psi_\text{in}}dt + \text{c.c.}. 
\end{equation}
As for the scattered energy, we specialized to the system field Hamiltonian (Eq. \eqref{eq:H_M-EM v4}) where we treat each frequency band separately. Again, the derivation follows in the same way for each term in Eq. \eqref{eq:H_M-EM v4}, with the total absorption given by
\begin{equation}
    \begin{split}
        \label{app:eq:Afreqband v1}
        \mathcal{A}=&-i\hbar\omega_{ba}\int\limits_{-\infty}^{\infty}\left\langle\frac{\boldsymbol{\mu}_{ab}\cdot \boldsymbol{\bar{E}}_{-}^{ba}(t)}{\hbar} \bar{\sigma}_{ab}(t)\right\rangle dt -i\hbar\omega_{cb}\int\limits_{-\infty}^{\infty}\left\langle\frac{\boldsymbol{\mu}_{bc}\cdot \boldsymbol{\bar{E}}_{-}^{cb}(t)}{\hbar} \bar{\sigma}_{bc}(t)\right\rangle dt\\
        &-i\hbar\omega_{da}\int\limits_{-\infty}^{\infty}\left\langle\frac{\boldsymbol{\mu}_{ad}\cdot \boldsymbol{\bar{E}}_{-}^{da}(t)}{\hbar} \bar{\sigma}_{ad}(t)\right\rangle dt -i\hbar\omega_{cd}\int\limits_{-\infty}^{\infty}\left\langle\frac{\boldsymbol{\mu}_{dc}\cdot \boldsymbol{\bar{E}}_{-}^{cd}(t)}{\hbar} \bar{\sigma}_{dc}(t)\right\rangle dt +\text{c.c.}.
    \end{split}
\end{equation}
In deriving Eq. \eqref{app:eq:Afreqband v1} we applied the slowly varying approximation (Eq. \eqref{eq:sigmabarapprox}) and suggestively added $\hbar$ in the numerator and denominator of each term. To push forward, we consider a general term in the absorption equation (Eq. \eqref{app:eq:Afreqband v1}) given by
\begin{equation}
T_{ij} = -i\hbar\omega_{ij}\int\limits_{-\infty}^{\infty}\left\langle\frac{\boldsymbol{\mu}_{ji}\cdot \boldsymbol{\bar{E}}_{-}^{ij}(t)}{\hbar} \bar{\sigma}_{ji}(t)\right\rangle dt + \text{c.c.},
\end{equation}
for $i>j$ and input the adjoint of $\fbar{ij}{+}$ (Eq. \eqref{eq:fbardef}) rearranged so that
\begin{equation}
    \frac{\boldsymbol{\mu}_{ji}}{\hbar}\cdot\bar{\boldsymbol{E}}_-^{ij}(t) = \fbar{ji}{-} + \eta_{ij}\bar{\psi}^\dagger_{ij}(0,t),
\end{equation}
then
\begin{equation}
    T_{ij} = -i\hbar\omega_{ij}\int\limits_{-\infty}^{\infty}\left\langle\fbar{ji}{-} \bar{\sigma}_{ji}(t)\right\rangle dt -i\hbar\omega_{ij}\int\limits_{-\infty}^{\infty}\left\langle\eta_{ij}\bar{\psi}_{ij}(0,t) \bar{\sigma}_{ji}(t)\right\rangle dt + \text{c.c.}.
\end{equation}
Working out the second term using the solution for $\bar{\psi}_{ij}(0,t)$ (Eq. \eqref{eq:psibar solution})
\begin{equation}
    \begin{split}
        &-i\hbar\omega_{ij}\int\limits_{-\infty}^{\infty}\left\langle\eta_{ij}\bar{\psi}_{ij}(0,t) \bar{\sigma}_{ji}(t)\right\rangle dt + \text{c.c.} \\
        &= -i\hbar\omega_{ij}\int\limits_{-\infty}^{\infty}\left\langle\eta_{ij}\hat{\psi}_{ij}^\dagger(0,t) \bar{\sigma}_{ji}(t)\right\rangle dt  +\hbar\omega_{ij}\frac{|\eta_{ij}|^2}{2v_{ij}}\int\limits_{-\infty}^{\infty}\left\langle \bar{\sigma}_{ii}(t)\right\rangle dt + \text{c.c.}\\
        &=\hbar\omega_{ij}\Gamma_{ij}^\text{nr}\int\limits_{-\infty}^{\infty}\left\langle \bar{\sigma}_{ii}(t)\right\rangle dt,
    \end{split}
\end{equation}
where the first term is always zero because the initial reservoir ket is the vacuum state so $\hat{\psi}_{ij}(0,t) \ket{\Psi_\text{in}}= 0$; then 
\begin{equation}
    T_{ij} = -i\hbar\omega_{ij}\int\limits_{-\infty}^{\infty}\left\langle\fbar{ji}{-} \bar{\sigma}_{ji}(t) - \bar{\sigma}_{ij}(t)\fbar{ij}{+}\right\rangle dt + \hbar\omega_{ij}\Gamma_{ij}^\text{nr}\int\limits_{-\infty}^{\infty}\left\langle \bar{\sigma}_{ii}(t)\right\rangle dt.
\end{equation}
Using the general form $T_{ij}$ for each term in Eq. \eqref{app:eq:Afreqband v1}, the absorption is given by
\begin{equation}
    \begin{split}
        \mathcal{A} =& \hbar\omega_{ba}\Gamma_{ba}^\text{nr}\int\limits_{-\infty}^{\infty}\left\langle \bar{\sigma}_{bb}(t)\right\rangle dt + \hbar\omega_{da}\Gamma_{da}^\text{nr}\int\limits_{-\infty}^{\infty}\left\langle \bar{\sigma}_{dd}(t)\right\rangle dt + \hbar\omega_{cb}\Gamma_{cb}^\text{nr}\int\limits_{-\infty}^{\infty}\left\langle \bar{\sigma}_{cc}(t)\right\rangle dt+ \hbar\omega_{cd}\Gamma_{cd}^\text{nr}\int\limits_{-\infty}^{\infty}\left\langle \bar{\sigma}_{cc}(t)\right\rangle dt\\
        &-i\hbar\omega_{ba}\int\limits_{-\infty}^{\infty}\left\langle\fbar{ab}{-} \bar{\sigma}_{ab}(t) - \bar{\sigma}_{ba}(t)\fbar{ba}{+}\right\rangle dt-i\hbar\omega_{da}\int\limits_{-\infty}^{\infty}\left\langle\fbar{ad}{-} \bar{\sigma}_{ad}(t) - \bar{\sigma}_{da}(t)\fbar{da}{+}\right\rangle dt\\
        &-i\hbar\omega_{cb}\int\limits_{-\infty}^{\infty}\left\langle\fbar{bc}{-} \bar{\sigma}_{bc}(t)- \bar{\sigma}_{cb}(t)\fbar{cb}{+}\right\rangle dt-i\hbar\omega_{cd}\int\limits_{-\infty}^{\infty}\left\langle\fbar{dc}{-} \bar{\sigma}_{dc}(t)- \bar{\sigma}_{cd}(t)\fbar{cd}{+}\right\rangle dt.
    \end{split}
\end{equation}
We expand each $\omega_{ij}$ for the last four terms and group terms proportional to each $\omega_i$; then
\begin{equation}
    \begin{split}
        \mathcal{A} =& \hbar\omega_{ba}\Gamma_{ba}^\text{nr}\int\limits_{-\infty}^{\infty}\left\langle \bar{\sigma}_{bb}(t)\right\rangle dt + \hbar\omega_{da}\Gamma_{da}^\text{nr}\int\limits_{-\infty}^{\infty}\left\langle \bar{\sigma}_{dd}(t)\right\rangle dt + \hbar\omega_{cb}\Gamma_{cb}^\text{nr}\int\limits_{-\infty}^{\infty}\left\langle \bar{\sigma}_{cc}(t)\right\rangle dt+ \hbar\omega_{cd}\Gamma_{cd}^\text{nr}\int\limits_{-\infty}^{\infty}\left\langle \bar{\sigma}_{cc}(t)\right\rangle dt\\
        &+i\hbar\omega_a \int\limits_{-\infty}^{\infty}\left\langle\fbar{ab}{-} \bar{\sigma}_{ab}(t) - \bar{\sigma}_{ba}(t)\fbar{ba}{+}+\fbar{ad}{-} \bar{\sigma}_{ad}(t) - \bar{\sigma}_{da}(t)\fbar{da}{+}\right\rangle dt\\
        &+i\hbar\omega_b \int\limits_{-\infty}^{\infty}\left\langle  \bar{\sigma}_{ba}(t)\fbar{ba}{+}-\fbar{ab}{-} \bar{\sigma}_{ab}(t)+\fbar{bc}{-} \bar{\sigma}_{bc}(t)- \bar{\sigma}_{cb}(t)\fbar{cb}{+}\right\rangle dt\\
        &+i\hbar\omega_d \int\limits_{-\infty}^{\infty}\left\langle \bar{\sigma}_{da}(t)\fbar{da}{+}-\fbar{ad}{-} \bar{\sigma}_{ad}(t)+\fbar{bc}{-} \bar{\sigma}_{dc}(t)- \bar{\sigma}_{cd}(t)\fbar{cd}{+}\right\rangle dt\\
        &+i\hbar\omega_c \int\limits_{-\infty}^{\infty}\left\langle \bar{\sigma}_{cb}(t)\fbar{cb}{+}-\fbar{bc}{-} \bar{\sigma}_{bc}(t)+ \bar{\sigma}_{cd}(t)\fbar{cd}{+}-\fbar{dc}{-} \bar{\sigma}_{dc}(t)\right\rangle dt.
    \end{split}
\end{equation}
We note that the the integrand of the last four terms are the same as the terms on the right side of  the equations of motion for each population term $\sbar{aa},\sbar{bb},\sbar{dd}$, and $\sbar{cc}$ given in Eq. \eqref{app:dynamics sbar v1}. Inputting each equation for the population equations of motion, the absorption reduces to
\begin{equation}
\label{eq:appA}
    \begin{split}
        \mathcal{A} &= \hbar\omega_{ba}\Gamma_{ba}^\text{nr}\int\limits_{-\infty}^{\infty}\left\langle \bar{\sigma}_{bb}(t)\right\rangle dt + \hbar\omega_{da}\Gamma_{da}^\text{nr}\int\limits_{-\infty}^{\infty}\left\langle \bar{\sigma}_{dd}(t)\right\rangle dt + \hbar\omega_{cb}\Gamma_{cb}^\text{nr}\int\limits_{-\infty}^{\infty}\left\langle \bar{\sigma}_{cc}(t)\right\rangle dt+ \hbar\omega_{cd}\Gamma_{cd}^\text{nr}\int\limits_{-\infty}^{\infty}\left\langle \bar{\sigma}_{cc}(t)\right\rangle dt\\
        &\hspace{4mm}+\hbar\omega_a \int\limits_{-\infty}^{\infty}\frac{d}{dt}\left\langle\sbar{aa}\right\rangle dt +\hbar\omega_b \int\limits_{-\infty}^{\infty}\frac{d}{dt}\left\langle\sbar{bb}\right\rangle dt+\hbar\omega_d \int\limits_{-\infty}^{\infty}\frac{d}{dt}\left\langle\sbar{dd}\right\rangle dt+\hbar\omega_c \int\limits_{-\infty}^{\infty}\frac{d}{dt}\left\langle\sbar{cc}\right\rangle dt\\
        &=\hbar\omega_{ba}\Gamma_{ba}^\text{nr}\int\limits_{-\infty}^{\infty}\left\langle \bar{\sigma}_{bb}(t)\right\rangle dt + \hbar\omega_{da}\Gamma_{da}^\text{nr}\int\limits_{-\infty}^{\infty}\left\langle \bar{\sigma}_{dd}(t)\right\rangle dt + \hbar\omega_{cb}\Gamma_{cb}^\text{nr}\int\limits_{-\infty}^{\infty}\left\langle \bar{\sigma}_{cc}(t)\right\rangle dt+ \hbar\omega_{cd}\Gamma_{cd}^\text{nr}\int\limits_{-\infty}^{\infty}\left\langle \bar{\sigma}_{cc}(t)\right\rangle dt,
    \end{split}
\end{equation}
where the integral over the time derivative of the population terms is zero because we have assumed that the boundary of integration is beyond the ``interaction interval'' where the interaction between the system and electromagnetic field is zero.

Next we check the consistency of our results by calculating the change in energy of the reservoir. Since each reservoir decay is independent, the total change in energy of the reservoir should be given by 
\begin{equation}
    \Delta E_\mathrm{R} = \Delta E_\mathrm{R}^{ba} + \Delta E_\mathrm{R}^{cb} + \Delta E_\mathrm{R}^{cd} + \Delta E_\mathrm{R}^{da},
\end{equation}
where $\Delta E_\mathrm{R}^{pq}$ is the change in energy of the reservoir corresponding to the decay from $\ket{p}\to\ket{q}$ (p>q). Following the derivation for the absorption \cite{PhysRevA.106.023115}, the change in energy of each reservoir is given by
\begin{equation}
    \begin{split}
        \Delta E_\mathrm{R}^{pq} &= \langle \Psi(t_F)|H_\mathrm{R}^{pq}|\Psi(t_F)\rangle - \langle \Psi(t_I)|H_\mathrm{R}^{pq}|\Psi(t_I)\rangle\\
        & = \int\limits_{t_I}^{t_F} dt \bra{\Psi(t_0)}\frac{d}{dt} H_\mathrm{R}^{pq}(t)\ket{\Psi(t_0)}\\
        & = -\hbar\int\limits_{-\infty}^{\infty} dt \bra{\Psi_{\text{in}}} \eta_{pq}^*\bar\sigma_{pq}(t)\frac{d}{dt}\bar\psi_{pq}(0,t)\ket{\Psi_{\text{in}}} + \text{c.c.},
    \end{split}
\end{equation}
where in the second line we moved to the Heisenberg picture and in the last line took the limits of integration to infinity in the interaction picture in which we work. Quoting the result for the dynamics of $\check \psi_{pq}(z,t)$ \cite{PhysRevA.106.023115}, we have the relation
\begin{equation}
    \frac{d}{dt}\bar\psi_{pq}(0,t) = -i\omega_{pq}\bar\psi_{pq}(0,t) - i\eta_{pq}\bar\sigma_{qp}(t)\delta(0).
\end{equation}
Inputting this result and using the formal solution given by Eq. \eqref{eq:psibar solution}, we find
\begin{equation}
    \begin{split}
        &\eta_{pq}^*\bar\sigma_{pq}(t)\frac{d}{dt}\bar\psi_{pq}(0,t) + \text{H.c.} \\
        &= -i\omega_{pq}\eta_{pq}^*\bar\sigma_{pq}(t)\bar\psi_{pq}(0,t) + \text{H.c.}\\
        &=-i\omega_{pq}\eta_{pq}^*\bar\sigma_{pq}(t)\hat \psi_{pq}(0,t) - \omega_{pq}\frac{\Gamma_{pq}^\text{nr}}{2}\bar\sigma_{pp}(t) + \text{H.c.},
    \end{split}
\end{equation}
which leads to
\begin{equation}
    \begin{split}
        \Delta E_\mathrm{R}^{pq} &= i\hbar\omega_{pq}\int\limits_{-\infty}^{\infty}dt\bra{\Psi_\text{in}}\eta_{pq}^*\bar\sigma_{pq}(t)\hat \psi_{pq}(0,t)\ket{\Psi_\text{in}} + c.c. + \hbar\omega_{pq}\Gamma_{pq}^\text{nr}\int dt \langle \bar\sigma_{pp}(t)\rangle\\
        &=\hbar\omega_{pq}\Gamma_{pq}^\text{nr}\int dt \langle \bar\sigma_{pp}(t)\rangle,
    \end{split}
\end{equation}
where in the second line we used that the initial state of the reservoir is vacuum and so $\hat \psi_{pq}(0,t)\ket{\Psi_\text{in}} = 0$. Combining all contributions from the different transitions and comparing to Eq. \eqref{eq:appA} for the absorption, we find an exact agreement, with $\Delta E_\mathrm{R} = \mathcal{A}$ and all the energy lost from the electromagnetic field going to the reservoir. 

\section{Absorption check}\label{app:Absorption check}
For simplicity, consider only one term in the expansion of the absorption corresponding to the transition from $\ket{b}\to \ket{a}$, which is given by
\begin{equation}
    \mathcal{A}_{ba}^\text{exact} = \int\limits_{-\infty}^\infty\bra{\Psi_\text{in}}\bar{\boldsymbol{E}}^{ba}_-(t)\cdot \boldsymbol{\mu}_{ab}\frac{d\bar{\sigma}_{ab}(t)}{dt}\ket{\Psi_\text{in}}dt + \text{c.c.}.
\end{equation}
Note that this equation for the absorption is exact with no approximation on the time derivative of the dipole moment. Continuing with the exact form of Eq. \eqref{eq:sigmabarapprox}, we expand $\mathcal{A}_{ba}$ as
\begin{equation}
\label{eq:C2}
    \begin{split}
        \mathcal{A}_{ba}^\text{exact} &= -i\omega_{ba}\int\limits_{-\infty}^\infty\bra{\Psi_\text{in}}\bar{\boldsymbol{E}}^{ba}_-(t)\cdot \boldsymbol{\mu}_{ab}\bar{\sigma}_{ab}(t)\ket{\Psi_\text{in}}dt + \text{c.c.}\\
        &+\int\limits_{-\infty}^\infty\bra{\Psi_\text{in}}\bar{\boldsymbol{E}}^{ba}_-(t)\cdot \boldsymbol{\mu}_{ab}e^{-i\omega_{ba}t}\frac{d\check{\sigma}_{ab}(t)}{dt}\ket{\Psi_\text{in}}dt + \text{c.c.},
    \end{split}
\end{equation}
where the first term is the equation we use for the calculated absorption in Eq. \eqref{app:eq:Afreqband v1} which we put as $\mathcal{A}_{ba}^\text{app.}$ and denotes the approximate calculation. Inputting the Fourier transform for the positive and negative frequency components into the second term on the right-hand side of Eq. \eqref{eq:C2}, the difference between the exact and approximate absorption is given by
\begin{equation}
    \begin{split}
        \mathcal{A}_{ba}^\text{exact} - \mathcal{A}_{ba}^\text{app.}& = \int\limits_{-\infty}^\infty\bra{\Psi_\text{in}}\int\limits_{0}^{\infty}\bar{\boldsymbol{E}}^{ba}_-(-\omega)e^{i\omega t}\frac{d\omega}{\sqrt{2\pi}}\cdot \boldsymbol{\mu}_{ab}e^{-i\omega_{ba}t}\int\limits_0^\infty(-i\omega^\prime)\check{\sigma}_{ab}(\omega^\prime)e^{-i\omega^\prime t}\frac{d\omega^\prime}{\sqrt{2\pi}}\ket{\Psi_\text{in}}dt + \text{c.c.}\\
        &=\int\limits_{0}^{\infty}d\omega\int\limits_{0}^{\infty}d\omega^\prime \bra{\Psi_\text{in}}\bar{\boldsymbol{E}}^{ba}_-(-\omega)\cdot \boldsymbol{\mu}_{ab}(-i\omega^\prime)\check{\sigma}_{ab}(\omega^\prime)\ket{\Psi_\text{in}}\delta(\omega - \omega_{ba} - \omega^\prime) + \text{c.c.} \\
        &=-i\int\limits_{0}^{\infty}d\omega (\omega - \omega_{ba}) \bra{\Psi_\text{in}}\bar{\boldsymbol{E}}^{ba}_-(-\omega)\cdot \boldsymbol{\mu}_{ab}\check{\sigma}_{ab}(\omega - \omega_{ba})\ket{\Psi_\text{in}} + \text{c.c.}.
    \end{split}
\end{equation}
Now $\boldsymbol{\mu}_{ab}\check{\sigma}_{ab}(\omega - \omega_{ba})$ is an operator evolving in a frame shifted by $\omega_{ba}$ and so is peaked when its argument is zero, i.e., when $\omega\approx\omega_{ba}$. On the other hand, $\bar{\boldsymbol{E}}^{ba}_-(-\omega)$ is peaked at the center frequency of the field, say $\omega_0$. Within the RWA, when $\omega_0\sim \omega_{ba}$ and has a bandwidth that is much less than the center frequency, clearly the term $(\omega - \omega_{ba})$ in the integral implies that most frequency contributions to the integral will be negligible. So to very good approximation, we have
\begin{equation}
    \mathcal{A}_{ba}^\text{exact} \approx \mathcal{A}_{ba}^\text{app.},
\end{equation}
as is argued in Section \ref{sec:Scattered and Absorbed energy}.

\section{Perturbation theory \label{app:Perturbation theory}}
In this section, we start with the formal solutions to the relevant system operators (for the general form see Eq. \eqref{eq:exact solution to sigma_ij}) that will contribute to the population $\sbar{cc}$ at fourth order:
\begin{subequations}
    \begin{gather}
        \sbar{ab} = \int\limits_{-\infty}^t G_{ba}(t - t_1)\left[\sbari{aa}\fhati{ba}{+} - \sbari{bb}\fhati{ba}{+} + \fhati{bc}{-}\sbari{ac}\right]dt_1\\
        \sbar{ac} = \int\limits_{-\infty}^t G_{ca}(t - t_1)\left[ \sbari{ab}\fhati{cb}{+} - \sbari{bc}\fhati{ba}{+}\right]dt_1,\\
        \sbar{bc} = \int\limits_{-\infty}^t G_{cb}(t - t_1)\left[\sbari{bb}\fhati{cb}{+} - \sbari{cc}\fhati{cb}{+} - \fhati{ab}{-}\sbari{ac} \right]dt_1,\\
         \sbar{bb} = \int\limits_{-\infty}^t G_{bb}(t - t_1)\left[\sbari{ba}\fhati{ba}{+} - \fhati{ab}{-}\sbari{ab} + \fhati{bc}{-}\sbari{bc} - \sbari{bc}\fhati{cb}{+} -i\Gamma_{cb}\sbari{cc}\right]dt_1,\\
         \sbar{cc} = \int\limits_{-\infty}^t G_{cc}(t - t_1)\left[\sbari{cb}\fhati{cb}{+} - \fhati{bc}{-}\sbari{bc} \right]dt_1.
    \end{gather}
\end{subequations}
We begin the perturbative iteration by inputting the only nonzero zeroth order solution solution, $\bar\sigma_{aa}^{(0)}(t) = \hat 1$, into the formal solution for $\sbar{ab}$, which is the only nonzero solution at first order. Then
\begin{equation}
    \begin{split}
        \bar\sigma_{ab}^{(1)}(t) &= \int\limits_{-\infty}^t G_{ba}(t - t_1)\fhati{ba}{+}dt_1,\\
        &=\int\limits_{0}^{\infty}G_{ba}(\tau)\hat{F}_+^{ba}(t - \tau)d\tau,\\
        &=\int \dbarw{\mathrm{I}}G_{ba}(\omega_\mathrm{I}) \hat{F}_+^{ba}(\omega_\mathrm{II})e^{-i\omega_\mathrm{I} t},
    \end{split}
\end{equation}
where in the second line we changed variables and in the last line we inputted the inverse Fourier transform for $\hat{F}_+^{ba}(t - \tau)$, and did the time integral by defining
\begin{equation}
    \begin{split}
        G_{ij}(\omega) &\equiv \int\limits_0^\infty G_{ij}(\tau)e^{i\omega \tau}d\tau\\
        & = \frac{1}{\omega_{ij} - \omega -i\frac{\Gamma_{i}}{2} - i\frac{\Gamma_{j}}{2}}.
    \end{split}
\end{equation}
We proceed by inputting the first order solution $\bar\sigma_{ab}^{(1)}(t)$ into the 
formal solution to $\sbar{bb}$ and $\sbar{ac}$ and repeat the same steps of inputting the inverse Fourier transform and evaluating the time integral, then
\begin{subequations}
    \begin{gather}
        \bar\sigma_{bb}^{(2)}(t) = \int\dbarw{\mathrm{I}}^\prime\dbarw{\mathrm{I}}G_{ba}^*(\omega_\mathrm{I}^\prime)G_{ba}(\omega_\mathrm{I})\hat{F}_-^{ab}(-\omega_\mathrm{I}^\prime)\hat{F}_+^{ba}(\omega_\mathrm{I})e^{-i(\omega_\mathrm{I} - \omega_\mathrm{I}^\prime)t}\\
        \bar\sigma_{ac}^{(2)}(t) = \int\dbarw{\mathrm{II}}\dbarw{\mathrm{I}}G_{ba}(\omega_\mathrm{I})G_{ca}(\omega_\mathrm{I} + \omega_\mathrm{II})\hat{F}_+^{ba}(\omega_\mathrm{I})\hat F_+^{cb}(\omega_\mathrm{II})e^{-i(\omega_\mathrm{I} + \omega_\mathrm{II})t},    
\end{gather}
\end{subequations}
where in determining $\bar\sigma_{bb}^{(2)}(t)$ we combined terms using
\begin{equation}
    G^*_{ba}(\omega_\mathrm{I}^\prime) - G_{ba}(\omega_\mathrm{I}) = \frac{G^*_{ba}(\omega_\mathrm{I}^\prime)G_{ba}(\omega_\mathrm{I})}{G_{bb}(\omega_\mathrm{I} - \omega_\mathrm{I}^\prime)}.
\end{equation}
The second-order terms $\bar\sigma_{bb}^{(2)}(t)$, $\bar\sigma_{ac}^{(2)}(t)$, together with $\bar\sigma_{aa}^{(2)}(t) = - \bar\sigma_{bb}^{(2)}(t)$, are the only nonzero terms at second-order. Since we are only interested in the final state population $\bar\sigma_{cc}^{(4)}(t)$ the only third order term we need to calculate is $\bar\sigma_{bc}^{(3)}(t)$, which is given by
\begin{equation}
    \bar\sigma_{bc}^{(3)}(t) = \int\dbarw{\mathrm{I}}^\prime\dbarw{\mathrm{I}}\dbarw{\mathrm{II}}G_{ba}^*(\omega_\mathrm{I}^\prime)G_{ba}(\omega_\mathrm{I})G_{ca}(\omega_\mathrm{I} + \omega_\mathrm{II})\hat{F}_-^{ab}(-\omega_\mathrm{I}^\prime)\hat{F}_+^{ba}(\omega_\mathrm{I})\hat{F}_+^{cb}(\omega_\mathrm{II})e^{-i(\omega_\mathrm{I} + \omega_\mathrm{II} - \omega_\mathrm{I}^\prime)t},
\end{equation}
where we combined terms using
\begin{equation}
    G^*_{ba}(\omega_\mathrm{I}^\prime) - G_{ca}(\omega_\mathrm{I} + \omega_\mathrm{II} ) = \frac{G^*_{ba}(\omega_\mathrm{I}^\prime)G_{ca}(\omega_\mathrm{I} + \omega_\mathrm{II} )}{G_{cb}(\omega_\mathrm{I} + \omega_\mathrm{II} - \omega_\mathrm{I}^\prime)}.
\end{equation}
Finally, we input $\bar\sigma_{bc}^{(3)}(t)$ into the right-hand side of the formal solution to $\bar\sigma_{cc}(t)$, which is given by
\begin{equation}
    \begin{split}
        \bar\sigma_{cc}^{(4)}(t) &= \int\dbarw{\mathrm{II}}^\prime\dbarw{\mathrm{I}}^\prime\dbarw{\mathrm{I}}\dbarw{\mathrm{II}}G_{ba}^*(\omega_\mathrm{I}^\prime)G_{ba}(\omega_\mathrm{I})G_{cc}(\omega_\mathrm{I} + \omega_\mathrm{II} - \omega_\mathrm{I}^\prime - \omega_\mathrm{II}^\prime)\left[G_{ca}(\omega_\mathrm{I} + \omega_\mathrm{II}) - G_{ca}^*(\omega_\mathrm{I}^\prime + \omega_\mathrm{II}^\prime)\right]\\
        &\hspace{40mm}\times\hat{F}_-^{bc}(-\omega_\mathrm{II}^\prime)\hat{F}_-^{ab}(-\omega_\mathrm{I}^\prime)\hat{F}_+^{ba}(\omega_\mathrm{I})\hat{F}_+^{cb}(\omega_\mathrm{II}) e^{-i(\omega_\mathrm{I} + \omega_\mathrm{II} - \omega_\mathrm{I}^\prime - \omega_\mathrm{II}^\prime)t}.
    \end{split}
\end{equation}

Inputting $\bar\sigma_{cc}^{(4)}(t)$ into Eq. \eqref{eq:pexcitation} for the total probability of excitation and doing both time integrals, we arrive at our final result
\begin{equation}
    \begin{split}
        p =& 2\pi\hspace{-1.5mm}\int\hspace{-1.5mm}\dbarw{\mathrm{II}}^\prime\dbarw{\mathrm{I}}^\prime\dbarw{\mathrm{I}}\dbarw{\mathrm{II}}G_{ba}^*(\omega_\mathrm{I}^\prime)G_{ba}(\omega_\mathrm{I})2\mathrm{Im}\left[G_{ca}(\omega_\mathrm{I} + \omega_\mathrm{II})\right]\delta(\omega_\mathrm{I} + \omega_\mathrm{II} - \omega_\mathrm{I}^\prime - \omega_\mathrm{II}^\prime)\\
        &\times\langle\hat{F}_-^{bc}(-\omega_\mathrm{II}^\prime)\hat{F}_-^{ab}(-\omega_\mathrm{I}^\prime)\hat{F}_+^{ba}(\omega_\mathrm{I})\hat{F}_+^{cb}(\omega_\mathrm{II})\rangle,
    \end{split}
\end{equation}
as in Eq. \eqref{eq:pwork}.

\section{Changing variables \label{app:changingvariables}}
We start with Eq. \eqref{eq:pexcfinal} for the total probability of excitation $\bar p$ and note that all integrals implicitly range from $0\to\infty$. In most cases, the transition frequencies ($\omega_{ba}, \omega_{ca}$ etc.) will be much larger than both the system linwidths ($\Gamma_{pq}$) and the incident field bandwidths. So to good approximation we are free to extend the lower limit of integration to $-\infty.$ With this modification we change variables according to
\begin{subequations}
\begin{gather*}
    \tilde{\omega}_\mathrm{I} = \frac{\1 + \2}{2},\hspace{5mm}\tilde{\omega}_\mathrm{II} = \frac{\1 - \2}{2},\hspace{5mm}\tilde{\omega}_\mathrm{I}^\prime = \frac{\1^\prime + \2^\prime}{2},\hspace{5mm}\tilde{\omega}_\mathrm{II}^\prime = \frac{\1^\prime - \2^\prime}{2},\\
    \int\limits_0^\infty d\1^\prime d\2^\prime d\1 d\2 \to  \int\limits_{-\infty}^\infty d\1^\prime d\2^\prime d\1 d\2 = 4\int\limits_{-\infty}^\infty d\tilde{\omega}_\mathrm{I}^\prime d\tilde{\omega}_\mathrm{II}^\prime d\tilde{\omega}_\mathrm{I}d\tilde{\omega}_\mathrm{II}.
\end{gather*}
\end{subequations}
Then the total excitation probability is given by
\begin{equation}
    \begin{split}
        \bar p & = \frac{\eta N}{(2\pi)^2}4\int\limits_{-\infty}^\infty d\tilde{\omega}_\mathrm{I}^\prime d\tilde{\omega}_\mathrm{II}^\prime d\tilde{\omega}_\mathrm{I}d\tilde{\omega}_\mathrm{II}G_{ba}^*(\tilde{\omega}_\mathrm{I}^\prime + \tilde{\omega}_\mathrm{II}^\prime)G_{ba}(\tilde{\omega}_\mathrm{I} + \tilde{\omega}_\mathrm{II})L(2\tilde{\omega}_\mathrm{I})\\
        &\hspace{30mm}\times
        \frac{\langle a_\mathrm{II}^\dagger(\tilde{\omega}_\mathrm{I}^\prime - \tilde{\omega}_\mathrm{II}^\prime)a_\mathrm{I}^\dagger(\tilde{\omega}_\mathrm{I}^\prime + \tilde{\omega}_\mathrm{II}^\prime)a_\mathrm{I}(\tilde{\omega}_\mathrm{I} + \tilde{\omega}_\mathrm{II})a_\mathrm{II}(\tilde{\omega}_\mathrm{I} - \tilde{\omega}_\mathrm{II})\rangle}{A_\mathrm{eff}^2}2\pi\delta(2(\tilde{\omega}_\mathrm{I} - \tilde{\omega}_\mathrm{I}^\prime))\\
        & = \frac{\eta N}{(2\pi)^2}2\int\limits_{-\infty}^\infty d\tilde{\omega}_\mathrm{I} d\tilde{\omega}_\mathrm{II}^\prime d\tilde{\omega}_\mathrm{II}G_{ba}^*(\tilde{\omega}_\mathrm{I} + \tilde{\omega}_\mathrm{II}^\prime)G_{ba}(\tilde{\omega}_\mathrm{I} + \tilde{\omega}_\mathrm{II})L(2\tilde{\omega}_\mathrm{I})\\
        &\hspace{30mm}\times\frac{\langle a_\mathrm{II}^\dagger(\tilde{\omega}_\mathrm{I} - \tilde{\omega}_\mathrm{II}^\prime)a_\mathrm{I}^\dagger(\tilde{\omega}_\mathrm{I} + \tilde{\omega}_\mathrm{II}^\prime)a_\mathrm{I}(\tilde{\omega}_\mathrm{I} + \tilde{\omega}_\mathrm{II})a_\mathrm{II}(\tilde{\omega}_\mathrm{I} - \tilde{\omega}_\mathrm{II})\rangle}{A_\mathrm{eff}^2}2\pi,
    \end{split}
\end{equation}
where in the second line we evaluated the delta function. We change variables again by setting $\omega_\mathrm{I} = \tilde{\omega}_\mathrm{I} + \tilde{\omega}_\mathrm{II}$, $\omega_\mathrm{I}^\prime = \tilde{\omega}_\mathrm{I} + \tilde{\omega}_\mathrm{II}^\prime$, and $\omega = \tilde\omega_{\mathrm{I}}$ ($d\tilde{\omega}_\mathrm{I} d\tilde{\omega}_\mathrm{II}^\prime d\tilde{\omega}_\mathrm{II} = d\1d\1^\prime d\omega$), then
\begin{equation}
    \begin{split}
        \bar p& = \frac{\eta N}{(2\pi)^2}\int\limits_{-\infty}^\infty d\1d\1^\prime d(2\omega)G_{ba}^*(\1^\prime)G_{ba}(\1)L(2\omega)\frac{\langle a_\mathrm{II}^\dagger(2\omega - \1^\prime)a_\mathrm{I}^\dagger(\1^\prime)a_\mathrm{I}(\1)a_\mathrm{II}(2\omega - \1)\rangle}{A_\mathrm{eff}^2}2\pi\\
        &=\eta N\int L(\omega) \left|\int d\1G_{ba}(\1)\frac{a_\mathrm{II}(\omega - \1)a_\mathrm{I}(\1)\ket{\Psi_\mathrm{in}}}{A_\mathrm{eff}}\right|^2 \frac{d\omega}{2\pi},
    \end{split}
\end{equation}
where in the last line we changed variables by setting $2\omega \to \omega$.

\section{Classical light: The CW limit}
\label{sec:app:Classical light: The CW limit}
Consider Eq. \eqref{eq:pulsed_classical_excitation_prob} for the excitation probability driven by two coherent fields
\begin{equation}
\label{eq:pulsed_classical_excitation_prob_appendix}
    \begin{split}
        \overline p^\mathrm{cl} = \eta \frac{|\alpha\one|^2}{A_\mathrm{eff}} \frac{|\alpha\two|^2}{A_\mathrm{eff}} \int L(\omega) \left|\int\limits G_{ba}(\omega_\mathrm{I})\varphi_\mathrm{I}(\omega_\mathrm{I})\varphi_\mathrm{II}(\omega - \omega_\mathrm{I})\dbarw{\mathrm{I}}\right|^2d\omega,
    \end{split}
\end{equation}
with general spectral amplitude functions $\varphi_J(\omega)$. To take the CW limit, we set each classical field to be oscillating at the respective center frequency $\bar\omega_J$ for a time $T$, then 
\begin{equation}
    \varphi_J(\omega) = \frac{1}{\sqrt{\Omega}}\mathrm{sinc}\left(\frac{(\omega - \bar\omega_J)\pi}{\Omega}\right),
\end{equation}
where we set $\Omega = 2\pi/T$ to be the effective bandwidth of the field. In the CW limit, $T\to\infty$ ($\Omega\to 0$) and the function $\varphi_J(\omega)$ is strongly peaked at $\bar\omega_J$, taking this limit and using the spectral amplitudes in Eq.  \eqref{eq:pulsed_classical_excitation_prob_appendix}, to good approximation, the excitation probability is given by
\begin{equation}
\label{eq:pulsed_classical_excitation_prob_appendix2}
    \begin{split}
        \overline p^\mathrm{cl}_\mathrm{cw} = \eta \frac{|\alpha\one|^2}{A_\mathrm{eff}} \frac{|\alpha\two|^2}{A_\mathrm{eff}} L(\bar\omega_\mathrm{I} + \bar\omega_\mathrm{II})|G_{ba}(\bar\omega_\mathrm{I})|^2\int \left|\int\limits \varphi_\mathrm{I}(\omega_\mathrm{I})\varphi_\mathrm{II}(\omega - \bar\omega_\mathrm{I})\dbarw{\mathrm{I}}\right|^2d\omega,
    \end{split}
\end{equation}
where the spectral response functions of the atom were evaluated at the center frequencies of the incident light. With the strongly peaked approximation, the frequency integrals in Eq. \eqref{eq:pulsed_classical_excitation_prob_appendix2} can be evaluated. First, $\varphi_\mathrm{II}$ is independent of $\bar\omega_\mathrm{I}$ and can pulled out of the first integral, squared, and then integrated to unity since the spectral amplitude functions are square normalized. The integral over $\bar\omega_\mathrm{I}$ is evaluated and equal to
\begin{equation}
    \left|\int\frac{ d\omega_\mathrm{I}}{\sqrt{2\pi}} \varphi_\mathrm{I}(\omega_\mathrm{I})\right|^2 = \frac{\Omega}{2\pi} = \frac{1}{T}.
\end{equation}
Putting everything together and dividing by another factor of $T$, the excitation probability \textit{rate} is given by
\begin{equation}
    r_\mathrm{cw}^\mathrm{cl} \equiv \frac{\overline p^\mathrm{cl}_\mathrm{cw}}{T} = F_\mathrm{I}F_\mathrm{II} \sigma(\bar\omega_\mathrm{I}, \bar\omega_\mathrm{II}),
\end{equation}
where we set $F_J = |\alpha_J|^2 / (A_\mathrm{eff}T)$ to be the effective photon flux for each field and
\begin{equation}
    \sigma(\omega_\mathrm{I}, \omega_\mathrm{II}) = \eta  L(\omega_\mathrm{I} + \omega_\mathrm{II})|G_{ba}(\omega_\mathrm{I})|^2
\end{equation}
is the classical two-photon cross-section in the CW limit. 

\section{Squeezing operator transformation \label{app:squeezing operator}}
We start with the squeezing operator given in Eq. \eqref{eq:squeezing operator} and define the operator
\begin{equation}
    \mathcal{O}^\dagger = \beta\int d\omega_\mathrm{I}d\omega_\mathrm{II}\gamma(\omega_\mathrm{I},\omega_\mathrm{II})a_\mathrm{I}^\dagger(\omega_\mathrm{I})a_\mathrm{II}^\dagger(\omega_\mathrm{II}),
\end{equation}
so that the squeezing operator is given by
\begin{equation}
    S = e^{\mathcal{O}^\dagger - \mathcal{O}}.
\end{equation}
Using the Baker-Hausdorff lemma \cite{sakurai1995modern}, we have
\begin{equation}
\label{eq:squeezed state transformation sum_v1}
        S^\dagger a_J(\omega_J)S = a_J(\omega_J) + \sum_{n=1}^\infty \frac{C_n^J(\omega_J)}{n!},
\end{equation}
where we set
\begin{equation}
\label{eq:recursive relation}
        C_{n+1}^J(\omega_J) = [\mathcal{O} - \mathcal{O}^\dagger,C_n^J(\omega_J)].
\end{equation}
Evaluating the commutator $C_1^J(\omega_J)$, we find
\begin{subequations}
    \begin{gather}
        C_1^J(\omega_J) = \beta\int d\omega_{\bar{J}} \alpha(\omega_J + \omega_{\bar{J}}^\prime)\phi(\omega_J, \omega_{\bar{J}}^\prime)a_{\bar J}^\dagger(\omega_{\bar{J}}^\prime),
    \end{gather}
\end{subequations}
where we take $\bar{J}$ to be the opposite of $J$. Then we approximate the result $C_1^J(\omega_J)$ using the fact that in the CW limit $ \alpha(\omega_J + \omega_{\bar{J}}^\prime)$ is strongly peaked when $\omega_J + \omega_{\bar{J}}^\prime = \bar\omega_J + \bar\omega_{\bar{J}}$; to good approximation
\begin{equation}
        C_1^J(\omega_J) \approx \beta\phi_J(\omega_J)\int d\omega_{\bar{J}}^\prime \alpha(\omega_J + \omega_{\bar{J}}^\prime)a_{\bar{J}}^\dagger(\omega_{\bar{J}}^\prime),
\end{equation}
where we used the definition in Eq. \eqref{eq:phi_J(w_J)}. Using the result for $C_1^J(\omega_J)$ and the recursive relation (Eq. \eqref{eq:recursive relation}), we evaluate the second commutator using the same approximation twice, and find
\begin{equation}
        C_2^J(\omega_J) \approx |\beta\phi_J(\omega_J)|^2\sqrt{\Omega_p}\int d\omega_J^\prime\alpha^*(\omega_J^\prime + \bar\omega_J + \bar\omega_{\bar{J}} - \omega_J)a_J(\omega_J^\prime),
\end{equation}
where we used 
\begin{equation}
    \int d\omega \alpha(\omega) = \sqrt{\Omega_p}.
\end{equation}
Continuing with each commutator we find 
\begin{equation}
\begin{split}
\label{eq:squeezed state transformation sum_v2}
        S^\dagger a(\omega_J) S &= a_J(\omega_J) + \sum_{n=1}^\infty \frac{|\beta \phi_J(\omega_J)\sqrt{\Omega_p}|^{2n}}{(2n)!}\frac{1}{\sqrt{\Omega_p}}\int d\omega_J^\prime\alpha^*(\omega_J^\prime + \bar\omega_J + \bar\omega_{\bar{J}} - \omega_J)a(\omega_J^\prime)\\
        &\hspace{35mm}+\frac{\beta \phi_J(\omega_J)}{|\beta \phi_J(\omega_J)|}\sum_{n=0}^\infty \frac{|\beta \phi_J(\omega_J)\sqrt{\Omega_p}|^{2n+1}}{(2n+1)!}\int d\omega_{\bar{J}}^\prime \alpha(\omega_J + \omega_{\bar{J}}^\prime)a_{\bar{J}}^\dagger(\omega_{\bar{J}}^\prime),
\end{split} 
\end{equation}
and using the Taylor series for $\text{sinh}(x)$ and $\text{cosh}(x)$, the squeezing transformation is equal to
\begin{equation}
\label{eq:sqtransform}
\begin{split}
        S^\dagger a_J(\omega_J)S &=  a_J(\omega_J) + \frac{(c_J(\omega_J) - 1)}{\sqrt{\Omega_p}}\int d\omega_J^\prime\alpha^*(\omega_J^\prime + \bar\omega_J + \bar\omega_{\bar{J}} - \omega_J)a_J(\omega_J^\prime)\\
        &\hspace{20mm}+ \frac{e^{i\theta_J(\omega_J)}s_J(\omega_J)}{\sqrt{\Omega_p}}\int d\omega_{\bar{J}}^\prime \alpha(\omega_J + \omega_{\bar{J}}^\prime)a^\dagger_{\bar{J}}(\omega_{\bar{J}}^\prime),
\end{split}
\end{equation}
where we set
\begin{equation}
\label{eq:scdef}
        s_J(\omega_J) = \text{sinh}(|\beta \phi_J(\omega_J)|\sqrt{\Omega_p}), \hspace{5mm}c_J(\omega_J) = \text{cosh}(|\beta \phi_J(\omega_J)|\sqrt{\Omega_p}),\hspace{5mm}e^{i\theta_J(\omega_J)} = \frac{\beta \phi_J(\omega_J)}{|\beta \phi_J(\omega_J)|},
\end{equation}
which following the properties of $\phi_J(\omega_J)$ satisfy
\begin{equation}
        s_{\bar{J}}(\bar\omega_J + \bar\omega_{\bar{J}} - \omega_J) = s_J(\omega_J)\hspace{5mm} \theta_{\bar{J}}(\bar\omega_J + \bar\omega_{\bar{J}} - \omega_J) = \theta_J(\omega_J).
\end{equation}
Although we do not include the calculation here, we 
find
\begin{subequations}
    \begin{gather}
        \left[S^\dagger a_J(\omega_J)S, S^\dagger a_{\bar{J}}^\dagger(\omega_{\bar{J}}^\prime)S\right] = \delta(\omega_J - \omega_{\bar{J}}^\prime)\delta_{J{\bar{J}}},
    \end{gather}
\end{subequations}
as expected, because the squeezing operator is unitary.

\section{Magnetic quantum number}
\label{app:Magnetic quantum number}
In this section we consider the matrix element between the atomic states $6\mathrm{S}_{1/2}$ $6\mathrm{P}_{1/2}$, and $7\mathrm{S}_{1/2}$, including the magnetic quantum number. We begin with the transition between $6\mathrm{S}_{1/2}$ and $6\mathrm{P}_{1/2}$ and expand the matrix element using the spherical basis
\begin{equation}
    \langle 6\mathrm{P}_{1/2} m^\prime|e\boldsymbol{r}|6\mathrm{S}_{1/2} m\rangle = \sum_{\eta = -1}^{\eta = 1}\hat{\boldsymbol{u}}_\eta \langle 6\mathrm{P}_{1/2} m^\prime|er^\eta|6\mathrm{S}_{1/2} m\rangle.
\end{equation}
The spherical basis is defined by
\begin{subequations}
\begin{gather}
     \hat{\boldsymbol{u}}_{\pm1} = \mp\frac{1}{\sqrt{2}}(\hat{\boldsymbol{x}} \pm i \hat{\boldsymbol{y}}),\hspace{5mm} \hat{\boldsymbol{u}}_0 =  \hat{\boldsymbol{z}},\hspace{5mm}  \hat{\boldsymbol{u}}_\eta = (-1)^\eta  \hat{\boldsymbol{u}}^*_{-\eta},\hspace{5mm}  \hat{\boldsymbol{u}}_\eta^*\cdot  \hat{\boldsymbol{u}}_{\eta^\prime} = \delta_{\eta \eta^\prime}\\
     A^{\pm1} = \mp\frac{1}{\sqrt{2}}(A^x \mp i A^y),\hspace{5mm} A^0 = A^z,
\end{gather}
\end{subequations}
with $\mathbf{A}$ being an arbitrary vector with Cartesian components labeled by superscripts $\{x,y,z\}$ and spherical components labeled by superscripts $\{-1,0,1\}$. The quantity $\langle 6\mathrm{P}_{1/2} m^\prime|er^\eta|6\mathrm{S}_{1/2} m\rangle$ is a rank one tensor that can be expanded using the Wigner–Eckart theorem in terms of the Wigner 3-j symbol \cite{messiah1962quantum}
\begin{equation}
    \langle 6\mathrm{P}_{1/2} m^\prime|er^\eta|6\mathrm{S}_{1/2} m\rangle = (-1)^{\frac{1}{2} - m^\prime}         
    \begin{pmatrix}
        \frac{1}{2} & 1 & \frac{1}{2}\\
        -m^\prime & \eta & m
    \end{pmatrix}
    \langle 6\mathrm{P}_{1/2}||er||6\mathrm{S}_{1/2}\rangle,
\end{equation}
where $\langle 6\mathrm{P}_{1/2}||er||6\mathrm{S}_{1/2}\rangle$ is the reduced matrix element.

Assuming the first transition is driven by linearly polarized light, we set $\boldsymbol{e}_\mathrm{I} = \hat{\boldsymbol{z}} = \hat{\boldsymbol{u}}_0$, then
\begin{equation}
\begin{split}
        \boldsymbol{e}_\mathrm{I}\cdot\langle 6\mathrm{P}_{1/2} m^\prime|e\boldsymbol{r}|6\mathrm{S}_{1/2} m\rangle &= \delta_{m^\prime m}(-1)^{\frac{1}{2} - m}         
    \begin{pmatrix}
        \frac{1}{2} & 1 & \frac{1}{2}\\
        -m & 0 & m
    \end{pmatrix}
    \langle 6\mathrm{P}_{1/2}||er||6\mathrm{S}_{1/2}\rangle\\
    & = \frac{\delta_{m^\prime m}}{\sqrt{6}}(-1)^{\frac{1}{2} - m}\langle 6\mathrm{P}_{1/2}||er||6\mathrm{S}_{1/2}\rangle,
\end{split}
\end{equation}
where in the first line we used the orthogonality of the Wigner-3j symbol and evaluated it for $m = \pm \frac{1}{2}$ in the second line. We note that for linearly polarized light, the magnetic quantum number of the excited intermediate state must match that of the initial ground state, as expected due to the conservation of angular momentum. Second, the absolute square of the transition matrix element is related to the reduced matrix element via
\begin{equation}
    |\boldsymbol{e}_\mathrm{I}\cdot\langle 6\mathrm{P}_{1/2} m^\prime|e\boldsymbol{r}|6\mathrm{S}_{1/2} m\rangle|^2 = \frac{\delta_{m^\prime m}}{6}|\langle 6\mathrm{P}_{1/2}||er||6\mathrm{S}_{1/2}\rangle|^2. 
\end{equation}
To set its value we evaluate the radiative decay rate of the excited intermediate state, given by
\begin{equation}
    \Gamma_{6\mathrm{P}_{1/2}6\mathrm{S}_{1/2}}^\mathrm{r} = \frac{\omega_{6\mathrm{P}_{1/2}6\mathrm{S}_{1/2}}^3}{3\pi\epsilon_0c^2\hbar}\sum_{m = -1/2}^{m = 1/2} |\langle 6\mathrm{P}_{1/2} m^\prime|e\boldsymbol{r}|6\mathrm{S}_{1/2} m\rangle|^2,
\end{equation}
where we sum over all the possible lower energy level states to which the state $|6\mathrm{P}_{1/2} m^\prime\rangle$ could decay. To evaluate the right-hand side we again make use of the Wigner–Eckart theorem 
\begin{equation}
    \begin{split}
        & \sum_{m = -1/2}^{m = 1/2} \langle 6\mathrm{P}_{1/2} m^\prime|e\boldsymbol{r}|6\mathrm{S}_{1/2} m\rangle\cdot \langle 6\mathrm{S}_{1/2}m|e\boldsymbol{r}|6\mathrm{P}_{1/2} m^\prime \rangle\\
        & = \sum_{m = -1/2}^{m = 1/2}\sum_{\eta = -1}^{\eta = 1}   
    \begin{pmatrix}
        \frac{1}{2} & 1 & \frac{1}{2}\\
        -m & \eta & m^\prime 
    \end{pmatrix}
    \begin{pmatrix}
        \frac{1}{2} & 1 & \frac{1}{2}\\
        -m & \eta & m^\prime
    \end{pmatrix}
    |\langle 6\mathrm{P}_{1/2}||er||6\mathrm{S}_{1/2}\rangle|^2\\
    & = \frac{1}{2}|\langle 6\mathrm{P}_{1/2}||er||6\mathrm{S}_{1/2}\rangle|^2,
    \end{split}
\end{equation}
where we made use of the double Wigner-3j orthogonality \cite{messiah1962quantum}; the total decay rate is given by
\begin{equation}
    \Gamma_{6\mathrm{P}_{1/2}6\mathrm{S}_{1/2}}^\mathrm{r} = \frac{\omega_{6\mathrm{P}_{1/2}6\mathrm{S}_{1/2}}^3}{3\pi\epsilon_0c^2\hbar}\frac{1}{2}|\langle 6\mathrm{P}_{1/2}||er||6\mathrm{S}_{1/2}\rangle|^2
\end{equation}
Combining the two results
\begin{equation}
\label{eq:matrixelement1}
    \begin{split}
    |\boldsymbol{e}_\mathrm{I}\cdot\langle 6\mathrm{P}_{1/2} m^\prime|e\boldsymbol{r}|6\mathrm{S}_{1/2} m\rangle|^2 & = \frac{\delta_{m^\prime m}}{6}|\langle 6\mathrm{P}_{1/2}||er||6\mathrm{S}_{1/2}\rangle|^2\\
    & = \frac{\delta_{m^\prime m}}{3} \left( \frac{\Gamma_{6\mathrm{P}_{1/2}6\mathrm{S}_{1/2}}^\mathrm{r} 3\pi\epsilon_0c^2\hbar}{\omega_{6\mathrm{P}_{1/2}6\mathrm{S}_{1/2}}^3}\right), 
    \end{split}
\end{equation}
which can be set by experiment. 

The second transition in the two-photon excitation involves the $6\mathrm{P}_{1/2}$ and $7\mathrm{S}_{1/2}$ energy levels. Since they both have the same total angular momentum label $j$, they both have the same magnetic quantum states. As a result if we set the second field to also be linearly polarized ($\boldsymbol{e}_\mathrm{II} = \hat{\boldsymbol{z}}$) all the results of the first transition carry over to the second. Skipping the details, we also have the relation
\begin{equation}
\label{eq:matrixelement2}
    |\boldsymbol{e}_\mathrm{II}\cdot\langle 7\mathrm{S}_{1/2} m^\prime|e\boldsymbol{r}|6\mathrm{P}_{1/2} m\rangle|^2  = \frac{\delta_{m^\prime m}}{3} \left( \frac{\Gamma_{7\mathrm{S}_{1/2}6\mathrm{P}_{1/2}}^\mathrm{r} 3\pi\epsilon_0c^2\hbar}{\omega_{7\mathrm{S}_{1/2}6\mathrm{P}_{1/2}}^3}\right), 
\end{equation}
which we use to set the magnitude of the transition matrix element.  Further, since the second field is also linearly polarized, the angular momentum of the second excited state must be the same as the intermediate state, and therefore the same as the ground state.  

To end our analysis, we point out that each transition matrix element is the same independent of what is the initial ground state. To take into account both excitation pathways we take the ground state to be in an equal mixture of the two states since they are degenerate, and sum over the two possibilities. But since the matrix elements are the same so are their respective excitation pathways. As a result, we effectively drop the magnetic quantum number and set the ground state to be $\ket{a} = \ket{6\mathrm{S}_{1/2}}$, the intermediate state to be $\ket{b} = \ket{6\mathrm{P}_{1/2}}$ and the two-photon excited state to be $\ket{c} = \ket{7\mathrm{S}_{1/2}}$; the corresponding matrix elements are set using Eq. \eqref{eq:matrixelement1} and \eqref{eq:matrixelement2}.  Finally, we set the fourth state to be $\ket{d} = \ket{7\mathrm{S}_{1/2}}$, and since light only decays to and from this state all that is required is the total decay rate, which is again set by experiment. 
\bibliography{bib}
\nocite{}
\end{document}